\documentclass[preprint2]{aastex}

\usepackage{lineno}
\usepackage{datetime}
\usepackage{color}

\newcommand{\F}{\textit{Fermi}}
\renewcommand{\P}{\textit{Planck}}
\newcommand{\g}{$\gamma$}

\newcommand{\nhi}{N(\mathrm{H\,\scriptstyle{I}})}

\newcommand{\nhd}{N(\mathrm{H_2})}
\newcommand{\wco}{W_\mathrm{CO}}
\newcommand{\hi}{\mathrm{H\,\scriptstyle{I}}}

\newcommand{\hd}{\mathrm{H}_2}
\newcommand{\xco}{X_\mathrm{CO}}
\newcommand{\tmap}{\tau_{353}}
\newcommand{\like}{\mathcal{L}}
\newcommand{\kms}{km~s$^{-1}$}
\newcommand{\zm}{$z_\mathrm{max}$}

\newcommand{\changes}{}
\newcommand{\chbis}{}

\slugcomment{Draft compiled on \usdate\today, at \currenttime}

\shorttitle{\F-LAT observations of HVCs and IVCs}
\shortauthors{\F-LAT collaboration}

\begin{document}


\title{\textit{Fermi}-LAT Observations of High- and Intermediate-Velocity Clouds: Tracing Cosmic Rays in the Halo of the Milky Way}


\author{
L.~Tibaldo\altaffilmark{1,2},
S.~W.~Digel\altaffilmark{1,3}, 
J.~M.~Casandjian\altaffilmark{4}, 
A.~Franckowiak\altaffilmark{1}, 
I.~A.~Grenier\altaffilmark{4}, 
G.~J\'ohannesson\altaffilmark{5}, 
D.J.~Marshall\altaffilmark{4}, 
I.~V.~Moskalenko\altaffilmark{1}, 
M.~Negro\altaffilmark{1,6}, 
E.~Orlando\altaffilmark{1}, 
T.~A.~Porter\altaffilmark{1}, 
O.~Reimer\altaffilmark{7,1}, 
A.~W.~Strong\altaffilmark{8}
}
\altaffiltext{1}{W. W. Hansen Experimental Physics Laboratory, Kavli Institute for Particle Astrophysics and Cosmology, Department of Physics and SLAC National Accelerator Laboratory, Stanford University, Stanford, CA 94305, USA}
\altaffiltext{2}{email: ltibaldo@slac.stanford.edu}
\altaffiltext{3}{email: digel@stanford.edu}
\altaffiltext{4}{Laboratoire AIM, CEA-IRFU/CNRS/Universit\'e Paris Diderot, Service d'Astrophysique, CEA Saclay, 91191 Gif sur Yvette, France}
\altaffiltext{5}{Science Institute, University of Iceland, IS-107 Reykjavik, Iceland}
\altaffiltext{6}{Istituto Nazionale di Fisica Nucleare, Sezione di Torino, I-10125 Torino, Italy}
\altaffiltext{7}{Institut f\"ur Astro- und Teilchenphysik and Institut f\"ur Theoretische Physik, Leopold-Franzens-Universit\"at Innsbruck, A-6020 Innsbruck, Austria}
\altaffiltext{8}{Max-Planck Institut f\"ur extraterrestrische Physik, 85748 Garching, Germany}

\begin{abstract}
{\changes
It is widely accepted that cosmic rays (CRs) up to at least PeV energies are Galactic in origin. Accelerated particles are injected into the interstellar medium where they propagate to the farthest reaches of the Milky Way, including a surrounding halo. The composition of CRs coming to the solar system can be measured directly and has been used to infer the details of CR propagation that are extrapolated to the whole Galaxy. In contrast, indirect methods, such as observations of \g-ray emission from CR interactions with interstellar gas, have been employed to directly probe the CR densities in distant locations throughout the Galactic plane. In this article we use 73 months of data from the \F~Large Area Telescope in the energy range between 300~MeV and 10~GeV  to search for \g-ray emission produced by CR interactions in several high- and intermediate-velocity clouds located at up to $\sim 7$~kpc above the Galactic plane. We achieve the first detection of intermediate-velocity clouds in \g~rays and set upper limits on the emission from the remaining targets, thereby tracing the distribution of CR nuclei in the halo for the first time. We find that the {\chbis \g-ray} emissivity {\chbis per H~atom} decreases with increasing distance from the plane at 97.5\% confidence level. This corroborates the notion that CRs at the relevant energies originate in the Galactic disk. The emissivity of the upper intermediate-velocity Arch hints at a 50\% decline of CR densities within 2~kpc from the plane. We compare our results to predictions of CR propagation models.
}
\end{abstract}



\section{INTRODUCTION}

{\changes
Cosmic rays (CRs) are a wide-spread phenomenon, pervading galaxies and intergalactic space.
They are directly detected within the solar system with spacecraft, balloon-borne, and ground-based instruments. Other methods to study CRs include observations of radio synchrotron emission and \g~rays produced in interactions of {\chbis CRs} with interstellar gas, radiation and magnetic fields. CRs consist of all known stable and long-lived isotopes ranging from the most abundant hydrogen and helium nuclei through the traces of very rare Actinides, and particles such as electrons, positrons, and antiprotons.
The energy density of CRs in the interstellar medium (ISM) 
is comparable to {\chbis those of} the Galactic interstellar radiation field, magnetic field, and turbulent motions of the interstellar gas. {\chbis CRs} are one of the essential components 
of the Galactic ecosystem \citep[e.g.,][]{ferriere2001}, 
{\chbis affecting} the thermal, chemical and magnetohydrodynamical state of the 
ISM \citep[e.g.,][]{ptuskin2006,indriolo2009}. Emission 
from CRs interacting within the Milky Way is also a source of foregrounds{\changes/backgrounds} for 
many observations spanning from radio to \g~rays.

{\changes Observations of X-ray and \g-ray emission from Galactic objects such as supernova remnants, pulsars and massive-star clusters reveal the presence of energetic particles, suggesting that efficient acceleration processes occur in their vicinities \citep[{\chbis for recent reviews see, e.g.,}][]{reynolds2008,bykov2014}. The Galactic origin of CRs is also supported by \g-ray observations of external galaxies, which indicate that the CR densities are not universal \citep[e.g.,][]{sreekumar1993} and are correlated with the star-formation properties of the galaxies \citep[e.g.,][]{LATSFGalaxies}.}
Also, the radial sizes of radio halos 
measured in external, edge-on spiral 
galaxies appear to be correlated with active star formation in their disks 
\citep{dahlem1995}.

{\chbis CR particles accelerated in sources in the Galactic disk are injected into the ISM where they propagate through a surrounding halo before escaping into the intergalactic space {\citep[e.g.,][]{strong2010}}.}
This paradigm is substantiated in models  
that can reproduce reasonably well all the related observables
\citep[{\chbis for a recent review see, e.g.,}][]{strong2007}. 
{\changes The propagation of particles in the ISM leads to the destruction of primary nuclei via spallation and gives rise to secondary nuclei and isotopes which are rare in nature, and other particles. The observed abundances of stable secondary CR nuclei (e.g., boron) and radioactive isotopes ($^{10}$Be, $^{26}$Al, $^{36}$Cl, $^{54}$Mn) allow the separate determination of the confinement halo size and diffusion coefficient \citep[e.g.,][]{strong1998}
which is somewhat model dependent. {\chbis Typical constraints on the halo height arising from such method} are 4--6~kpc
{\chbis \citep[e.g.,][]{moskalenko2001,trotta2011}}.}

Observations of edge-on external galaxies show directly the existence of radio halos extending to kpc distances perpendicularly to their disks \citep[e.g.,][]{dahlem1994}, and sometimes up to $\sim 15$~kpc \citep{irwin1999}.
Radio observations of large-scale synchrotron emission from the Milky Way were also interpreted in the context of diffusion models as 
indicating a vertical 
extent of the halo of $4$~kpc to $10$~kpc 
\citep[e.g.,][]{strong2000}, with values of $\sim 10$~kpc favored based on a 
recent analysis including \textit{WMAP} {\changes and 408~MHz observations}  \citep{orlando2013}. {\changes The discrepancies between {\chbis halo sizes} determined with different methods are not necessarily surprising. The method based on the direct detection of CR nuclei provides the size of the effective volume filled with CRs that {are still} confined to the Milky Way. 
In contrast, observations of synchrotron emission show larger halos because part of the emission may be generated by particles that are leaking into the intergalactic space, i.e., that are already beyond the ``point of no return''.}
%
%

{\changes The Large Area Telescope (LAT) on the {\it Fermi Gamma-ray Space Telescope} 
mission \citep{atwood2009} has been surveying the sky since 2008 {\chbis in} a wide energy range of 20~MeV to $ > 300$~GeV.
These observations {\chbis provide information about CR particles} (mostly protons and electrons) in distant locations that can be used to study CR propagation \citep[e.g.,][]{LATdiffpapII}.
Measurements 
with the LAT confirmed earlier observations \citep[e.g.,][]{strong1988} that the radial gradient  of the \g-ray emissivity (\g-ray emission 
rate per hydrogen atom) in the ISM outside the solar circle is much smaller than that in the density of putative CR sources \citep{abdo2010cascep,ackermann20113quad}. This can be interpreted as
another hint of a $\sim 10$~kpc CR confinement halo, {\chbis because it facilitates radial diffusion of CRs and compensates for the paucity of CR sources in the outer Galaxy.}

In this work we propose a more direct method to constrain the propagation of 
CRs 
in the halo of the Milky Way with LAT data.  Although the interstellar gas in 
the Milky Way is largely confined to the Galactic plane with a scale height for 
atomic hydrogen of $\lesssim 300$~pc across much of the disk 
\citep{kalberla2009}, certain populations of 
interstellar clouds are known to exist at much greater distances from the 
plane.
So-called high-velocity clouds (HVCs) and intermediate-velocity clouds (IVCs)
are interstellar clouds of atomic hydrogen, $\hi$, with line-of-sight 
velocities inconsistent with Galactic rotation \citep[see, 
e.g.,][]{hvcassl2005}.
The demarcation between HVCs and IVCs is conventionally taken to be Doppler 
shifted 
velocity with absolute value between $80$~km~s$^{-1}$ and 
$100$~km~s$^{-1}$ with respect to the local standard of rest. 

The distances and 
metallicities of HVCs/IVCs can be inferred from spectroscopic observations of 
foreground and background stars.  HVCs typically have metallicities less 
than Solar and are generally considered to be remnants of the {\chbis formation of the Milky Way},  
having existed in the 
halo of the {\chbis Galaxy} since the disk formed, or old tidal streams extracted 
from nearby dwarf galaxies  \citep{wakker2001b}. Some IVCs have metallicities 
much 
closer to Solar and these may be gas ejected from the plane, e.g., by 
Galactic fountains \citep{wakker2001b}.  No molecular gas, e.g., CO emission,
has been detected in HVCs \citep{wakker1997b}, and IVCs are also typically free 
of CO emission as well.

These clouds have low column densities, $\nhi \sim 
10^{19}-10^{20}$ cm$^{-2}$, but can subtend hundreds of square degrees. The 
exposure of the LAT sky survey is deep enough for HVCs/IVCs to conceivably 
be detected in diffuse \g-ray emission and used to estimate the CR densities in 
the Galactic halo. These clouds are {\chbis not actively forming stars and are believed} to be free of internal sources of {\chbis CRs} and hence any {\g-ray} emission from them samples the large-scale distribution of Galactic CRs.

{\changes The article is structured as follows.}
In Section~\ref{targets} we describe our criteria for 
selecting a sample of HVCs/IVCs for study.
In Section~\ref{data} we discuss the 
\g-ray and multiwavelength survey data we use.  Section~\ref{ismmaps} provides 
the details of how we derive the $\hi$ maps,
in particular how we fit spectral components with different line-of-sight 
velocities to achieve kinematic separation of interstellar gas by distance along 
the line of sight, and other maps of ISM tracers to accurately account for 
foreground gas in the local neighborhood.  In Section~\ref{analgamma} we 
introduce the \g-ray analysis procedure to {\chbis measure or} constrain the CR densities in the 
targets, and the procedure we implemented to estimate systematic uncertainties. 
In Section~\ref{resdiscussion} we discuss the results for the individual regions 
that we studied and the implications for the propagation of CRs in the halo of 
the Milky Way.  The conclusions are in Section~\ref{conclusions}{\chbis.} 
Appendix~\ref{dustresults} provides more technical information about {\chbis the derivation of}  
the maps of the ISM,
including an iterative approach to unbiasing the maps of the column 
density in the dark neutral medium (DNM). {\changes Appendix~\ref{jackknifeerrs} describes how we computed uncertainties and upper limits with the jackknife method for {\chbis evaluating} the systematic uncertainties.}

\section{TARGETS}\label{targets}
\citet{wakker2001} reports the most complete census to date of IVCs and 
HVCs, including lower and upper limits on their distances (which we refer to as distance brackets),
from spectrophotometric distances of foreground and background stars.  
In order to be able to constrain the 
CR densities in well-defined regions of the Milky Way halo, we select as 
targets for our analysis clouds in \citet{wakker2001} for which
distance (altitude) brackets have been determined. We summarize the selected targets in 
Table~\ref{targettable}, grouped in three regions of interest (ROIs) as used 
for the analysis later. In each case the ROIs enclose the parts of 
the 
clouds associated with the foreground and background stars used for determining 
the distance brackets.

\begin{deluxetable}{lccccccc}
\tabletypesize{\scriptsize}
\tablecaption{Target regions and complexes\label{targettable}}
\tablewidth{0pt}
\tablehead{
\colhead{Region} & \multicolumn{2}{c}{ROI} & \colhead{Complex} &\colhead{$d$} & 
\colhead{$z$} & \colhead{Metallicity} & \colhead{$v_\mathrm{LSR}$}  \\
\colhead{} & \colhead{Center $(l,b)$} & \colhead{Pixels} & \colhead{} &
\colhead{(kpc)} & \colhead{(kpc)} &\colhead{(relative to solar)} & \colhead{(\kms)}
}
\startdata
A & $(155\arcdeg,37\arcdeg)$ & $161,81$ & Low-Latitude IV Arch & 0.9--1.8 & 
0.6--1.2 & $1.0\pm0.5$ &$-90 <v_\mathrm{LSR} < -40$\\
 & & & Complex~A & 4.0--9.9 & 2.6--6.8& 0.02--0.4 & $v_\mathrm{LSR}<-90$
\\
\tableline
B & $(182\arcdeg,57\arcdeg)$ & $125,141$ &  Lower IV Arch & 0.4--1.9 & 0.4--1.7
& $\sim 1$ & $-60 <v_\mathrm{LSR} < -15$\\
 & & &  Upper IV Arch & 0.8--1.8 & 0.7--1.7& $\sim 1$ &$v_\mathrm{LSR}<-60$\\
\tableline
C & $(251\arcdeg,69\arcdeg)$ & $85,117$ & IV Spur & 0.3--2.1 & 0.3--2.1& \nodata & $-90 
<v_\mathrm{LSR} < -20$\\
\enddata
\tablecomments{The ROIs are defined as rectangular regions in a \textit{plate 
carr\'ee} projection in Galactic coordinates, centered at each ROI center 
position and with the number of pixels of $0\fdg25$ width given above. See 
text for details.}
\tablecomments{The brackets in $d$ and $z$, {\changes and the metallicities} are taken from 
\citet{wakker2001}. {\chbis The metallicity of the IV Spur is unknown.}}
\tablecomments{The ranges in velocities with respect to the local standard of 
rest, $v_\mathrm{LSR}$, are the preliminary boundaries that we use in 
Section~\ref{himaps} to construct the $\nhi$ maps.}
\end{deluxetable}

Region~A encompasses an IVC known as the low-latitude intermediate velocity 
(IV) Arch, and an HVC known 
as Complex~A.  The low-latitude IV Arch has a mass of $1.5-6 \times 10^5 \;
M_\odot$ at a distance of $0.9-1.8$~kpc from the Sun, corresponding to a distance above the plane of $z=0.6-1.2$~kpc. 
According to \citet{wakker2001} the distance (comparable to that of the 
Perseus arm of the Galaxy), the metallicity (close to solar), and negative 
velocities in the range of $-20$~km~s$^{-1}$ to $-30$~km~s$^{-1}$ (higher than the Perseus arm 
itself in this direction) are all compatible with the characteristics of a 
high-interarm cloud that is part of the return flow of gas injected from the 
disk into the halo by a fountain process. Complex~A is a distant 
HVC with the narrowest distance bracket\footnote{HVC Complex~C has a broader 
distance bracket of $3.7-11.2$~kpc \citep{wakker2007}. We leave it for 
future studies.} of $4.0-9.9$~kpc 
($z=2.6-6.8$~kpc), which implies a mass of $(0.3-2) \times 10^{6} \; M_\odot$. 
Its low metallicity suggests that this HVC is a relic of the ancient 
Universe, a clump of gas that never underwent star formation and is now falling 
into the Milky Way.

Region~B encompasses two IVCs that are together known as the IV Arch. The gas 
in the IV Arch is seen in two distinct velocity ranges: $> 
-60$~km~s$^{-1}$ (the lower IV Arch) and $<-60$~km~s$^{-1}$ (the upper IV 
Arch\footnote{The nomenclature refers to the absolute value of the velocity.}). 
We consider in this analysis only the portion of the IV Arch seen at Galactic 
longitudes $l>150\arcdeg$, for which there is a distance bracket for the lower 
part of the Arch of $0.4-1.9$~kpc ($z=0.4-1.7$~kpc). At lower longitudes the 
distance is less reliably determined, and part of the lower IV Arch may be 
associated with a CO core with distance $<0.3$~kpc from the solar system 
\citep{wakker2001}. On the other 
hand, the upper IV Arch has a distance bracket of $0.8-1.8$~kpc 
($z=0.7-1.7$~kpc). Therefore, whether the two velocity 
components of the IV Arch are also physically separated entities is uncertain. The total 
mass of the IV Arch is of the order of several $10^5 \; M_\odot$, and its 
metallicity is not well determined, but possibly close to solar 
\citep{wakker2001}.

Region C encompasses the IV Spur, which is an extension of the IV Arch at lower 
velocities and higher Galactic latitudes. The distance bracket is $0.3-2.1$~kpc 
($z=0.3-2.1$~kpc), which implies a mass of $0.2-2.8 \times 10^5 \; M_\odot$. 
The metallicity is unknown.

All the target objects discussed reside in a relatively small sector of the 
Milky Way with $130\arcdeg \lesssim l \lesssim 280\arcdeg$ and $b>25\arcdeg$. 
This is in
fact one of the regions richest in IVCs and HVCs, and for which 
\citet{wakker2001} could get access to stellar probes observations.

Furthermore, in recent years distance brackets have been determined for 
additional, lower-mass IVCs and HVCs \citep[e.g.,][]{wakker2008}. Such clouds have 
ratios of mass over distance square of the order of $\lesssim 10^2 \; 
M_\odot$~kpc$^{-2}${. Therefore, for normal \g-ray emission rates per H~atom, 
as measured in the Galactic disk, we expect total fluxes of $\lesssim 10^{-10}$~cm$^{-2}$~s$^{-1}$ at energies $>100$~MeV.} These fluxes are 
one order of magnitude or more below the LAT sensitivity level
\citep{ackermann2012LATperf}. Therefore, we do not expect to be able to use 
LAT observations of these clouds to constrain their CR densities,  
and we do not include them among our targets. Conversely, the clouds 
selected for the 
analysis all have mass over distance square ratios of the order of $\gtrsim 
10^3 \; 
M_\odot$~kpc$^{-2}$, and hence expected \g-ray fluxes $\gtrsim 
10^{-9}$~cm$^{-2}$~s$^{-1}$, in the range detectable by the LAT.  

We have not included among our targets Complex~GCP, also known as the Smith 
Cloud, because this HVC is seen at Galactic latitudes $|b|<20\arcdeg$ toward 
the inner Galaxy, where the diffuse \g-ray foregrounds from the Galactic disk 
are brighter and more challenging to model.  We leave Complex~GCP for future 
studies. We note that 
\citet{drlica2014} and \citet{nichols2014} searched for \g-ray 
emission associated with Complex~GCP due to exotic processes related to 
hypothetical dark matter particles, and did not report on any significant detections. 

\section{DATA}\label{data}

\subsection{\g-Ray Data}\label{gdata}
{\changes The 
LAT is a pair-conversion \g-ray telescope for the energy range 20~MeV to 
$>300$~GeV. The instrument is described in \citet{atwood2009}, and its on-orbit 
performance discussed in \citet{ackermann2012LATperf}.} We use 
LAT observations accumulated from the beginning of the 
Science phase of the mission, MET\footnote{\F{} Mission Elapsed Time, measured in seconds since 1 January 2000.} 
239500801 (2008 August 4), up to MET 431481603 (2014 September 4), for a total 
of 73 months. We use the P7REP dataset \citep{bregeon2013FS} and select 
``Clean'' class events in order to limit the contamination from backgrounds due 
to residual CRs misclassified as \g~rays in the study of large-scale diffuse 
features. 

We apply further restrictions to the dataset used:
\begin{itemize}
 \item We select only time intervals when the LAT was operated in normal 
Science mode (e.g., excluding calibration data) and when the data quality was 
considered good (e.g., excluding solar flares which caused pile-up phenomena in the 
detector).
 \item We retain only candidate {\g~rays} detected at inclinations from the Earth 
zenith $<95\arcdeg$, in order to limit the contamination from the bright \g-ray 
emission produced by CR interactions in the Earth's atmosphere that forms 
extended features in the sky.
 \item We consider only candidate \g~rays with measured energies between 
300~MeV and 10~GeV. Below 300~MeV the LAT angular resolution rapidly 
deteriorates \citep{ackermann2012LATperf}, which makes separating different 
components of diffuse \g-ray emission based on their morphologies more difficult. Additionally, 
below $\sim 300$~MeV one would need to model the residual emission from CR interactions
in the upper atmosphere.  This can introduce a faint diffuse glow at moderately high declinations, which would need to be 
modeled, and would 
further complicate the study of 
faint extended features especially near the North Celestial pole where our
targets are located. Above 10 GeV the photon counts are sparse and the data are 
not very useful to constrain CR interactions in 
clouds owing to their spectrum {being} softer than {inverse-Compton (IC) emission from Galactic CR electrons} and isotropic backgrounds 
\citep[e.g.,][]{LATdiffpapII}. Limiting the energy range also helps to attenuate 
the systematic 
uncertainties due to the spectral models assumed for the various components of 
the \g-ray sky.  
\end{itemize}

For the \g-ray analysis described in this section and elsewhere in the article 
we used the official LAT \textit{Science Tools}, version 
v09r34p01\footnote{The \textit{Science Tools} are available along with LAT 
data from the \F~Science Support Center, \url{http://fermi.gsfc.nasa.gov/ssc}.}.

\subsection{Interstellar Medium Tracer Data}\label{ismdata}

If the CR densities are uniform within a cloud, the \g-ray 
intensities are simply proportional to the target gas column 
densities. Thus, in order to model \g-ray emission from IVCs and HVCs, and 
foreground emission from local gas we rely on a set of ISM tracers described here.

We employ as tracer of atomic gas the Leiden-Argentine-Bonn (LAB) all-sky survey of the 21-cm $\hi$ 
emission line by \citet{kalberla2005}. The survey provides the $\hi$ 
line brightness temperatures over a grid of $0\fdg5$ (with an effective 
angular resolution of $\sim 0\fdg6$), and covers velocities with respect 
to the local standard of rest from $-450$~km~s$^{-1}$ to $+400$~km~s$^{-1}$, at 
a resolution of 1.3~km~s$^{-1}$. Data were corrected for stray radiation 
reaching a residual noise contamination $<0.09$~K for most directions in 
the sky. We use data from \citet{kalberla2005} to {derive} $\nhi$ 
column-density maps as described in Section~\ref{himaps}.

Molecular hydrogen cannot be traced directly in its most common cold state. 
The most widely used surrogate tracer is the 2.6~mm line of $^{12}$CO. For this 
work we use a map of CO brightness temperature integrated over Doppler 
velocities, $\wco$, derived from the Center for Astrophysics composite CO 
survey \citep{dame2001} supplemented with higher-latitude
observations of the Ursa Major region \citep{devries1987}. The CO data, 
denoised using the moment-masking technique \citep{dame2011}, were used to 
construct the $\wco$ map on a $0\fdg25$ grid. The $\wco$ intensities are 
assumed to be proportional to the molecular hydrogen column densities $\nhd$.

Only ROI A contains significant CO emission 
(Figure~\ref{fig:comap}).
\begin{figure}[!htbp]\begin{center}
\includegraphics[width=0.5\textwidth]{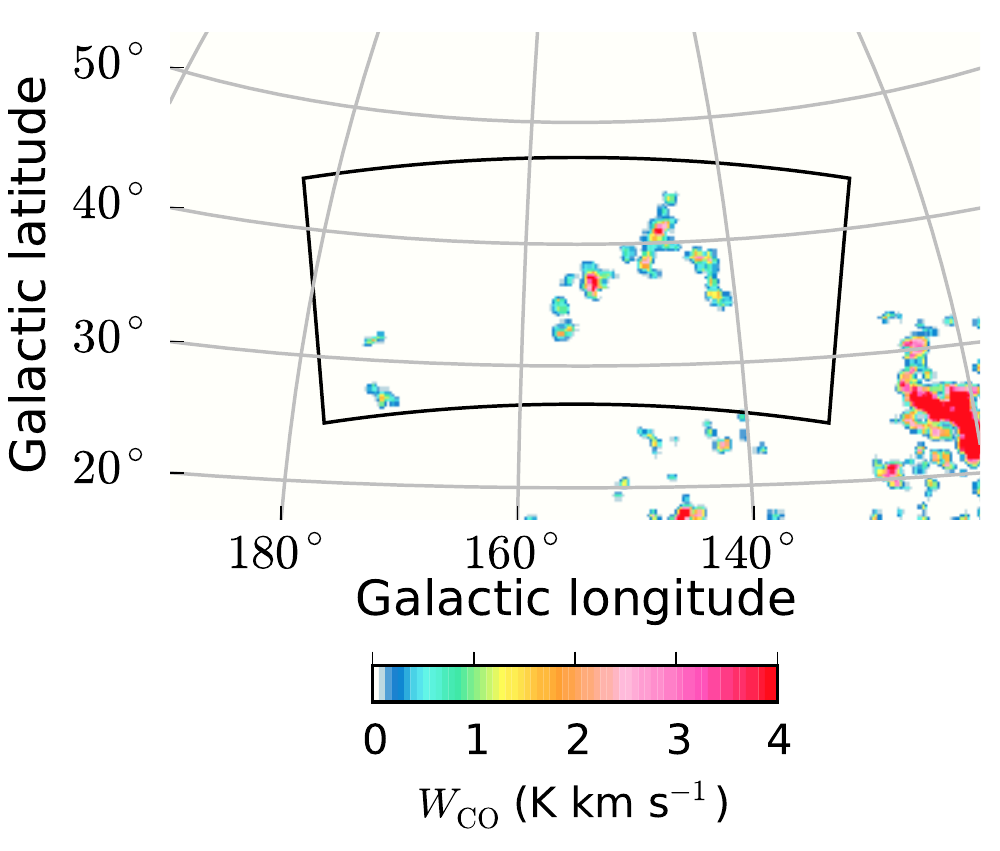}
\caption{$\wco$ intensities in the direction of ROI~A. The map is
shown in zenithal equidistant projection. Blank pixels are set to zero. The 
black line shows the border of the ROI.
}\label{fig:comap}
\end{center}\end{figure}
The CO emission detected is associated with the Ursa Major molecular clouds, 
located at {a} few hundred parsecs from the solar system \citep[e.g.,][]{devries1987}. 
Note that, according to the results from the component separation of
the all-sky survey performed 
using the \P-HFI instrument \citep{planck2013co}, there is no significant CO 
emission in our ROIs besides what is included in this CO map.

Various observations indicate that a linear combination of $\nhi$ and $\wco$ 
maps does not properly account for the totality of the ISM neutral gas, but 
large fractions of the gas can be in a DNM phase, also 
known as dark gas 
\citep[e.g.,][]{magnani2003,grenier2005,abdo2010cascep,
langer2010,planckfermi2014cham}. The 
DNM may be composed of molecular gas not associated with CO, revealed, e.g., by 
alternative tracers like CH \citep[e.g.,][]{magnani2003}, that is likely 
to exist because the CO molecule is more prone to photodissociation than $\hd$ 
in the external layers of molecular clouds 
\citep[e.g.,][]{wolfire2010,smith2014}. 
There can also be a contribution from atomic gas overlooked because of the 
approximations applied in deriving column densities from the $\hi$ line 
intensities \citep{fukui2014}. On large scales the DNM in the Milky Way is best 
traced by dust 
thermal emission or extinction and {\g~rays}, which reveal correlated excesses on 
top of the components traced by $\hi$ and CO {that can be interpreted as a contribution from undetected, but otherwise normal, neutral interstellar matter}
\citep[e.g.,][]{grenier2005,abdo2010cascep,planckfermi2014cham}. 

In order to trace the DNM in this work we rely on the all-sky model of thermal 
dust emission by the \citet{planck2013dustmaps} based on \P\ and {\it IRAS} data (\P\ public data release \texttt{R1.20}). 
We test both the map of 
optical depth at 353~GHz, $\tmap$, and the map of total dust 
radiance, $R$. The angular resolution of these maps is 5\arcmin. They are used 
to construct maps of the DNM as described in 
\ref{dustfit}.

\section{CONSTRUCTION OF THE INTERSTELLAR MEDIUM TRACER MAPS}\label{ismmaps}

For each ROI we construct maps of the $\nhi$ column densities for 
{different complexes along the line of sight}, and maps of the DNM column densities. 
The maps are constructed over regions that include at least 5\arcdeg{} 
boundaries around the ROI in order to properly take  into account the  
point-spread function (PSF) of the LAT in fits of the models to the \g-ray data.  

\subsection{Construction of the $\hi$ Maps}\label{himaps}
The column densities of atomic gas, $\nhi$, in IVCs and HVCs, and those in the 
local foreground gas can be separated based on Doppler velocities {in order to separate contributions from CR interactions with gas in different locations in the Galaxy}.
To do so we adapt the procedure described in 
\citet{abdo2010cascep}. The procedure used in this work has three steps. 
\begin{itemize}
 \item[i] For each ROI we define some preliminary velocity boundaries that 
separate {different components along the line of sight}, namely low-velocity (local) {gas, 
and gas in HVCs/IVCs} (Table~\ref{targettable}).
 \item[ii] For each line of sight on the grid of the LAB survey we adjust the velocity 
boundaries so that they fall at the closest minimum (or inflection point) in 
the brightness temperature profile. The column density $\nhi$ associated with 
each component is 
proportional to the integral of the brightness temperature over velocity within 
the adjusted velocity boundaries. 
 \item[iii] For each line of sight on the $\hi$ grid we fit to the brightness 
temperature profile a combination of Gaussian functions centered at the profile 
maxima (or inflection points if a maximum is not found for a velocity 
component) and use the results of the fit to correct the column densities of 
each component for the spillover from/into the adjacent components. 
\end{itemize}
The procedure is illustrated for an example line of sight in 
Figure~\ref{fig:hilines}.
\begin{figure}[!htbp]\begin{center}
\includegraphics[width=0.5\textwidth]{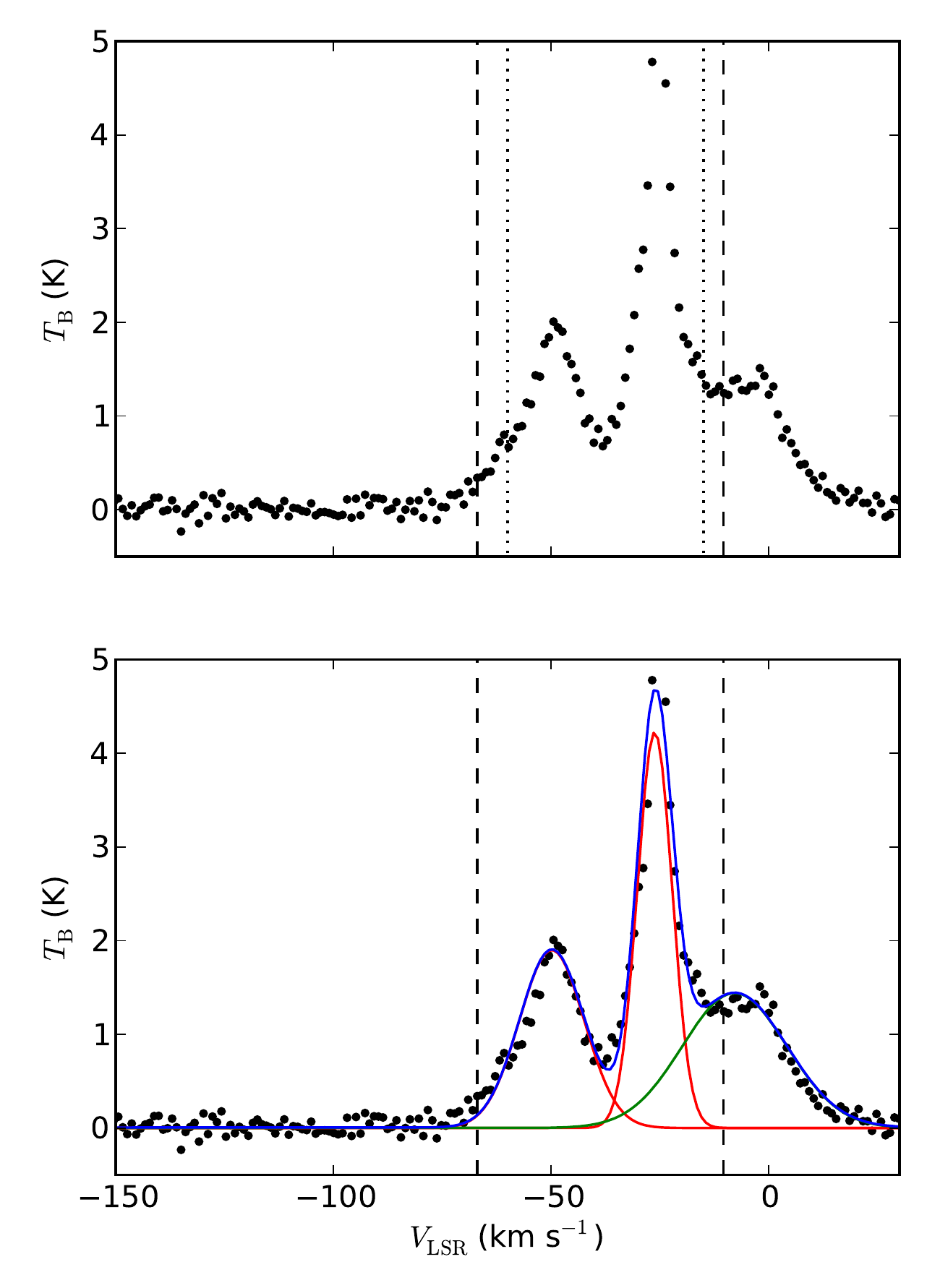}
\caption{$\hi$ brightness temperature, $T_\mathrm{B}$, as a 
function of Doppler velocity with respect to the local standard 
of rest, $V_\mathrm{LSR}$, for the line of sight at 
$l=172\fdg5$, $b=70\fdg5$. The upper panel illustrates how 
the preliminary velocity boundaries (dotted vertical lines) 
are adjusted to fall at the nearest minima or inflection 
points of the $T_\mathrm{B}$ profile (dashed vertical 
lines). The bottom panel illustrates the fitting procedure 
to correct the $\nhi$ column densities for spillover from 
one region to the next: in red Gaussian functions associated with peaks of the 
IV component, in green the Gaussian function associated with the peak of the 
low-velocity (local) component, in blue the total (no 
peaks were detected in this direction for the {high-velocity} component).}\label{fig:hilines}
\end{center}\end{figure}
Technical 
details of the procedure are illustrated {\chbis by}
\citet[][2.1.2, Appendix~B]{mythesis}.

In region A we set the preliminary velocity boundaries at $-40$~\kms\ to 
separate local gas from the low-latitude IV Arch, and at $-90$~\kms\ to 
separate the low-latitude IV Arch from Complex~A. At latitudes $b<24\arcdeg$ 
there is a significant contribution at higher velocities from gas in the disk 
of 
the Galaxy belonging to the Perseus and outer arms. {\chbis Because} 
\citet{abdo2010cascep} {\chbis found} that the \g-ray emissivity of the gas in these 
arms is compatible with the local emissivity, we assign all of this gas to the local 
component. We note that in any case this low-latitude region is not included in 
the 
ROI, but only used as part of the border to ensure a proper convolution of the 
model with the LAT point-spread function.

In region B we set the preliminary velocity boundaries at $-15$~\kms\ to 
separate local gas from the lower IV Arch, and at $-60$~\kms\ to separate the 
lower from upper IV Arch. In the case of region C we set the preliminary 
velocity boundaries at $-20$~\kms\ to separate local gas from the IV Spur, and 
at $-90$~\kms\ to separate the IV Spur from {high-velocity} gas; no significant emission is 
present at high velocities and the corresponding map is not used in the construction of the 
\g-ray model.

The Gaussian fits reproduce the measured brightness temperature profiles within 
$\sim 5\%$ on average (for the ratio of the integral of the absolute values of the differences between 
measured and fitted temperature along the line of sight to the integral of measured 
temperatures). The fit is  
significantly worse along a few lines\footnote{Out of 11421, 34001, 32421 
analyzed in region A, B, and C, respectively.} of sight with differences up to 
$\sim 
25\%$. The magnitude of the correction for the spillover is on average $3\%$ 
of the uncorrected column densities.  (The average is the largest for ROI~B, 
where it is $\sim 6\%$.) Even though the correction is only approximate, 
owing to the imperfect modeling of the temperature profiles by the Gaussian 
functions, the average magnitudes of the errors are $<0.5\%$ of the 
total column densities. In a small subset of the lines of sight ($\sim 0.5\%$) 
the fits 
do not converge properly, mostly due to the faintness of $\hi$ emission in 
these high-latitude regions. In such cases we do not apply any correction for 
spillover. 

We have calculated the $\nhi$ column densities under the approximation of small 
optical depth. An approximate correction for $\hi$ optical depth based on the 
assumption of a uniform spin temperature has often been used in the literature
\citep[e.g.,][]{abdo2010cascep,planckfermi2014cham}, but the low column 
densities in the regions studied make the effect of the correction small. 
Assuming a uniform spin temperature of 80~K \citep[as in][]{planck2011halo} 
would increase the column 
densities on average at most by 5\% for the local component in region A, and 
less for other components and regions. We also note that, by using dust maps to 
trace gas missed by the $\nhi$ and $\wco$ maps as described in 
\ref{dustfit}, we are effectively accounting also for $\hi$ densities 
underestimated due to 
the 
approximation of small optical depth.

We replace the $\nhi$ column density seen in ROI A at $l=138.5\arcdeg$, 
$b=37\arcdeg$ with the average of neighboring lines of sight owing to an 
anomalous temperature profile indicative of {an} artifact in the LAB data.

The resulting $\nhi$ maps are shown in Figures~\ref{fig:himapsa}, 
\ref{fig:himapsb}, and~\ref{fig:himapsc}.
\begin{figure*}[!htbp]\begin{center}
\includegraphics[width=1\textwidth]{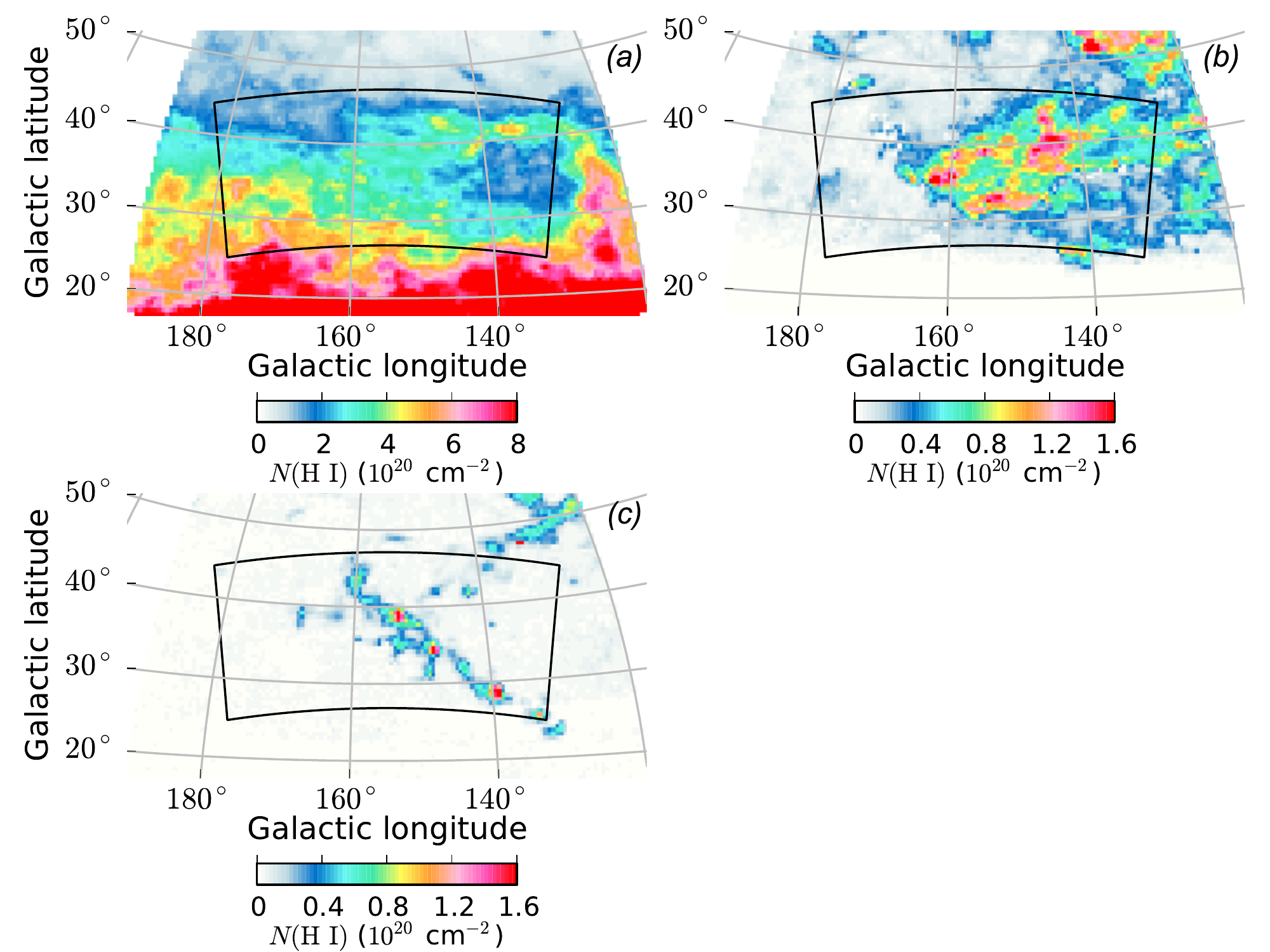}
\caption{$\nhi$ column density maps for ROI~A: a) 
Low-velocity (local) component, b) the 
low-latitude IV Arch, c) Complex A. The black lines show
the border of ROI~A, as used in the analysis. The maps are 
shown in zenithal equidistant projection.}\label{fig:himapsa}
\end{center}\end{figure*}
\begin{figure*}[!htbp]\begin{center}
\includegraphics[width=1\textwidth]{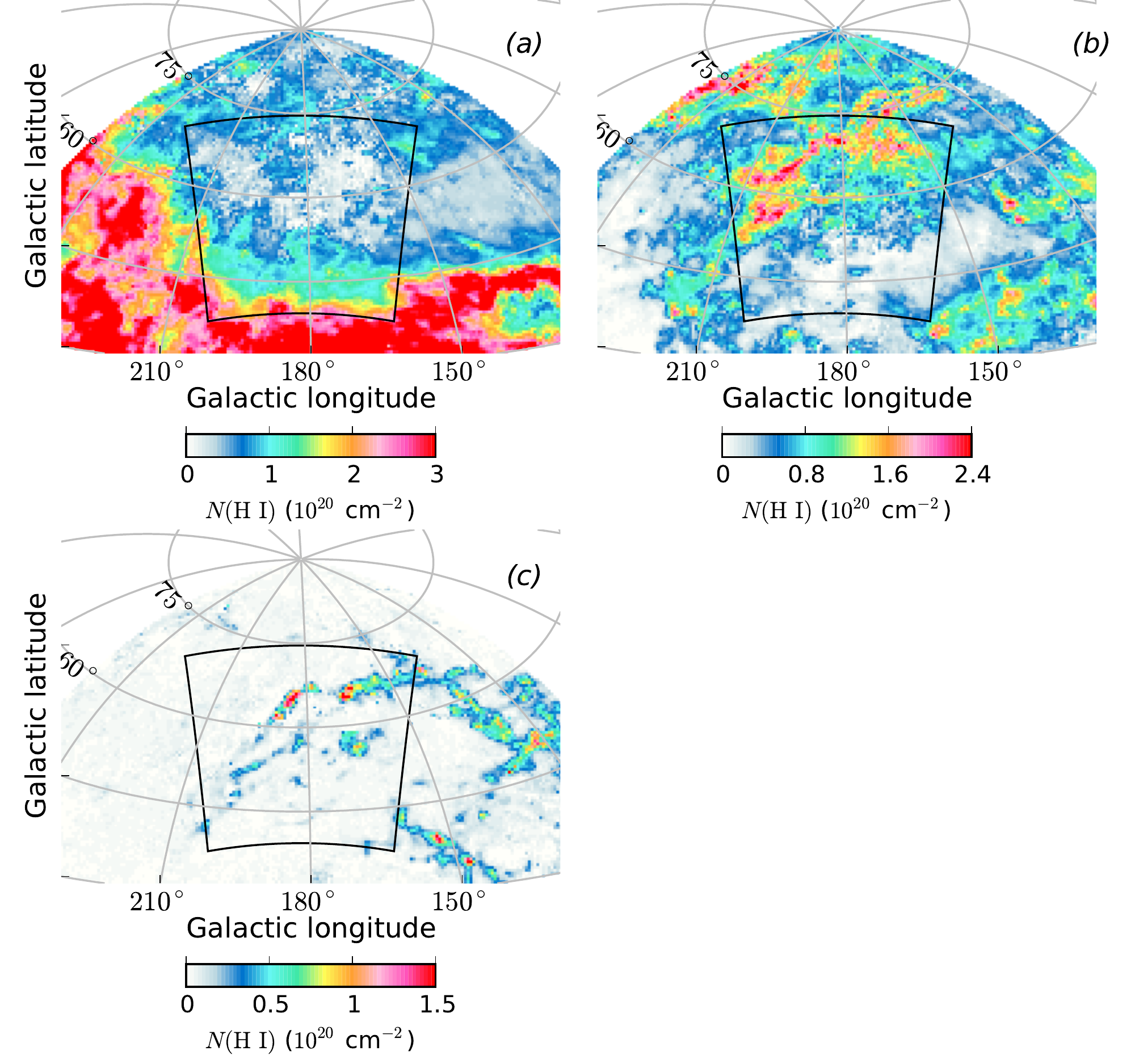}
\caption{$\nhi$ column density maps for ROI~B: a) 
Low-velocity (local) component, b) the 
lower IV Arch, c) the upper IV Arch. The black lines show
the border of ROI~B, as used in the analysis. The maps are 
shown in zenithal equidistant projection.}\label{fig:himapsb}
\end{center}\end{figure*}
\begin{figure*}[!htbp]\begin{center}
\includegraphics[width=1\textwidth]{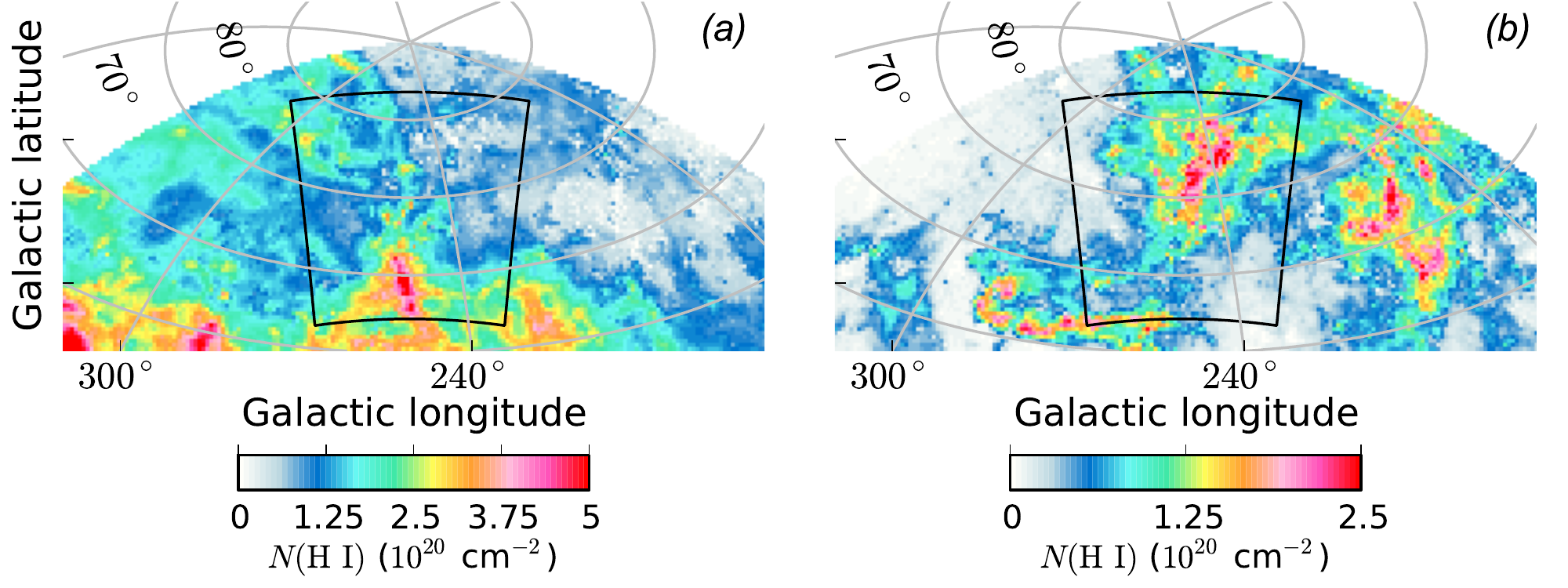}
\caption{$\nhi$ column density maps for ROI~C: a) 
Low-velocity (local) component, b) the 
IV Spur. There is no significant $\hi$ emission at high velocities in the ROI. 
The black lines show
the border of ROI~C, as used in the analysis. The maps are 
shown in zenithal equidistant projection.}\label{fig:himapsc}
\end{center}\end{figure*}
Figure~\ref{fig:himapsb}~(a) and (b) show some level of anticorrelation between 
higher/lower $\nhi$ intensity regions. This is due to some genuine 
anticorrelation between the intensities of $\hi$ emission in velocity space. 
At the same time, the lower IV Arch shows a bright feature at $\sim 
-30$~\kms\ that blends into the local $\hi$ components, which makes the 
separation particularly challenging. This is addressed by the spillover 
correction.

\subsection{Construction of the DNM Maps}\label{dustfit}
Since the work of \citet{grenier2005} {on the 
construction of models of interstellar \g-ray emission, dust residuals, i.e., total dust 
opacity or extinction minus the best-fit linear 
combination of $\hi$ and $\wco$ maps, have been used to trace the DNM, neutral interstellar matter not properly traced by the linear combination of $\nhi$ and $\wco$.} However, the original 
implementation of this method has two limitations: 1) the fit of 
the dust map is biased by the missing DNM component; 2) the effect of the 
assumptions on the stated errors in the fit of the dust maps is overlooked.

Recently, the \citet{planckfermi2014cham} performed an analysis of \P\ and 
LAT data of the Chameleon interstellar complex where they refined the 
original method by: 1) developing a ``circular'' fitting method, where dust 
residuals are used to model the DNM in the \g-ray analysis,  and \g-ray 
residuals to model the DNM in the dust analysis iteratively until a stable 
solution is found; 2) exploring several assumptions on the errors used in the 
fits and folding the differences into the errors on the final 
results.

The method developed by the \citet{planckfermi2014cham} derives a map of the DNM column 
densities from {\g~rays} to use in fitting the dust maps. Owing to the low gas column densities (on average one order of 
magnitude lower than in the Chameleon complex) and the limited statistics of 
the \g-ray dataset, it was not applicable to our 
ROIs. We therefore developed an alternative method to acknowledge and address the 
issue of the missing DNM component in the dust fit that does not rely on \g-ray 
data and is described in the rest of the section.

Owing to the limited angular resolution of \g-ray 
data and the other ISM tracers, and to the input projections accepted by the 
LAT \textit{Science Tools}, we reproject the maps from 
the \citet{planck2013dustmaps} onto the 
grid chosen for the \g-ray analysis in each ROI.

Under the assumptions that dust grains are well mixed with gas in the cold 
and warm phases of the ISM {\chbis \citep[e.g.,][]{bohlin1978}}, and that the properties of dust grains and photon fields are uniform within each phase and complex, we build a model for the dust map 
$D$ {\changes(i.e., {\chbis the map of} optical depth at 353~GHz, $\tmap$, or radiance, $R$)} by assuming 
that either quantity is proportional to the atomic and molecular column 
densities as traced by $\nhi$ and $\wco$, respectively, plus an isotropic term 
representing any zero-level shift, for example, due to residual contamination 
from the cosmic infrared background.
\begin{equation}
 M^0(l,b)=\sum_{\imath=1}^3 y^0_{\hi, \imath} \cdot \nhi(l,b)_i + 
y^0_\mathrm{CO} \cdot \wco(l,b) + D^0_\mathrm{iso}
\end{equation}
{Here and in the following equations  we indicate with $y_{\hi, \imath}$ the dust specific power (opacity) per hydrogen atom, and with $y_\mathrm{CO}$ the specific power (opacity) per $\wco$ unit in the molecular phase. We} use indices $\imath=1,2,3$ to label the 
complexes separated kinematically along the line of sight as described in 
Section~\ref{himaps}.
The model $M^0$ is fit to the dust map $D$ using the minimum $\chi^2$ method. 
The $\chi^2$ is defined, for a generic model $M$, as
\begin{equation}\label{chisqdust}
 \chi^2 = \sum_{l,b} \frac{[D(l,b)-M(l,b)]^2}{\sigma^2(l,b)}.
\end{equation}
Following the work of the \citet{planckfermi2014cham}, we consider different 
possible assumptions on the uncertainties $\sigma(l,b)$. Our baseline 
assumption is that the uncertainties are mainly due to imperfections of 
the model {\chbis because of the assumption of} proportionality between 
radiance/opacity and gas column densities. In the absence of a reliable estimate of the uncertainties on $M$, we assume the 
uncertainties to be $\sigma(l,b) \propto M(l,b)$. We empirically consider two alternative hypotheses to assess the impact of this choice:
\begin{itemize}
\item $\sigma(l,b) \propto D(l,b)$, which tests how the agreement between $M$ and $D$ affects the results of the fits;
\item $\sigma(l,b) \propto \sigma_\tau(l,b)$, where $\sigma_\tau(l,b)$ is the 
map of statistical uncertainties on $\tmap$ provided by 
the \citet{planck2013dustmaps} (for fits of $\tmap$ only).
\end{itemize}
{\chbis We initially assume that the} proportionality coefficients are 30\% for the model or the dust map itself, and 
1 
when using $\sigma_\tau(l,b)$. These are adjusted later as a part of the 
iterative procedure to obtain a reduced $\chi^2=1$. The $\chi^2$ is minimized
to determine the best-fit coefficients, {$\tilde{y}^0$} and 
{$\tilde{D}^0_\mathrm{iso}$}.

From the results we derive a dust residual map
\begin{eqnarray}
 D^0_\mathrm{res}(l,b) = D(l,b) + \nonumber \\ \hspace{-36pt} - \left[ \sum_{\imath=1}^3 
\tilde{y}^0_{\hi, \imath} \cdot \nhi(l,b)_i + 
\tilde{y}^0_\mathrm{CO} \cdot \wco(l,b) + 
\tilde{D}^0_\mathrm{iso}\right]
\end{eqnarray}
The map is then filtered to extract the significant 
positive residuals associated with the DNM. The residual map is convolved with 
a Gaussian kernel with standard deviation of 1\arcdeg\ in order to suppress 
statistical 
fluctuations while retaining large-scale real features. Then we build a 
histogram of the residuals and we fit a Gaussian curve to the core of the 
distribution, within $\pm 2$~standard deviations of the distribution itself. We 
set a threshold at 
2 standard deviations of the fitted Gaussian curve, and produce the filtered 
residual map 
$\bar{D}^0_\mathrm{res}$ by retaining pixels of the original residual map 
when the values in the convolved residual map are above the threshold, and 
setting 
other pixels to 0.

Then
we proceed iteratively to refine the model. At each iteration 
$\alpha = 1, 2, \ldots$:
\begin{itemize}
 \item {\chbis The model is defined:}
 \begin{eqnarray}\label{eq:dustmodel}
  M^\alpha(l,b) & = & \sum_{\imath=1}^3 y^\alpha_{\hi, \imath} \cdot \nhi(l,b)_i + \nonumber \\
&+& y^\alpha_\mathrm{CO} \cdot \wco(l,b) + \nonumber \\
&+& y^\alpha_\mathrm{res} \cdot \bar{D}^{\alpha-1}_\mathrm{res}(l,b) + 
D^\alpha_\mathrm{iso};
 \end{eqnarray}
 \item We fit the model $M^\alpha$ to the dust map $D$ minimizing the $\chi^2$ 
as defined in Eq.~\ref{chisqdust};
 \item {\chbis The dust residual map is built:}
  \begin{eqnarray}
   D^\alpha_\mathrm{res}(l,b) = D(l,b) + \nonumber \\ \hspace{-54pt}- \left[ \sum_{\imath=1}^3 
\tilde{y}^\alpha_{\hi, \imath} \cdot \nhi(l,b)_i + 
\tilde{y}^\alpha_\mathrm{CO} \cdot \wco(l,b) + 
\tilde{D}^\alpha_\mathrm{iso}\right]
  \end{eqnarray}
  where $\tilde{y}^\alpha$ and $\tilde{D}^\alpha_\mathrm{iso}$ are the best-fit 
coefficients;
 \item We filter
$D^\alpha_\mathrm{res}$ to extract the significant positive 
residuals $\bar{D}^\alpha_\mathrm{res}$ following the procedure described for 
iteration 0.
\end{itemize}
The iterative procedure is stopped when the two following conditions are 
simultaneously satisfied:
1) the differences between all the best-fit coefficients in 
iteration $\alpha$ and iteration $\alpha-1$ are compatible with 0 within 
$1\sigma$ statistical uncertainties from the fit; 2) the difference between 
minimum $\chi^2$ in iteration $\alpha$ and iteration $\alpha-1$ is $<1$.

After the first pass just described, the scaling coefficients for the 
uncertainties 
$\sigma(l,b)$ are re-defined, so that the reduced $\chi^2 = 1$ in the latest 
iteration. Requiring the reduced $\chi^2$ to equal 1 ensures that the 
statistical uncertainties from the fit are truly representative of the 
differences between dust maps and models. This, in turn, also makes the 
criteria to stop the iteration of the fitting procedure more meaningful, even 
if only in a data-driven fashion. Using the new scaling coefficients we perform 
a second pass, after which the final reduced $\chi^2$ values are found to be 1 
within 1\% in all ROIs and variants of the maps. {\changes The distribution of the residuals in the final iteration does not show any significant negative values, which demonstrates how the procedure is appropriately addressing the initial bias due to the missing template for the DNM (see Appendix~\ref{dustresults} for further details).}
The scaling coefficients for the errors on $\tmap$ provided 
by the \citet{planck2013dustmaps} are found to be systematically greater than 1. The 
error map represents only statistical uncertainties from a fit where the dust 
thermal emission spectral energy distribution was fitted using a modified 
blackbody 
spectrum. Hence, a scaling coefficient greater than 1 may capture the presence of 
systematic errors, or discrepancies between the distribution in the 
sky of dust optical depth and our simple model based on the assumption that the 
optical depth scales with gas column densities.

{\chbis Because} the \citet{planck2013dustmaps} recommends the use of $R$ as the best 
dust column-density tracer at high latitudes, we take as our baseline DNM 
templates those obtained from $R$ assuming that the error is proportional to 
the model. These maps are shown in Figure~\ref{fig:dnmmaps}.
\begin{figure*}[!htbp]\begin{center}
\includegraphics[width=1\textwidth]{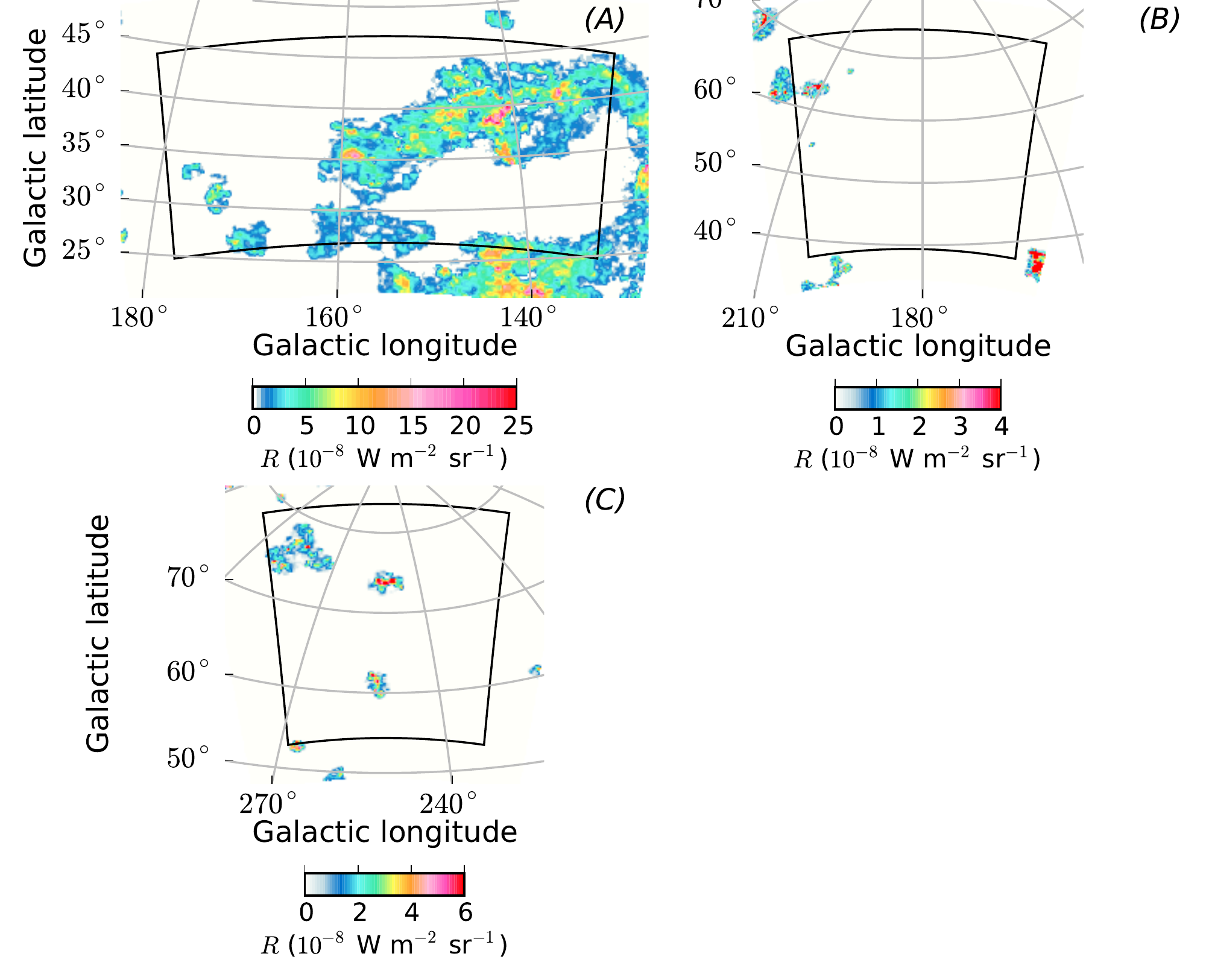}
\caption{DNM templates built from radiance $R$, as determined from the 
iterative 
fitting procedure for the three ROIs A, B, and C. The black lines show
the borders of the ROIs. The errors to define the $\chi^2$ for the fit were 
assumed to be proportional to the model in Eq.~\ref{eq:dustmodel}, with the 
proportionality constant 
defined so that the reduced $\chi^2$ in the final fit is 1. The maps are 
shown in zenithal equidistant projection.}\label{fig:dnmmaps}
\end{center}\end{figure*}
We note that the DNM component is particularly abundant in ROI~A, where it 
follows a known interstellar structure dubbed the North Celestial Pole (NCP) 
Loop \citep{meyerdierks1991}, possibly related to the Ursa Major molecular 
complex \citep[e.g.,][]{pound1997}.
Further results from the analysis described in this section are discussed in 
Appendix~\ref{dustresults}.

\section{\g-RAY ANALYSIS}\label{analgamma}

\subsection{Model Building}\label{modelbuild}
The \g-ray emission measured using the LAT can be modeled as the sum of 
diffuse components and discrete sources. Diffuse emission is produced in the 
interstellar space of the Milky Way by 
energetic interactions of CRs with gas, via nuclear collisions leading to 
\g-ray production mainly through neutral pion decay and by electron 
Bremsstrahlung, and with low-energy radiation fields, due to {IC} scattering by electrons.

Multiple studies of interstellar \g-ray emission have shown that CR fluxes can 
be approximated as uniform on the scales of interstellar complexes {\changes located away from potential CR sources}
\citep[e.g.,][]{abdo2010cascep,planckfermi2014cham}.
Under this hypothesis, the \g-ray intensities from 
interactions with gas are proportional to the target gas column 
densities.
We can therefore model the diffuse intensity as a combination of the gas tracer maps 
described in Section~\ref{ismmaps}. We assume the spectral shape to be that 
measured using 
LAT data for the local interstellar space \citep{casandjian2014} in order to 
perform the fit over a broad energy range. This assumption will be 
critically evaluated against the data in the assessment of systematic 
uncertainties ({Sec.}~\ref{sys_eval}). For each tracer map we allow a 
free scaling coefficient in the model, to define the overall CR flux in 
that component\footnote{HVCs and IVCs span volumes much larger than those of 
interstellar complexes studied so far using LAT data. Were there any 
variations of CR densities within the clouds, the scaling coefficient would 
represent an average CR flux over it.}.

The IC component is modeled based on the work by \citet{depalma2013}. They 
considered 8 models of IC emission calculated using the GALPROP\footnote{\url{http://galprop.stanford.edu}} code 
{\changes \citep{moskalenko2000,strong2000,porter2008}} for different values of some 
model parameters, and tuned them to the LAT data over the whole sky. The work 
by 
\citet{depalma2013}, originally based on the LAT P7 data, was recently {\chbis updated} 
for a P7REP data selection consistent with our dataset.
For the IC component we allow in 
the fit a free scaling coefficient, as well as a log-parabolic spectral correction to mitigate the effect of uncertainties in the local electron spectra. Among the models in \citet{depalma2013}, we arbitrarily select 
that with the CR source distribution based on the work by \citet{lorimer2004iaus}, 
a maximum height of the CR confinement halo of 10~kpc and a $\hi$ spin 
temperature of 150~K. The other models presented in \citet{depalma2013} will be 
used later in Section~\ref{sys_eval} to assess the impact of this choice on the results.

Finally, there is a diffuse component with almost isotropic distribution over 
the sky interpreted as the superposition of extragalactic diffuse \g-ray 
emission and residual backgrounds from CR interactions in the LAT misclassified 
as \g~rays. This is modeled using a spectral shape derived from the LAT data {\chbis(see \citealt{depalma2013} for the details of the procedure)}
consistently with our IC models, separately for \g~rays converting in the front 
and back sections of the LAT tracker, which are characterized by different 
residual background rates relative to their acceptances.
Owing to limitations in the assumption that the residual CR background can be treated as an isotropic \g-ray term, we allow {a free scaling coefficient in the fit} also for this term. 
Note that, due to the limited size of the ROIs and the quasi-isotropic distribution of IC \g-ray emission in the models, both IC and isotropic components have to be considered as effective background templates in our fits and the respective scaling factors are not the object of any interpretation in this work.

We incorporate point sources from the 3FGL list {\citep{3fgl}}. For each ROI we include in the model all the sources within 
10\arcdeg\ from the ROI border. The source positions are fixed at their catalog 
values. We assume for each source the spectral shape given in the catalog as 
well. For sources outside the ROI but within 10\arcdeg\ from the ROI border all 
the spectral parameters are fixed at their catalog values. For sources inside 
the ROI we leave all the spectral parameters free if the average significance 
$S$ of the source in 3FGL is $> \theta_1$, just the global normalization free if 
$\theta_2< S < \theta_1$, and all the parameters fixed otherwise. The 
thresholds $\theta_1$ and $\theta_2$ are chosen for each ROI so that the number 
of free parameters associated with sources is $<35$ to prevent convergence 
problems in the fitting procedure. The pairs of thresholds 
$(\theta_1,\theta_2)$ chosen are $(15\sigma,9\sigma)$, $(30\sigma,13\sigma)$, and
$(10\sigma,5\sigma)$, for regions A, B, and C, 
respectively. We verified that none of the 3FGL sources are positionally coincident with structures in the $\nhi$ maps in Figures~\ref{fig:himapsa}, \ref{fig:himapsb}, and \ref{fig:himapsc}. Following a visual inspection of the residual maps after a 
preliminary analysis we left free in addition to those sources also 
3FGL~J0809.5+5342 in ROI~A and 3FGL~J1159.2$+$2914 in ROI~C. We will show later in {F}igures~\ref{fig:gammaA}, \ref{fig:gammaB}, and \ref{fig:gammaC} that additional two years of data in our analysis with respect to 3FGL do not introduce any additional sources that significantly affect the results of the analysis.

To summarize, for each ROI the model for the \g-ray intensities is described
by Equation~\ref{eq:model}.
\begin{eqnarray}\label{eq:model}
\hspace{-24pt} I_\gamma(l,b,E) &= &\sum_{\imath=1}^3 \left[ 
x_{\mathrm{HI},\imath} \cdot \nhi(l,b)_\imath \; q_\mathrm{LIS}(E)\right] + \nonumber \\
& + & x_\mathrm{CO} \cdot \wco(l,b) \; q_\mathrm{LIS}(E) + \nonumber \\
&+& x_\mathrm{DNM} \cdot \bar{D}(l,b) \; q_\mathrm{LIS}(E) +\nonumber \\
 & + & x_\mathrm{IC} \cdot \mathrm{IC}(l,b,E) \left(\frac{E}{E_0}\right)^{\alpha_\mathrm{IC}+\beta_\mathrm{IC} \ln \frac{E}{E_0}}+ \nonumber \\
& + & x_\mathrm{Iso} \cdot
I_\mathrm{iso}(E) +\mathrm{SRC}
\end{eqnarray}
There, $q_\mathrm{LIS}(E)$ stands for the tabulated emissivity spectrum of 
$\hi$ in the local interstellar space from \citet{casandjian2014}. 
$\nhi_\imath$ are the $\hi$ column 
density maps that we constructed in Section~\ref{himaps}, so that, for each 
complex 
$x_{\mathrm{HI}}$ is the scaling factor with respect to the local emissivity. 
{Kinematically-separated c}omplexes along the line of sight are numbered as 1) low-velocity, i.e., local 
gas, {2) and 3) IVCs and HVCs}.
$\wco$ is the CO 
intensity map described in Section~\ref{ismdata} (used only for ROI~A); hence 
$x_\mathrm{CO}$ can be interpreted as the scaling factor that transforms the 
local $\hi$ emissivity into \g-ray emissivity per $\wco$ intensity unit in the 
region studied. Similarly, $\bar{D}$ is one of the templates of the DNM 
constructed in Section~\ref{dustfit}, and $x_\mathrm{DNM}$ is the scaling factor 
that 
transforms the local $\hi$ emissivity into \g-ray emissivity per dust radiance 
(or optical depth) unit. IC represents the IC emission model:
$x_\mathrm{IC}$ is an overall global scaling factor, $\alpha_\mathrm{IC}$ and $\beta_\mathrm{IC}$ are the parameters of the log-parabolic spectral correction, while $E_0$ is a reference energy which was set to 700~MeV (the choice is arbitrary {\chbis because} the parameter is degenerate with the others). Finally, $I_\mathrm{iso}$ is the tabulated 
isotropic background spectrum (different for front- and back-converting events), with $x_\mathrm{Iso}$ as common scaling factor,
and SRC represents the {\chbis model for individual sources} described above.

\subsection{Analysis Procedure and Baseline Results}\label{basefit}

Using the dataset described in Section~\ref{gdata}, for each ROI in 
Table~\ref{targettable} we build binned count cubes with $0\fdg25$ 
spacing in angle and 15 logarithmically spaced energy bins from 300~MeV to 
10~GeV for front- and back-converting events. The model in 
Eq.~\ref{eq:model} is fit to the data by using the binned likelihood analysis 
procedure implemented in the \textit{Science Tools} that takes into account 
the LAT exposure and PSF. We employ a joint 
likelihood technique to independently treat the front- and back-converting 
events in the fit in order to exploit the better angular resolution of the 
former. We use the \texttt{P7REP\_V15\_CLEAN} LAT instrument response 
functions. 

The maximum likelihood values of the coefficients of the interstellar emission 
models for the baseline 
analysis are reported in Table~\ref{tab:basefitres}.
In Figures~\ref{fig:gammaA}, 
\ref{fig:gammaB}, and \ref{fig:gammaC} we show the count 
maps and residual maps summed over the entire 
energy range.
\begin{figure*}[!htbp]\begin{center}
\includegraphics[width=1\textwidth]{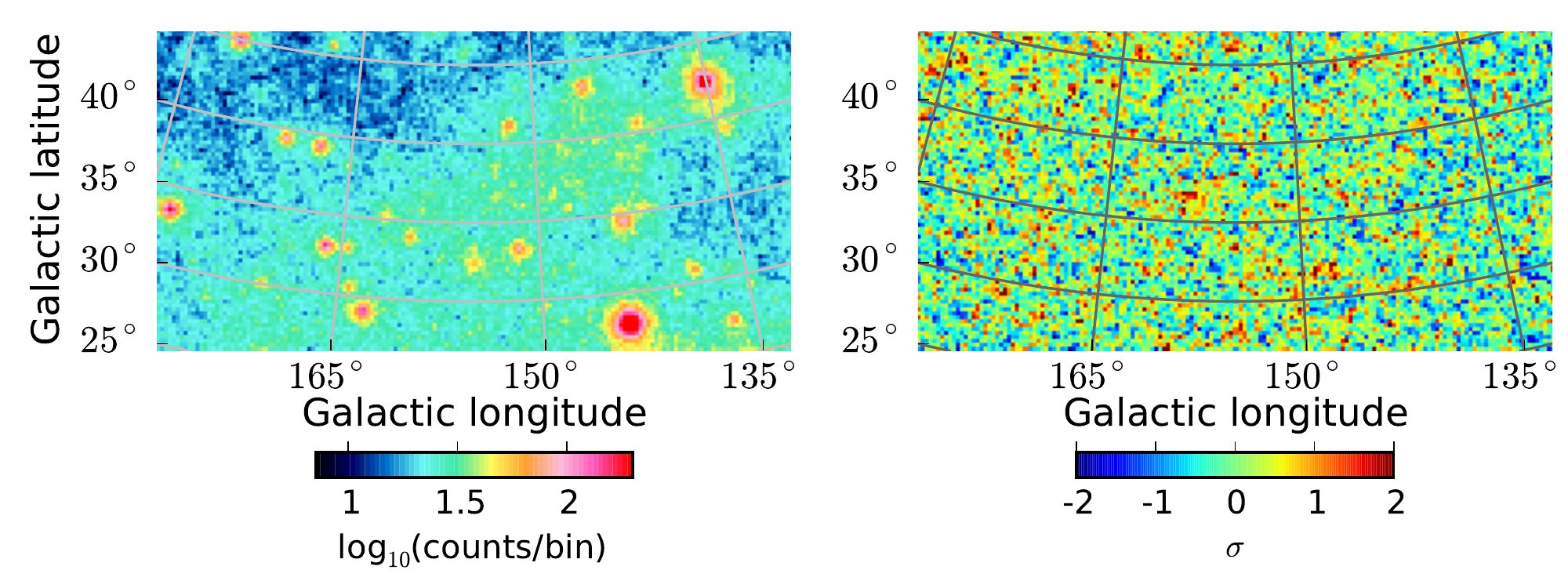}
\caption{Counts (left) and residuals after subtraction of 
the best-fit baseline model (right) for ROI~A. Counts and residuals are 
summed over the entire energy range from 300~MeV to 10~GeV. Residuals  are 
expressed in units of approximate standard deviations, calculated as counts 
minus model-predicted counts divided by the square root of counts. The maps are 
shown in the \textit{plate carr\'ee} projection centered at the ROI 
center used for the \g-ray analysis, and are smoothed for display using a 
Gaussian kernel of $\sigma=0\fdg5$.}\label{fig:gammaA}
\end{center}\end{figure*}
\begin{figure*}[!htbp]\begin{center}
\includegraphics[width=0.9\textwidth]{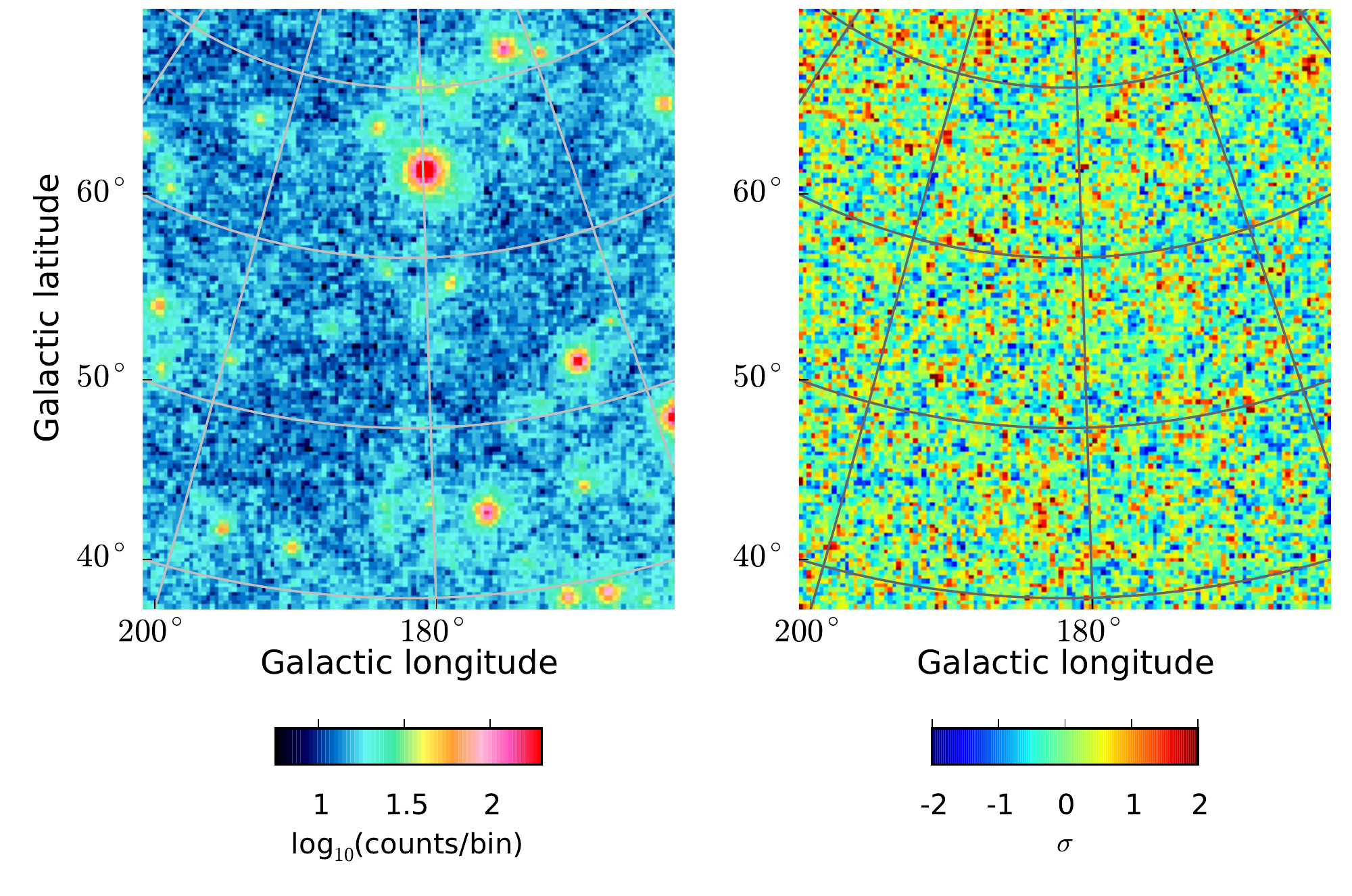}
\caption{Counts (left) and residuals after subtraction of 
the best-fit baseline model (right) for ROI~B. See the caption of 
Fig.~\ref{fig:gammaA} for details.}\label{fig:gammaB}
\end{center}\end{figure*}
\begin{figure*}[!htbp]\begin{center}
\includegraphics[width=0.7\textwidth]{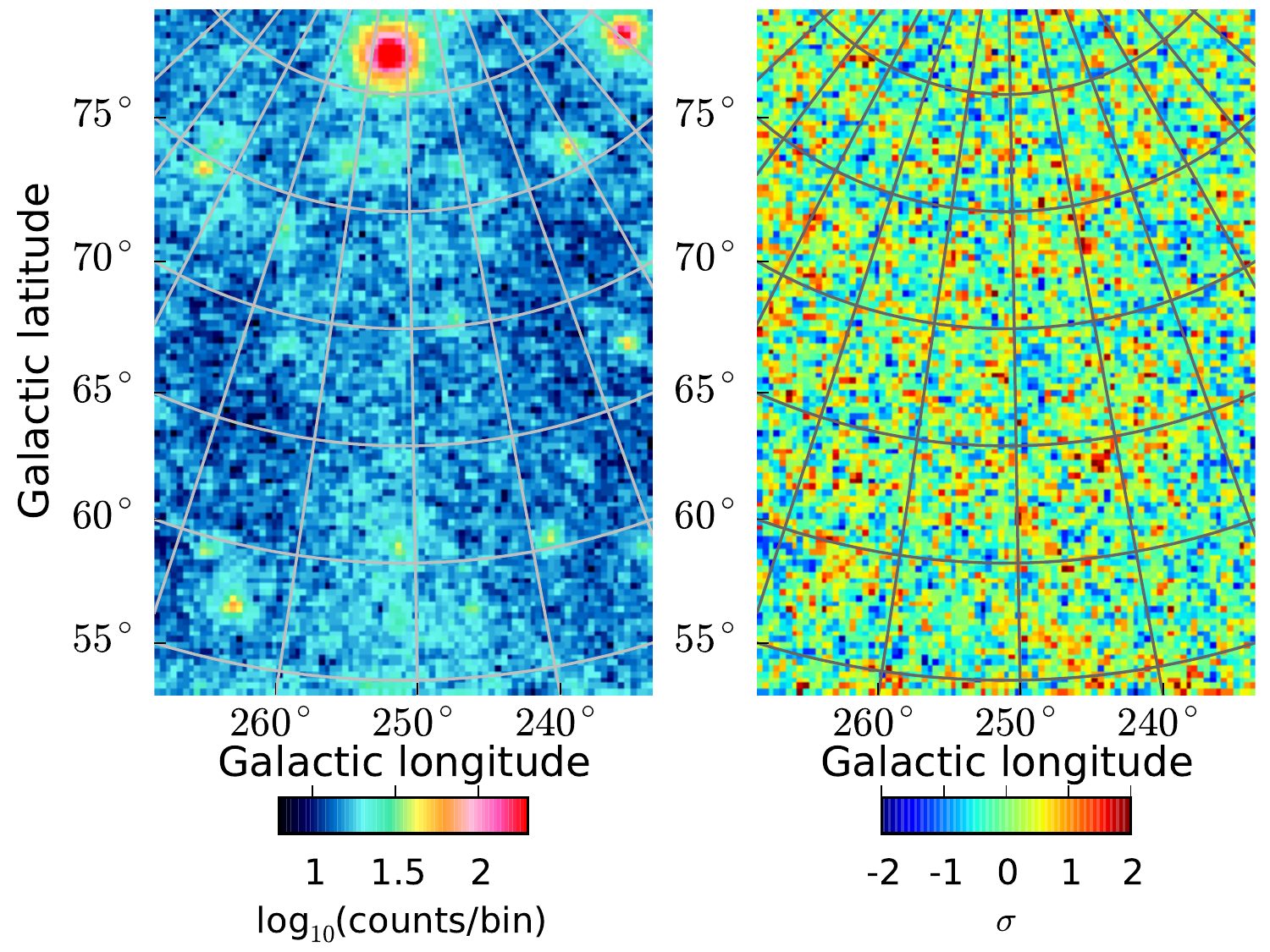}
\caption{Counts (left) and residuals after subtraction of 
the best-fit baseline model (right) for ROI~C. See the caption of 
Fig.~\ref{fig:gammaA} for details.}\label{fig:gammaC}
\end{center}\end{figure*}
The residual maps do not show any large-scale structures correlated with the 
templates used to model the \g-ray emission. Some localized residuals may hint 
at missing point sources not included in 3FGL, which was developed using the 
first 48 
months of LAT observations, as opposed to the 73 months used in our analysis.
\begin{deluxetable}{lccc}
\tabletypesize{\scriptsize}
\tablecaption{Interstellar best-fit coefficients from the 
baseline analysis\label{tab:basefitres}}
\tablewidth{0pt}
\tablehead{
\colhead{Parameter}  & \colhead{ROI~A} &\colhead{ROI~B} & 
\colhead{ROI~C}
}
\startdata
$x_{\mathrm{HI},1}$ & $0.999\pm0.017$ & $0.94\pm0.07$&$1.08\pm0.04$\\
$x_{\mathrm{HI},2}$ & $0.94\pm0.16$& $1.08\pm0.04$&$0.68\pm0.06$\\
$x_{\mathrm{HI},3}$ & $0.00^{+0.06}_{-0.0}$ & $0.00^{+0.10}_{-0.0}$&\nodata\\
$x_\mathrm{CO}$\tablenotemark{a} &$2.3\pm0.2$& \nodata&\nodata\\
$x_\mathrm{DNM}$\tablenotemark{b}  & $0.24\pm0.03$& 
$0.00^{+0.11}_{-0.0}$&$0.34\pm0.09$\\
$x_\mathrm{IC}$ &$1.19\pm0.05$&$1.31\pm0.05$&$1.82\pm0.09$\\
$\alpha_\mathrm{IC}$ & $-0.158\pm0.014$& $-0.06\pm0.02$& $0.02\pm0.05$ \\
$\beta_\mathrm{IC}$ & $0.045\pm0.008$& $0.02\pm0.01$& $-0.026\pm 0.014$ \\
$x_\mathrm{Iso}$ & $0.978\pm0.017$&$1.03\pm0.02$&$0.96\pm0.04$\\
\enddata
\tablecomments{See Eq.~\ref{eq:model} for a definition of the model parameters. 
The numbering scheme for the $x_{\mathrm{HI},\imath}$ coefficients is: 1) 
low-velocity, i.e., local gas, {2) and 3) IVCs and HVCs}.}
\tablecomments{Asymmetric error bands {\chbis are} computed using \texttt{Minos} when the best-fit value is at the boundary imposed during the likelihood optimization process.}
\tablenotetext{a}{10$^{20}$ cm$^{-2}$ (K \kms)$^{-1}$}
\tablenotetext{b}{10$^{32}$ sr W$^{-1}$}
\end{deluxetable}

For each target (HVC or IVC) in Table~\ref{targettable} we calculate the test 
statistic TS, defined as
\begin{equation}\label{eq:tsdef}
 \mathrm{TS}=2\ln\left(\frac{\like}{\like_0}\right),
\end{equation}
where $\like$ is the maximum likelihood for the model that includes the source 
of 
interest, while $\like_0$ is the maximum likelihood of the {\chbis minimal} model where the 
contribution from the source is set to null.
TS is expected to be distributed as a $\chi^2$ with a 
number of degrees of freedom given by the difference of the number of free 
parameters in the two models \citep[e.g.,][]{mattox1996}. We consider 
a target 
cloud as detected if $\mathrm{TS}>25$, which formally corresponds to a 
$5\sigma$ 
confidence level that the likelihood improvement is not due to statistical 
fluctuations. {\chbis As detailed in \citet{protassov2002}, the distribution of TS may deviate from a $\chi^2$ if the minimal model corresponds to a point on the topological border of the phase space of the parameters of the model including the source, which applies to our situation}. However, 
the TS values for our detected targets all greatly exceed the threshold.

The TS values are reported in Table~\ref{targeTSul}. For target HVCs and IVCs that are not 
detected we calculate a 95\% confidence level {\chbis (c.l.)} upper limit on $x_\mathrm{HI}$ 
based on a Bayesian integration of the likelihood profile as applied for 
measurements in the 2FGL catalog \citep[][3.5]{2FGL}.
\begin{deluxetable}{lccc}
\tabletypesize{\scriptsize}
\tablecaption{Target detections and upper limits\label{targeTSul}}
\tablewidth{0pt}
\tablehead{
\colhead{Region}  & \colhead{Complex} &\colhead{TS} & 
\colhead{95\% c.l. upper limit}
}
\startdata
A & Low-Latitude IV Arch & 127& \nodata\\
&  Complex~A & 0 & 0.21\\
\tableline
B &  Lower IV Arch & 750 & \nodata
\\
 & Upper IV Arch & 0 & 0.23 \\
\tableline
C & IV Spur & 154&\nodata\\
\enddata
\tablecomments{See Eq.~\ref{eq:tsdef} for the definition of TS.}
\tablecomments{The upper limits are expressed as a fraction of the emissivity of 
local $\hi$.}
\end{deluxetable}

{In Fig.~\ref{imIVgamma} we show the \g-ray emission components associated with detected IVCs, evaluated by subtracting from the total \g-ray counts the components attributed to individual sources and foreground/background interstellar emission by the likelihood fit. Contours from the $\nhi$ maps inferred from the 21-cm observations in Figures~\ref{fig:himapsa}, \ref{fig:himapsb}, and~\ref{fig:himapsc} are overlaid. The morphologies of the \g-ray detected signals follow closely the known $\hi$ structures in the low-latitude IV Arch and lower IV Arch. In the case of the IV Spur a cluster of \g~rays is associated with the densest part of the cloud, but the visual comparison is hampered by limited statistics in the \g-ray dataset.}
\begin{figure*}[!htbp]\begin{center}
\begin{tabular}{cc}
\includegraphics[width=0.5\textwidth]{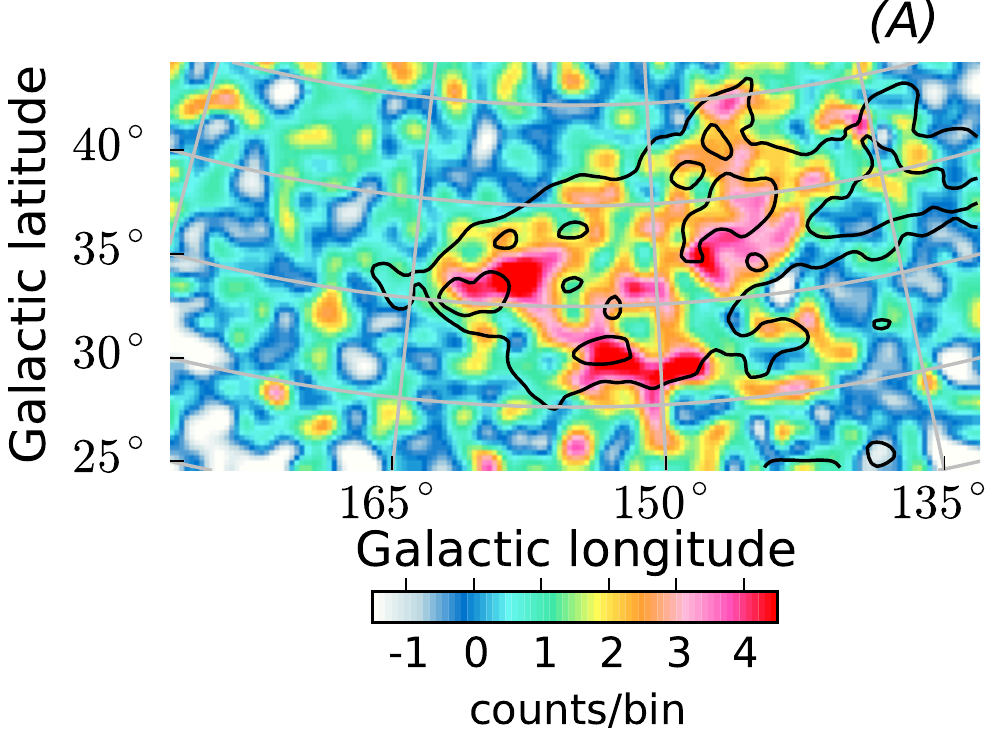}&
\includegraphics[width=0.5\textwidth]{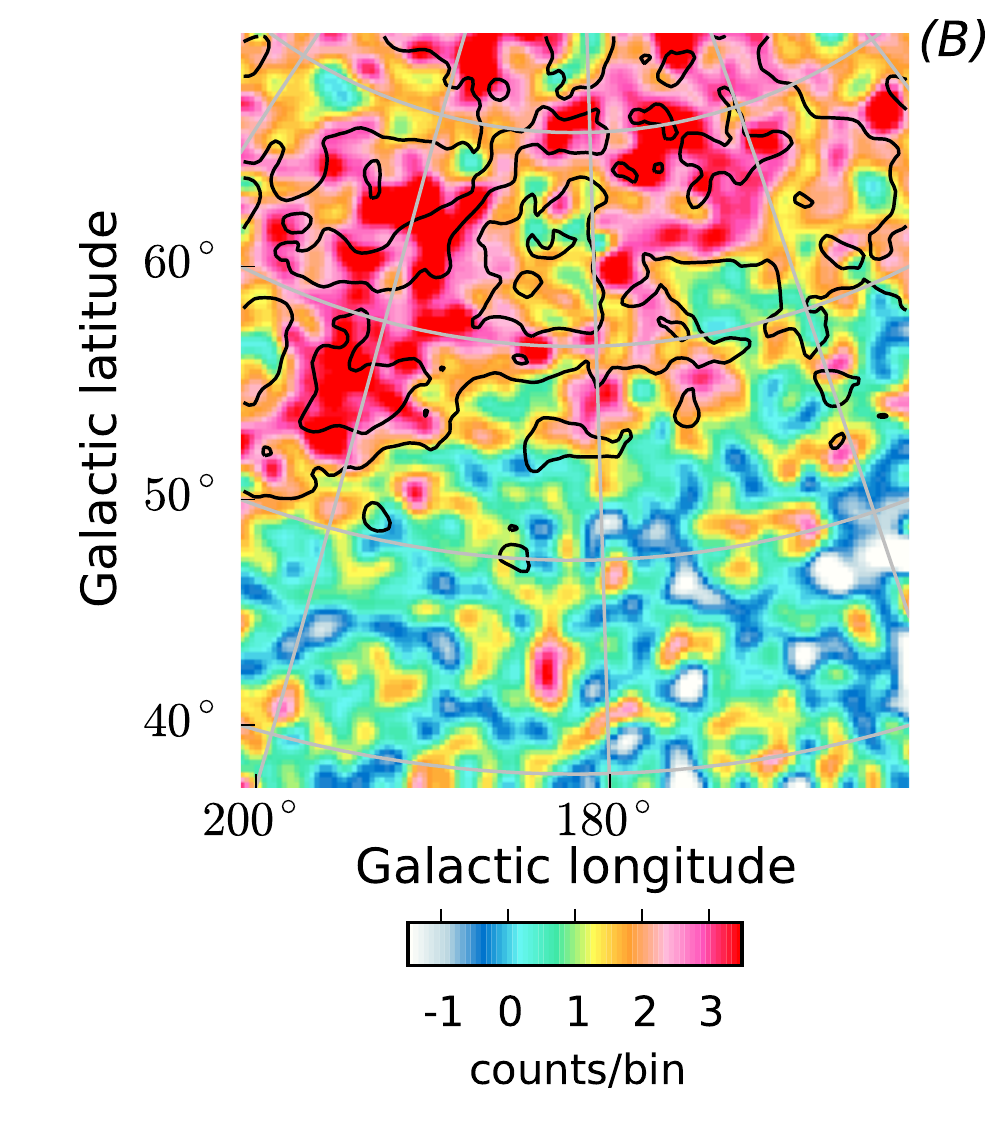}\\
\includegraphics[width=0.5\textwidth]{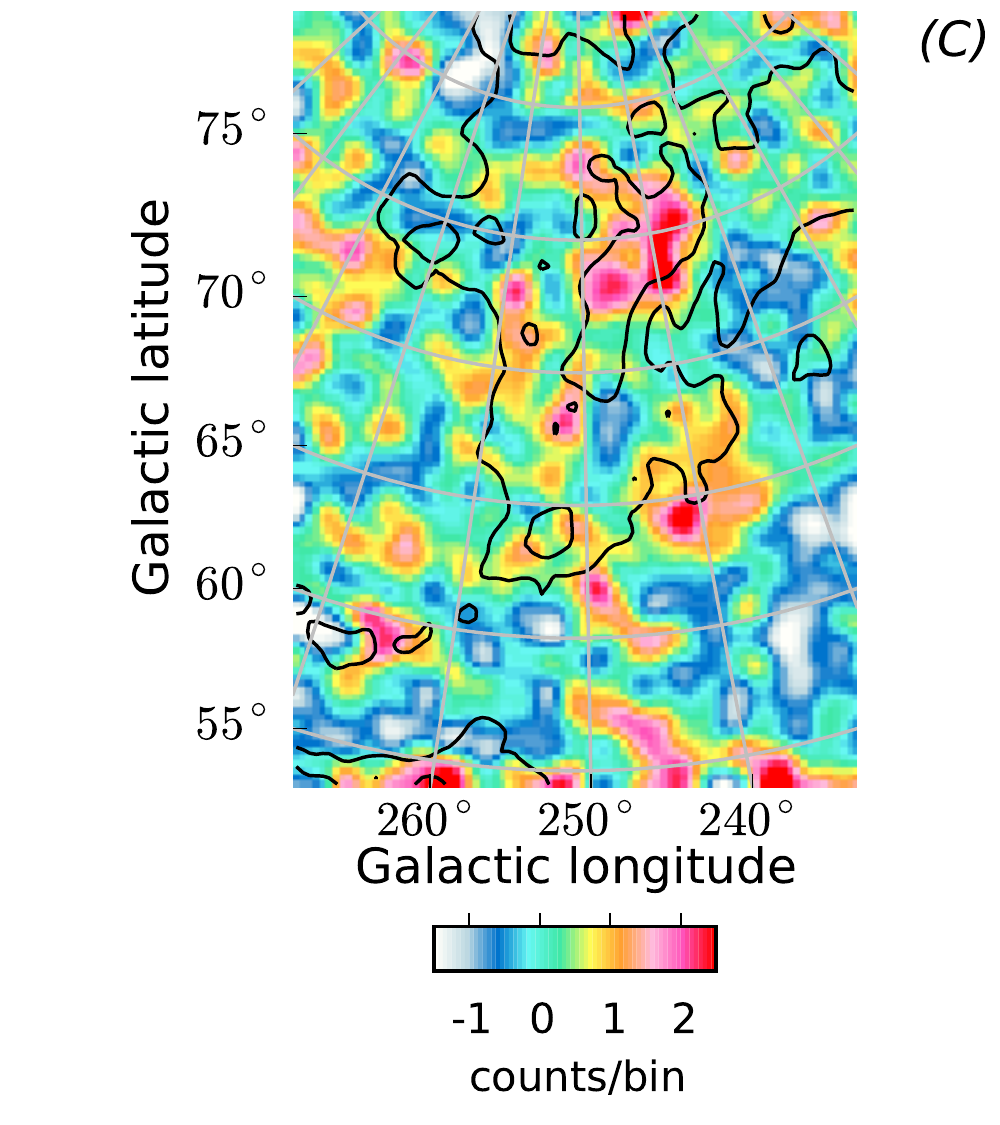}&
\end{tabular}
\caption{{{\g-ray} emission associated with the detected targets: total counts minus the contributions attributed by the fit to all emission components in Eq.~\ref{eq:model} but A) the low-latitude IV Arch, B) the lower IV Arch, C) the IV Spur. The maps were smoothed using a $2\arcdeg$ Gaussian kernel to suppress statistical fluctuations and enhance extended features. The overlaid contours of $\hi$ column density have levels  A) $0.6 \times 10^{20}$~atoms~cm$^{-2}$ and $1.2 \times 10^{20}$~atoms~cm$^{-2}$, B) $0.8 \times 10^{20}$~atoms~cm$^{-2}$ and $1.6 \times 10^{20}$~atoms~cm$^{-2}$, C) $0.9 \times 10^{20}$~atoms~cm$^{-2}$ and $1.8 \times 10^{20}$~atoms~cm$^{-2}$.}}\label{imIVgamma}
\end{center}\end{figure*}

\subsection{Spectra of the Detected Clouds}\label{cloudspectra}
A key assumption of our model is that the emission of the gaseous components has 
a spectrum 
described by the emissivity spectrum of local gas $q_\mathrm{LIS}$ from 
\citet{casandjian2014}, modified only by a global scaling factor over the whole 
energy range from 300~MeV to 10~GeV. While this assumption is expected to hold 
for the local (low-velocity) gas, there may be spectral 
variations for clouds at great distance from the disk.

For clouds detected in our analysis we can use the data to verify how well the 
$q_\mathrm{LIS}$ spectrum reproduces the observed one. To do so we perform the 
\g-ray analysis independently in 6 broad energy bins logarithmically spaced 
from 300~MeV to 10~GeV. In these fits the spectral parameters are frozen to 
their determination over the entire 
energy range except for the normalizations.
Figure~\ref{fig:hiloc_spec} shows as a reference the scaling factors of local 
$\hi$ emissivity in the three regions studied.
\begin{figure}[!htbp]\begin{center}
\includegraphics[width=0.5\textwidth]{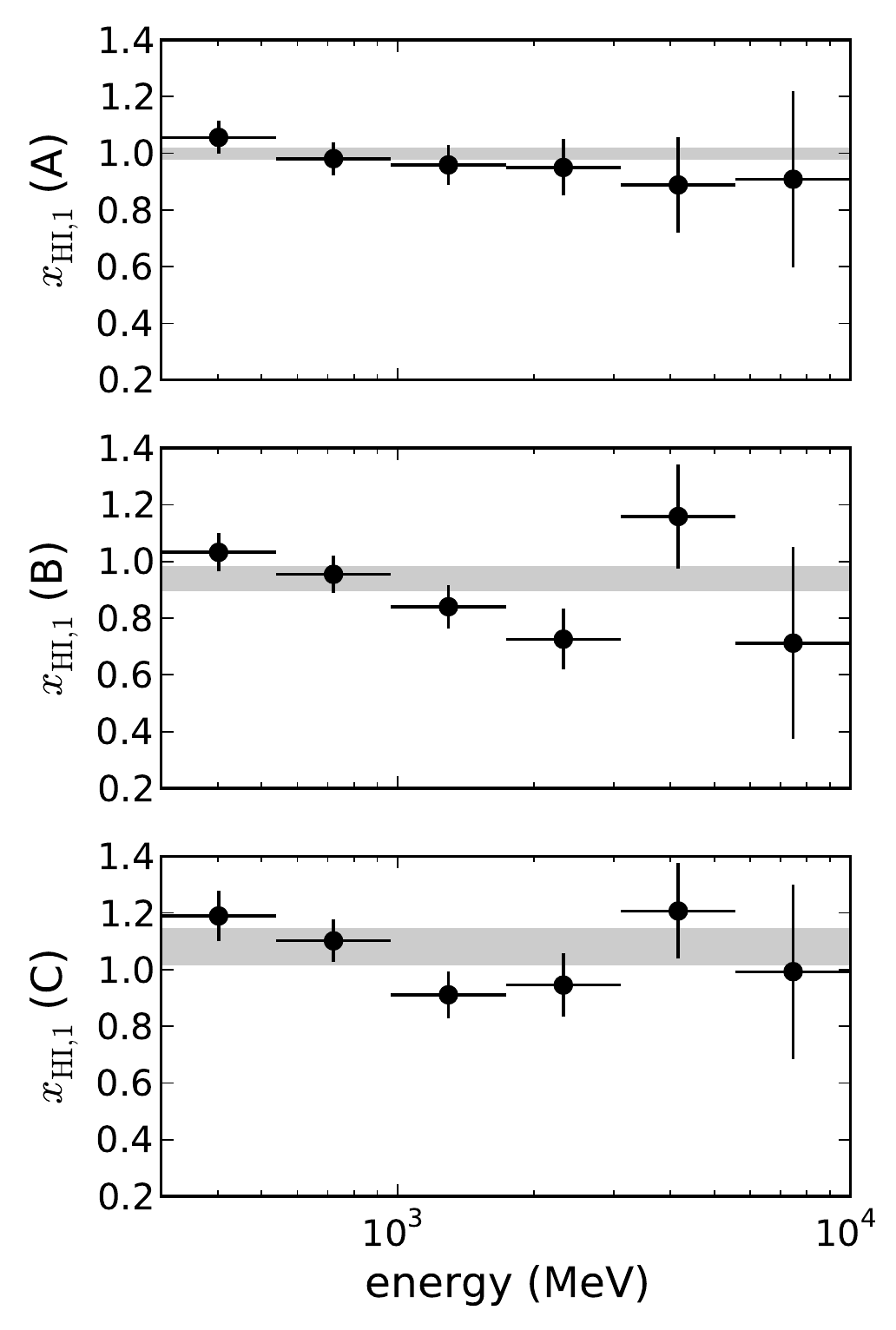}
\caption{Scaling factor of local $\hi$ emissivity, 
$x_{\mathrm{HI},1}$, as a function of energy for the 
three regions A, B and C. The shaded bands indicate the 
values obtained from the analysis over the entire energy 
range {\chbis with their $1\sigma$ statistical uncertainties}.}\label{fig:hiloc_spec}
\end{center}\end{figure}
Figure~\ref{fig:hiiv_spec} shows the scaling factors for the emissivity of all 
the detected IVCs with respect to local.
\begin{figure}[!htbp]\begin{center}
\includegraphics[width=0.5\textwidth]{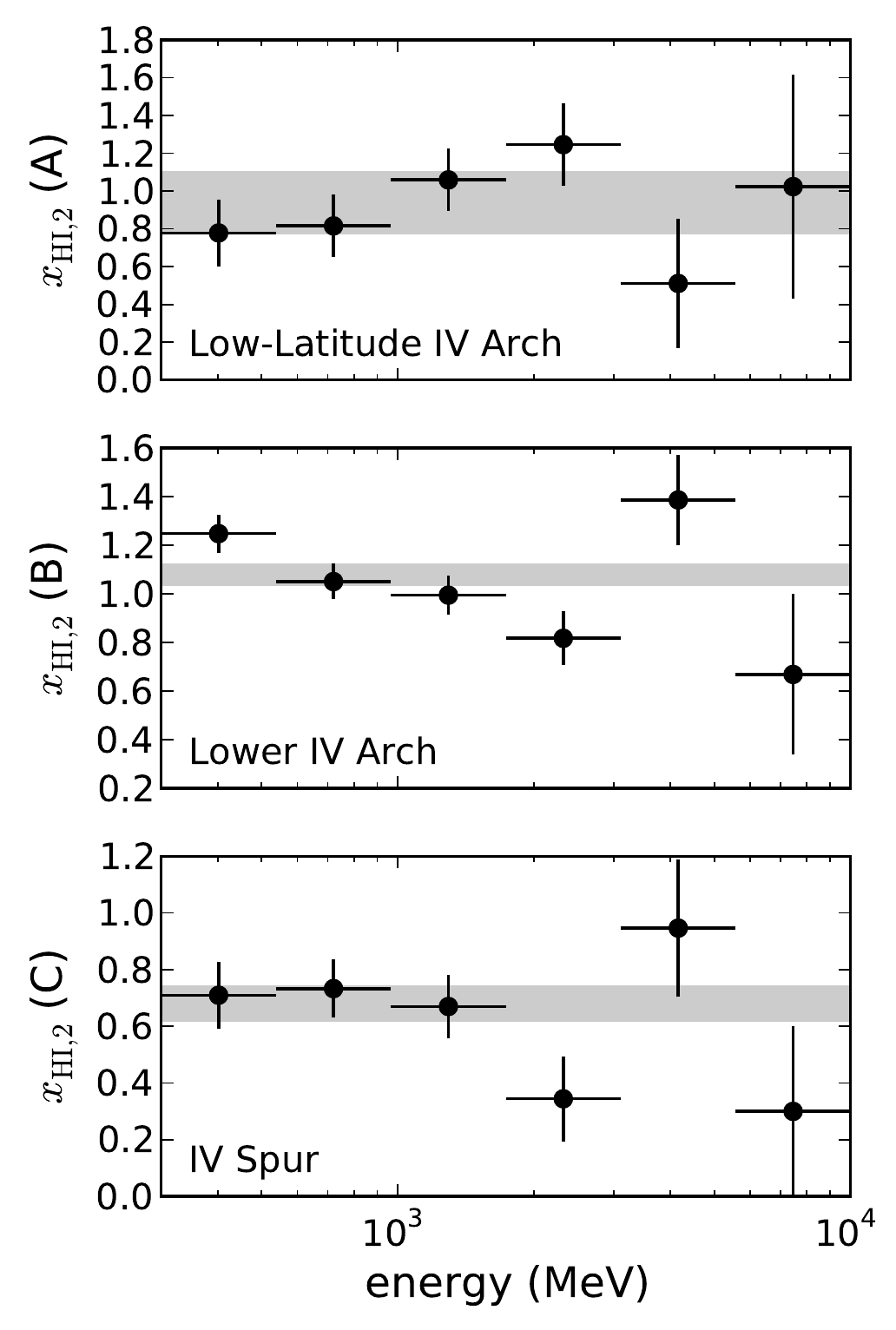}
\caption{Scaling factors of $\hi$ emissivity in three IVCs, 
$x_{\mathrm{HI},2}$, as a function of energy in regions A, B and C. The 
shaded bands indicate the values obtained from the analysis over the entire energy 
range {\chbis with their $1\sigma$ statistical uncertainties}.}\label{fig:hiiv_spec}
\end{center}\end{figure}

The scaling factors are all consistent within statistical uncertainties with 
the value determined from the fit over the entire energy range. In region A the residuals show a negative trend that 
indicates
a spectrum softer than the average in the local ISM. {\changes Similar trends may be present for both the local ISM and IV gas in region B.}
It is not clear if this is real or due to any small imperfections in the modeling, but further investigation on this aspect is beyond the scope of this work and left for future studies.  {\changes For the scope of our analysis it is justified to use the global scaling factors over the whole energy range {\chbis because} they are compatible within~1$\sigma$ with the weighted average of the values for the separate energy bins.}

\subsection{Assessment of Systematic Uncertainties}\label{sys_eval}

\subsubsection{Uncertainties of inputs to the 
\g-ray interstellar emission model}\label{modelunc}
Some inputs to the model in Eq.~\ref{eq:model} are subject to 
important uncertainties, most notably the DNM column densities, IC emission, 
and 
the spectrum of undetected clouds. We explored the associated systematic 
uncertainties by changing {\chbis the modeling assumptions} 
and evaluating the impact on the best-fit parameters for the \g-ray analysis.

For each region we modified the baseline DNM template by using the alternative 
templates from Section~\ref{dustfit} based on different assumptions for the 
errors in the determination of the DNM template, and replacing dust radiance 
with optical depth.

We also changed the IC model by using alternative models as described in 
\citet{depalma2013}, which included assuming maximum heights of the CR
confinement halo of 4~kpc and 10~kpc, and replacing the CR source distribution 
 inferred from pulsar observations \citep{lorimer2004iaus} with that by 
\citet{case1998} based on supernova remnant observations\footnote{We note that 
the derivation of the Galactic supernova remnant distribution in 
\citet{case1998} is subject to uncertainties and discordant with some later 
works \citep[e.g.,][]{green2014}. In our study, though, we use it only as a way 
to probe the uncertainties due to the modeling of CR propagation relying on the 
previous work by \citet{LATdiffpapII,depalma2013}.}. These are two of the most 
important input parameters to GALPROP to determine the morphology and spectrum 
of IC emission \citep{LATdiffpapII}.

Finally, we have shown in Section~\ref{cloudspectra} that for detected targets 
the 
assumption that the emission spectrum follows the emissivity in the local 
ISM is appropriate, but we cannot do the same for targets that 
are not significantly detected, i.e., Complex A and the Upper IV Arch. We 
explored possible variations with respect to the local emissivity spectrum based 
on the 
set of propagation models in \citet{LATdiffpapII}. We calculated the ratios of 
the emissivity spectra in those models at all locations in the outer Galaxy 
(where our clouds are located) over the emissivity spectrum at the solar 
circle. 
We parametrized the extrema of these ratios with two functions of energy
\begin{equation}\label{emf1-}
 f_\mathrm{1}^-=\left(\frac{E}{1\;\mathrm{GeV}}\right)^{-0.1},
\end{equation}
and
\begin{equation}\label{emf2-} 
f_\mathrm{2}^-=\left(\frac{E}{1\;\mathrm{GeV}}\right)^{
-[0.08+0.07\ln(E/1\;\mathrm{GeV})]}.
\end{equation}
Eq.~\ref{emf1-} captures the softening due to energy-dependent escape that is 
expected near the boundaries of the confinement region in the outer Galaxy for sources located mostly in the inner Galaxy. In addition, 
Eq.~\ref{emf2-} captures a low-energy hardening predicted by 
high-reacceleration models in \citet{LATdiffpapII} at larger Galactocentric 
radii\footnote{{ Our choice to include this effect in the evaluation of systematic uncertainties is conservative, {\chbis because} the low-energy hardening is a feature that appears only in models with high reacceleration, while models with convection or pure diffusion models are also viable \citep{LATdiffpapII}.}}. Note that both functions indicate a softer spectrum. In order to be 
conservative we explore the extreme assumptions that the emissivity spectrum 
may actually be harder at the locations of the undetected targets by inverting 
the signs of the exponents
\begin{equation}\label{emf1+}
 f_\mathrm{1}^+=\left(\frac{E}{1\;\mathrm{GeV}}\right)^{+0.1},
\end{equation}
and
\begin{equation}\label{emf2+} 
f_\mathrm{2}^+=\left(\frac{E}{1\;\mathrm{GeV}}\right)^{
+[0.08+0.07\ln(E/1\;\mathrm{GeV})]}.
\end{equation}
We use these functions to modify the emissivity spectra of the 
undetected clouds, 
by
multiplying the local emissivity spectrum 
$q_\mathrm{LIS}$ in Eq.~\ref{eq:model} by the correction function and scaling 
it 
so that the emissivity integrated in the 300~MeV to 10~GeV range is the same as 
for 
$q_\mathrm{LIS}$. In this way, we can still interpret the upper limits on 
$x_\mathrm{HI}$ for the undetected components as scaling factors with respect to the 
local emissivity, and, at the same time, explore the impacts on the other fit 
parameters.

We repeat the baseline analysis described in Section~\ref{basefit} for all the 
combinations of the model variations described above, and, for each parameter, 
we take the extreme variations with respect to the baseline fit as our estimate of the 
systematic uncertainty related to the model. In the cases of upper limits for 
undetected targets we use the worst, i.e., largest, upper limits in the 
discussion that follows.

We summarize the scaling factors for $\hi$ emissivity in the different regions 
and complexes in Table~\ref{hiscaling}, including the spread due 
to inputs to the \g-ray interstellar emission model as 
determined through the methodology described in this subsection.  
In general, the emissivity spread for the detected complexes is driven by the choice of the DNM template, and, to a lesser extent, the choice of IC model for local gas (which is less structured, hence more degenerate with IC emission). For non-detected complexes the systematic uncertainties are dominated by the choice of the emissivity spectrum. The results in ROI B show a stronger dependence on the assumed IC model, which affects also the determination of the lower IV Arch emissivity and the upper limit on the emissivity of the upper IV Arch at 2 statistical $\sigma$ level.
\begin{deluxetable}{lccccc}
\tabletypesize{\scriptsize}
\tablecaption{$\hi$ Emissivity scaling factors\label{hiscaling}}
\tablewidth{0pt}
\tablehead{
\colhead{Region}  & \colhead{Complex} &\colhead{Scaling Factor} & 
\colhead{Stat\tablenotemark{a}} & \colhead{Sys (Model)\tablenotemark{b}} & \colhead{Sys (Jackknife)\tablenotemark{c}}
}
\startdata
A & Local & 0.99 & 0.02 & $^{+0.04}_{-0.14}$ &0.08\\
& Low-Latitude IV Arch & 0.94& 0.15 & $^{+0.26}_{-0.01}$ & 0.08\\
&  Complex~A & \multicolumn{2}{c}{$<0.21$} & $<0.22$ & $<0.001$\\
\tableline
B & Local & 0.94 & 0.07 & $^{+0.13}_{-0.22}$ & 0.08\\
&  Lower IV Arch & 1.08 & 0.04 & $^{+0.10}_{-0.08}$ & 0.09
\\
 & Upper IV Arch & \multicolumn{2}{c}{$< 0.23$} & $<0.45$ & $< 0.01$ \\
\tableline
C & Local & 1.08 & 0.04 & $^{+0.05}_{-0.09}$ & 0.09\\
& IV Spur & 0.68 & 0.06 & $^{+0.09}_{-0.03}$ & 0.06\\
\enddata
\tablecomments{Upper limits are provided at 95\% confidence level.}
\tablenotetext{a}{$1\sigma$ statistical uncertainty {\changes from the likelihood analysis}.}
\tablenotetext{b}{Systematic spread from varying some inputs to the \g-ray 
interstellar emission model (see Sec.~\ref{modelunc}).}
\tablenotetext{c}{{\changes$1\sigma$ uncertainty or 95\% c.l. upper limit from the distributions obtained with the jackknife test (see Sec.~\ref{jackknife}).}}
\end{deluxetable}

\subsubsection{Jackknife Tests}\label{jackknife}

So far we have characterized the statistical uncertainties in the fit 
parameters in Section~\ref{basefit} and systematic uncertainties related to some 
inputs to the model in Section~\ref{modelunc}. However, other potential 
sources of systematic uncertainties are related to {\chbis the} assumption 
that 
the emissivity scaling coefficients for each component are 
constant across a given ROI, and to the presence of regions with 
large deviations or
mismodeled individual sources driving the fits.

To characterize the magnitude of these uncertainties we perform jackknife 
tests where we repeat the analysis described in Section~\ref{basefit} 150~times for 
each ROI, each time masking a circular region with random center within the ROI 
and covering {\changes 20\% of the ROI area. The mask size was chosen to be much larger than the 68\% event containment radius of the LAT for 
front-converting photons at the lowest end of the energy range at 300~MeV ($\sim 
1\fdg5$) and large enough to probe possible mild variations of the CR densities across the complexes studied.}

Figures~\ref{fig:jackknifeA}, \ref{fig:jackknifeB}, and \ref{fig:jackknifeC} show the 
distributions of the $\hi$ emissivity scaling coefficients for the three ROIs. 
The distributions all peak {\changes within $1\sigma$ from}
the best-fit values obtained from the baseline analysis and have standard 
deviations smaller than {\changes or comparable to} the $1\sigma$ statistical uncertainties from the \g-ray fits. This shows 
that from a statistical point of view the obtained parameter {\changes values} are representative of the entire ROIs.
\begin{figure*}[!htbp]\begin{center}
\includegraphics[width=\textwidth]{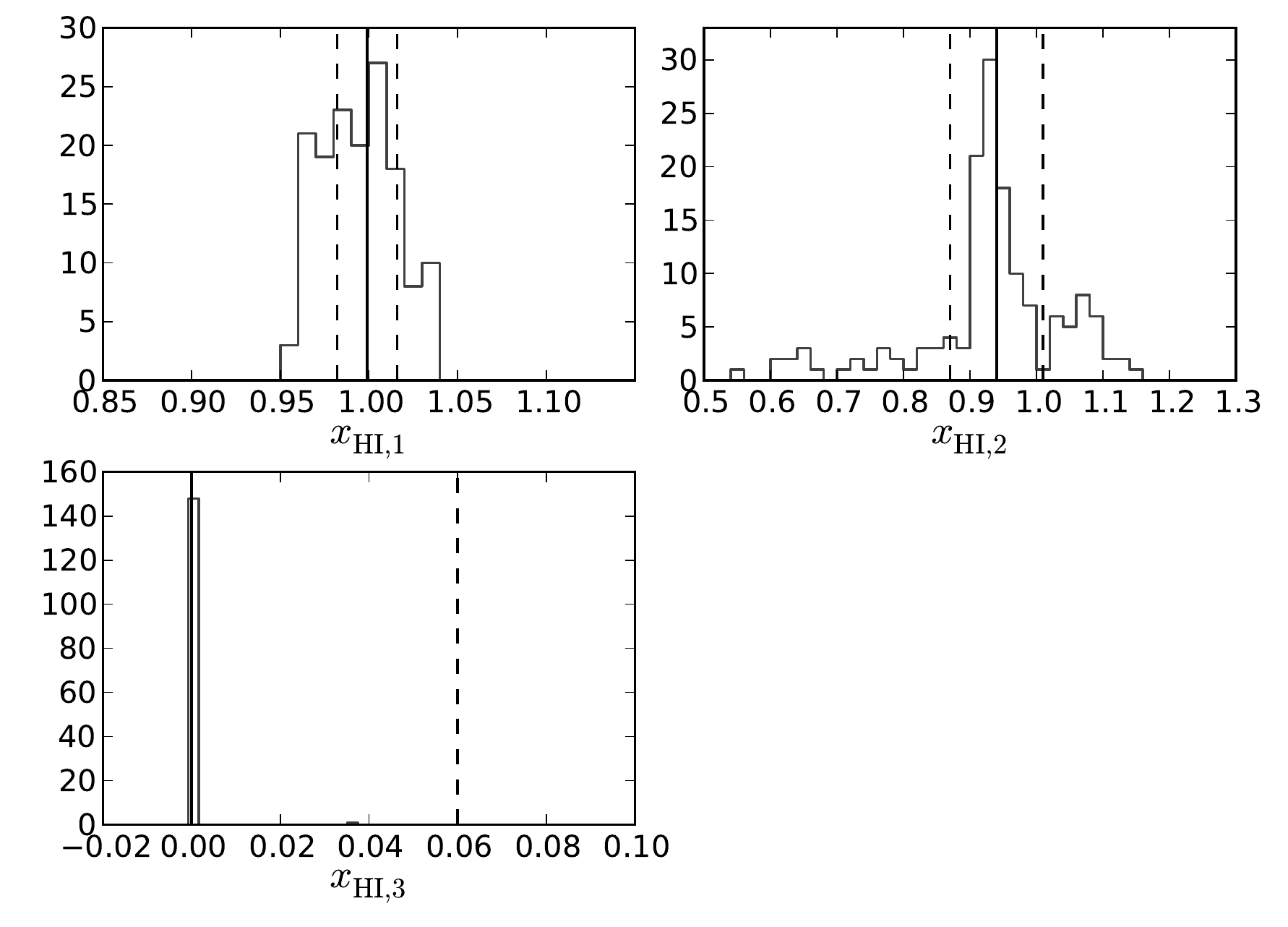}
\caption{Distribution of the $x_\mathrm{HI}$ scaling 
coefficients in ROI~A from the jackknife tests. In each panel the solid 
vertical line marks the best-fit value from the baseline 
analysis, and the dashed lines delimit its $\pm1\sigma$ 
statistical uncertainty interval.}\label{fig:jackknifeA}
\end{center}\end{figure*}
\begin{figure*}[!htbp]\begin{center}
\includegraphics[width=\textwidth]{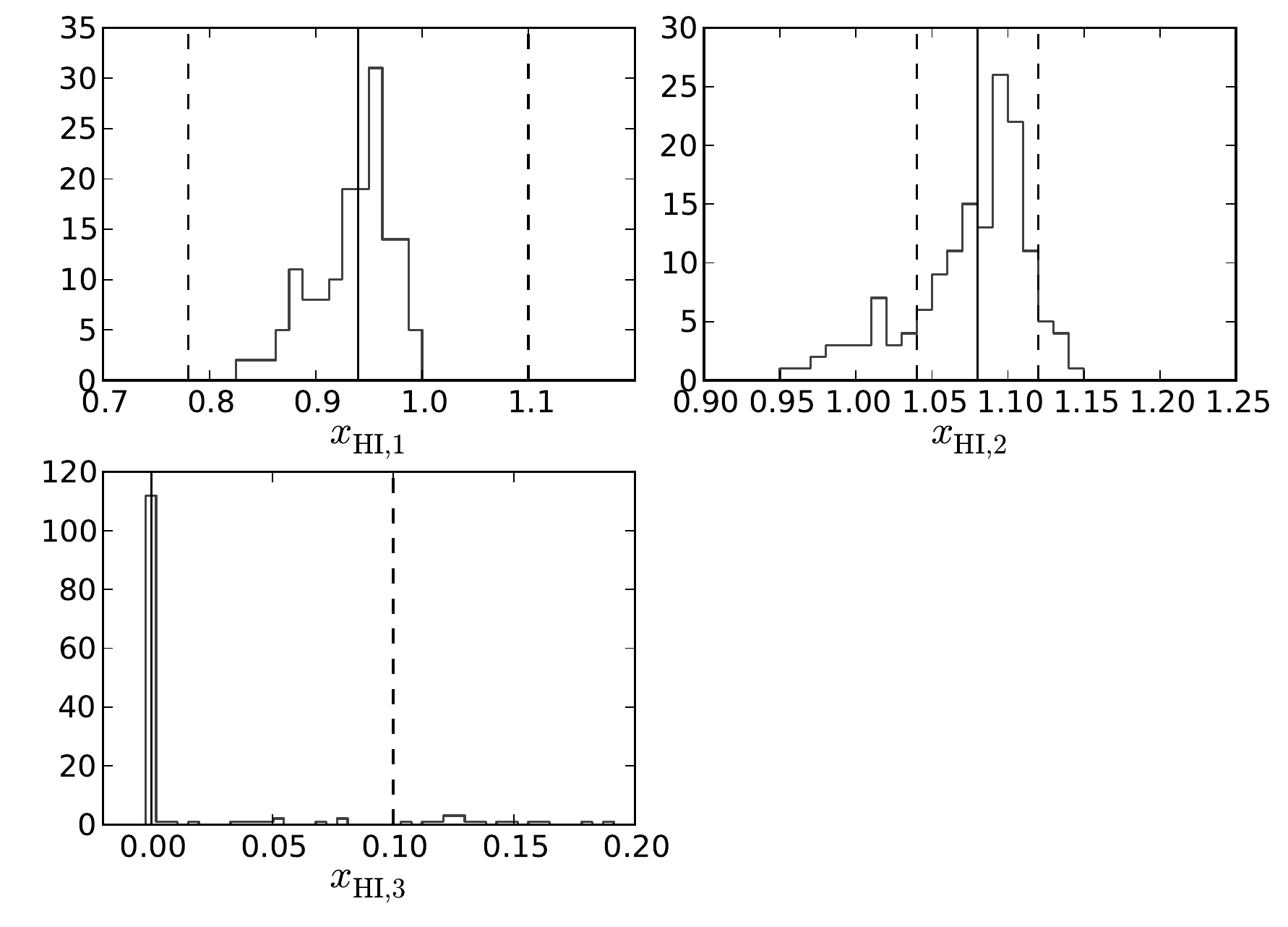}
\caption{Distribution of the $x_\mathrm{HI}$ scaling 
coefficients in ROI~B from the jackknife tests. See 
caption of Fig.~\ref{fig:jackknifeA} for details.}\label{fig:jackknifeB}
\end{center}\end{figure*}
\begin{figure*}[!htbp]\begin{center}
\includegraphics[width=\textwidth]{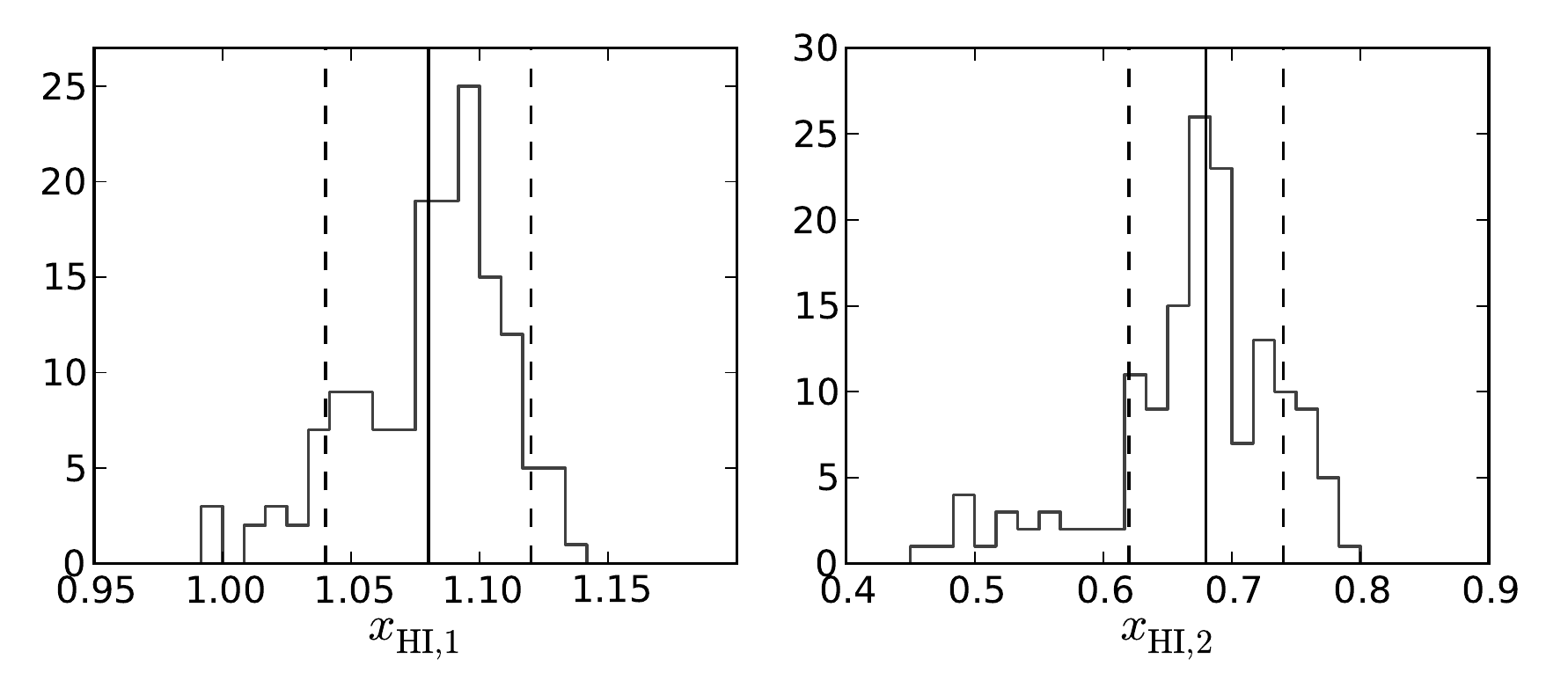}
\caption{Distribution of the $x_\mathrm{HI}$ scaling 
coefficients in ROI~C from the jackknife tests. See 
caption of Fig.~\ref{fig:jackknifeA} for details.}\label{fig:jackknifeC}
\end{center}\end{figure*}

In ROI~A, Figure~\ref{fig:jackknifeA}, the distribution of the scaling 
coefficients for Complex~A, $x_{\mathrm{HI},3}$, narrowly peaks at zero, showing that this parameter is always at the boundary of the allowed values. In the rest of the work, we will make use only of the upper limit on the Complex~A 
emissivity that, on the contrary, is a robustly determined quantity. Similar considerations hold for the upper IV Arch in ROI B.
{\changes In all ROIs, {\chbis most notably in ROI~A,} the distribution of the scaling 
coefficients $x_{\mathrm{HI},2}$  for the IVCs show tails at low or high
values, which {\chbis suggests} that emissivities lower or higher than average are possible within the gas 
encompassed by the respective templates.

To take into account these uncertainties in our estimates of the emissivities we compute, based on the distributions of the fit parameters in Figures~\ref{fig:jackknifeA}, \ref{fig:jackknifeB}, and \ref{fig:jackknifeC}, the $1\sigma$ confidence intervals for detected complexes and 95\% c.l. upper limits for complexes that are not significantly detected, as reported in Table~\ref{hiscaling} (details of the derivation are given in Appendix~\ref{jackknifeerrs}). For the detected complexes the confidence intervals are of the same magnitude as the statistical errors.  Hence we will combine them with the other uncertainties for the discussion that follows. In the case of the complexes that are not significantly detected the sources of uncertainty probed through the jackknife tests (such as possible variations of the CR densities across the complexes studied) have an impact on the emissivity upper limits much smaller than those due to uncertainties in the analysis model inputs and those encoded by the statistical error from likelihood fit, so they will be neglected.
}

\section{RESULTS AND DISCUSSION}\label{resdiscussion}

\subsection{The \g-Ray Emissivities of HVCs and IVCs}

The $\hi$ emissivity scaling coefficients in Table~\ref{hiscaling} can be 
converted into emissivity values. To do so we multiply them by the 
integrated emissivity between 300~MeV and 10~GeV from \citet{casandjian2014}, 
i.e., $7.1 \times 10^{-27}$~s$^{-1}$~sr$^{-1}$~H$^{-1}$. Note that this rests 
on the measurement described in Section~\ref{cloudspectra} that the shape of 
the emissivity spectrum 
is the same as in the local ISM for detected 
complexes, and is valid within the extreme variations parametrized in 
Eq.~\ref{emf1-}, \ref{emf2-}, \ref{emf1+}, \ref{emf2+} for those that 
are not significantly detected.

The conversion into emissivities requires taking into account one 
additional source of systematic uncertainty, i.e., the characterization of 
the LAT instrument response, and in particular its effective area. Based on the 
estimate of the systematic uncertainties of the LAT effective area in 
\citet{ackermann2012LATperf}, the systematic error on the local emissivity by 
\citet{casandjian2014} integrated between 300~MeV and 10~GeV is $0.6 \times 
10^{-27}$~s$^{-1}$~sr$^{-1}$~H$^{-1}$.

Note also that in many HVCs and IVCs the ratio of the ionized gas mass {to} the 
neutral atomic gas mass is much larger than in the local ISM 
and close to unity \citep[e.g.,][]{wakker2008}. If the column densities of 
ionized gas are correlated with those of neutral gas the scaling factors in the 
\g-ray fits will be overestimated to encompass the emission from 
ionized gas. In this respect our estimates of scaling factors and emissivities 
should be regarded as an upper limit to the real values. {The same consideration applies to the potential presence of CO-dark molecular gas in HVCs/IVCs.} Aside from the emissivity scaling coefficient slightly larger than~1 for the lower IV Arch there is no clear indication of such effect from our analysis.

The final results for the emissivities of the targets are summarized in 
Table~\ref{targetem}. For the purpose of interpretation we will rely on those 
findings and on the emissivity scaling factors in Table~\ref{hiscaling}, which 
to a good approximation are not 
subject to uncertainties due to the LAT instrument effective area.
\begin{deluxetable}{lccccccc}
\tabletypesize{\scriptsize}
\tablecaption{Emissivities of HVCs and IVCs\label{targetem}}
\tablewidth{0pt}
\tablehead{
\colhead{Region} & \colhead{Complex} & \colhead{$z$} & 
\colhead{\g-ray emissivity} & 
\colhead{Stat\tablenotemark{a}} & \colhead{Sys 
(LAT)\tablenotemark{b}} & \colhead{Sys (model)\tablenotemark{c}} & \colhead{Sys (Jackknife)\tablenotemark{d}}\\
\colhead{} & \colhead{} & \colhead{(kpc)} & \multicolumn{5}{c}{($10^{-27}$ 
s$^{-1}$ sr$^{-1}$ H$^{-1}$)} 
}
\startdata
A &  Low-Latitude IV Arch & 0.6--1.2 & 6.7 & 1.1 & 0.6 & $^{+1.8}_{-0.1}$ & 0.6\\
 & Complex~A & 2.6--6.8 & \multicolumn{5}{c}{$<1.7$}
\\
\tableline
B &   Lower IV Arch & 0.4--1.7 & 7.7& 0.3& 0.7& $^{+0.7}_{-0.6}$ & 0.6
\\
 &   Upper IV Arch & 0.7--1.7 & \multicolumn{5}{c}{$<3.5$} \\
\tableline
C  & IV Spur & 0.3--2.1 & 4.8 & 0.4 & 0.4 & $^{+0.6}_{-0.2}$ & 0.4\\
\enddata
\tablecomments{$z$ brackets are replicated from Table~\ref{targettable} for the 
reader's convenience. Data sources are listed there.}
\tablecomments{\g-ray emissivities are integrated between 300~MeV and 10~GeV.}
\tablecomments{We provide 95\% confidence level upper limits for the 
worst-case scenario from the model variations considered and 
taking into account uncertainties due to the LAT effective area.}
\tablenotetext{a}{$1\sigma$ statistical uncertainty {\changes from the likelihood analysis}.}
\tablenotetext{b}{Error from uncertainties in the LAT effective area.}
\tablenotetext{c}{Systematic spread from varying some inputs to the \g-ray 
interstellar emission model (see Sec.~\ref{sys_eval}).}
\tablenotetext{d}{{\changes$1\sigma$ uncertainty or 95\% c.l. upper limit from the distributions obtained with the jackknife test (see Sec.~\ref{jackknife}).}}
\end{deluxetable}

The \g-ray emissivities are a proxy for the CR densities in the complexes 
studied. Owing to the cross sections for \g-ray production in nucleon-nucleon 
inelastic collisions \citep[e.g.,][]{kamae2006}, the \g-ray energies studied 
between 300 MeV and 10 GeV constrain the CR densities from {\changes energies of $\sim 3$~GeV/nucleon up to $\sim 200$~GeV/nucleon}. Note that, {\changes while we do not expect significant changes in the He/H ratio because both elements are primordial, the abundances of metals, i.e., elements heavier than He, are known to vary in HVCs and IVCs, and} depending 
on the metallicity of target gas the emissivity per H atom can vary up to 
$< 5\%$ {\changes of the total} \citep[e.g.,][]{mori2009}. Furthermore, the total emissivity depends also on the CR composition and spectra of the different elements {\changes at a level up to $20\%$ to $30\%$ of the total emissivity per H atom} \citep{kachelriess2014}. Part of the emissivity variations 
(Table~\ref{targetem}) or, equivalently, differences in the emissivity scaling 
factors (Table~\ref{hiscaling}) therefore can be attributed to variations in 
the CR elemental composition and of their spectra, as well as to the uncertain metallicity of the targets \citep{wakker2001}.

Our measurement of the local gas emissivity in each ROI agrees well with unity, 
indicating the robustness of the analysis procedure and assumptions. 
{\changes The emissivity of the IV Arch indicates a $\sim 50\%$ decrease of the CR densities with respect to the local value  within 2~kpc from the Galactic plane,} while the HVC Complex A provides the strongest 
limit on the CR densities at a few kpc above the disk. The 
implications of these results are discussed in the following sections. 

\subsection{Testing the Origin of Cosmic Rays in the Galactic Disk}

If CRs originate in the disk of the Milky Way and diffusively propagate in 
the surrounding halo we expect the CR densities, and hence the \g-ray emissivities, 
to decrease with distance $z$ from the disk itself\footnote{{Owing to the distances to other massive star-forming galaxies, any contributions from CRs escaping from even the closest systems would be too small to affect this conclusion and also smaller that the uncertainty of our emissivity measurements.}}. We test this hypothesis 
against our data in Table~\ref{hiscaling} by 
computing the Kendall $\tau$ rank correlation coefficient. Using the emissivity 
scaling coefficients in Table~\ref{hiscaling}, and adopting for $z$ the 
mean within the bracket for targets in Table~\ref{targettable}, and $z=0$~kpc 
for the local gas in each ROI, we obtain $\tau=-0.54$. The negative value 
indicates, indeed, a negative correlation of emissivity with $|z|$.

We calculate the significance of this trend by comparing the $\tau$ 
of the actual data with a distribution of the $\tau$ coefficients obtained from 
the null hypothesis of no correlation between emissivity and $z$.  We generate null 
hypothesis datasets starting from the real dataset using the 
following procedure:
\begin{itemize}
 \item we draw a set of emissivity scaling coefficients with normal 
distribution centered at the measured values and with $\sigma$ equal to the {\changes sum in quadrature of
statistical uncertainties and systematic uncertainties from the jackknife tests} in Table~\ref{hiscaling} for significantly detected 
targets and local foreground gas in each ROI;
 \item we add upper limits at the values measured in 
Table~\ref{hiscaling} for targets which are not significantly detected;
 \item we randomly shift both scaling factors and upper limits with uniform 
probability distribution within the systematic uncertainty brackets {\changes from the model input variations} in 
Table~\ref{hiscaling} (assuming the bracket to be symmetric for upper limits);
\item we draw a set of $z$ values with uniform probability distribution within 
the 
brackets in Table~\ref{targettable}, and add three elements with $z=0$ to 
represent the local foreground gas in each ROI;
 \item we randomly shuffle in an independent fashion the emissivity scaling 
factors and $z$ sets so obtained to be consistent with the null hypothesis of 
no 
correlation between the two quantities. 
\end{itemize}
The distribution of 30000 null hypothesis datasets shows that there is a $2.5\%$ 
chance probability to obtain a $\tau$ coefficient $<-0.54$.

Therefore, our measurements provide evidence at 97.5\% confidence 
level that the \g-ray emissivity per H atom 
decrease{s} with distance from the disk of 
the Galaxy. {\changes This corroborates the notion that CRs at the relevant energies from GeV to TeV originate in the Galactic disk by directly tracing the distribution of CRs in the halo for the first time. Our result goes far beyond the test of the Galactic origin of CRs proposed by \citet{ginzburg1973}, i.e., measurement of the \g-ray emissivity of Magellanic Clouds, that was performed by \citet{sreekumar1993}.}

We note that we also expect a decrease in gas metallicity, 
hence emissivity per H atom, with distance from the disk if the most distant 
targets like Complex~A are primeval gas falling into the Milky Way 
\citep{wakker2001}. However, the expected effect of at most $\sim 5\%$ on the 
\g-ray emissivities \citep{mori2009} is not sufficient to explain the magnitude 
of the variations observed between the scaling coefficients in 
Table~\ref{hiscaling}. Changes in the spectra of the different CR elements could also cause emissivity variations. However, the magnitude of this effect is estimated to be {\changes at most $20\%$ to $30\%$ of the total emissivity per H~atom (taking into account He in both CRs and the ISM)}, lower than the decrease of emissivity of $>50\%$ for the Upper IV Arch and of $> 75\%$ for Complex A that we measured.

Additionally, in the region of the outer Galaxy where our targets are located 
the distance to the center of the Milky Way increases with increasing distance 
from the disk as well. While some propagation models advocate a sizable decrease 
of CR densities with increasing Galactocentric radius, observations 
\citep{abdo2010cascep,ackermann20113quad} indicate that the gradient of CR 
densities is marginal up to 15~kpc from the Galactic center. We conclude that 
the observed emissivity decrease as a function of $z$ is more likely due to 
diffusion in the halo. A quantitative comparison with propagation models 
follows in the next section.   

\subsection{Comparison with CR Propagation Models}
We compare our results with predictions from GALPROP.
We will rely on the set 
of GALPROP models in \citet{LATdiffpapII} that have been compared 
extensively to direct CR measurements and LAT data, and which include 
a number of variations of the most relevant and uncertain  
parameters of the calculations. {\changes In \citet{LATdiffpapII} the CR transport equations were solved for the case of cylindrical symmetry}, based on the assumption that CRs are injected 
in the disk of the Milky Way and propagate in a cylindrical volume with boundaries 
at maximum radius from the center of the Galaxy $R_\mathrm{max}$ and maximum 
height from the disk $z_\mathrm{max}$ {\chbis where the CR particle number densities are required to go to zero}.

Figure~\ref{fig:emgalprop} shows the emissivity scaling factors from 
Table~\ref{hiscaling} against predictions from the models in 
\citet{LATdiffpapII}.
\begin{figure*}\begin{center}
\begin{tabular}{cc}
 \includegraphics[width=0.5\textwidth]{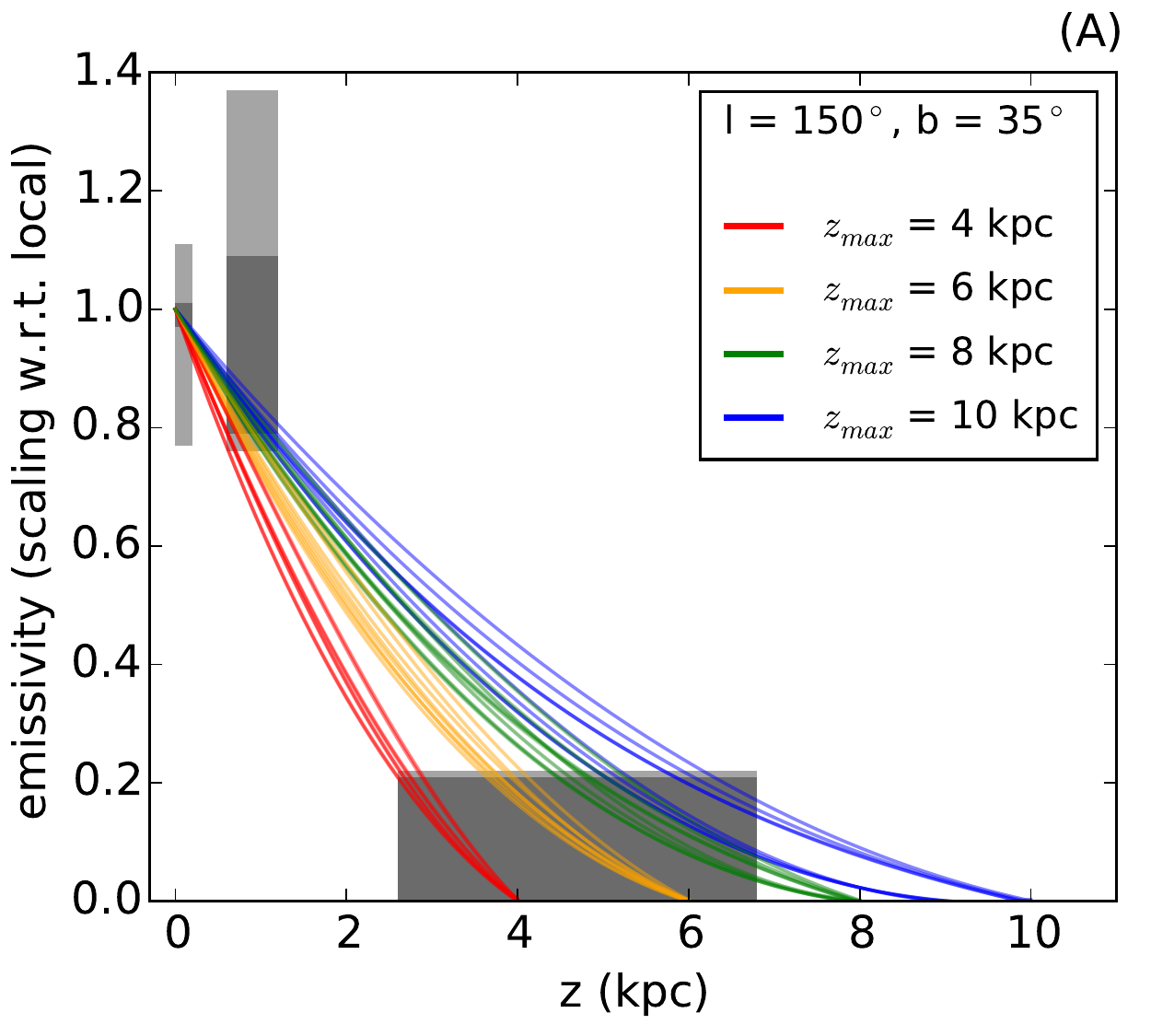}&
 \includegraphics[width=0.5\textwidth]{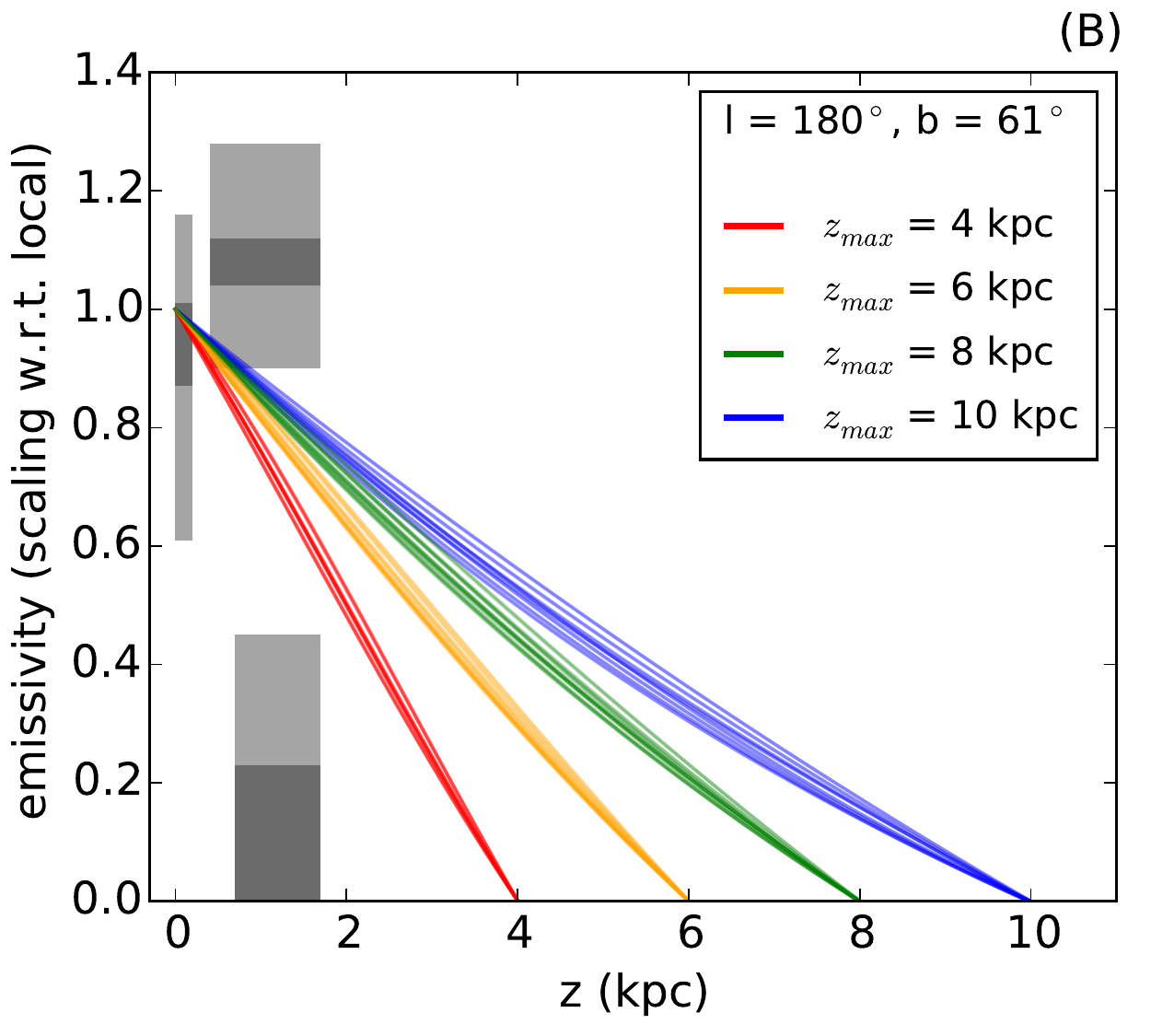}\\
 \includegraphics[width=0.5\textwidth]{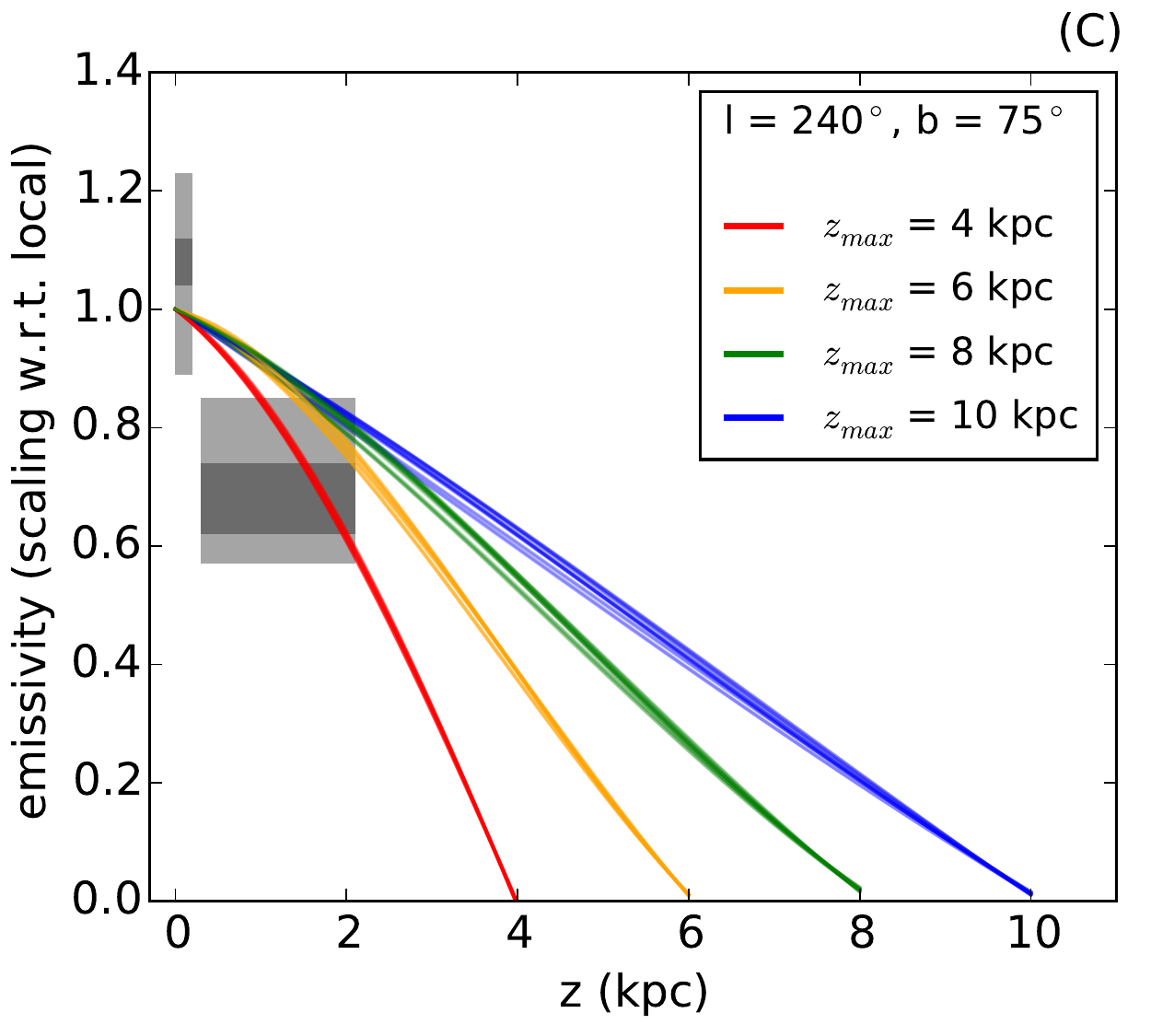}
\end{tabular}
 \caption{For ROIs A, B, and C we compare the emissivity 
scaling factors in Table~\ref{hiscaling} (gray rectangles) with predictions 
from the models in \citet{LATdiffpapII} (curves). 
The horizontal widths of the rectangles indicate the $z$ brackets of 
target IVCs and HVCs, {\changes i.e., the range between lower and upper limits on their 
altitudes} (Table~\ref{targettable}). The 
dark gray rectangles have vertical size corresponding to the statistical 
uncertainties, while for the light gray rectangles the vertical size 
encompasses the
sum {\changes in quadrature of statistical uncertainties and systematic uncertainties from the jackknife tests, augmented by the systematic spread from variations to the analysis model inputs}
from Table~\ref{hiscaling}. The emissivity of local gas 
is assigned to the range from $z=0$~kpc to $z=0.3$~kpc (disk). The model curves 
from 
\citet{LATdiffpapII} were calculated for the line of 
sight indicated in the legend of each panel, approximately 
corresponding to the column density peaks of the target 
complexes. The curves are color-coded based on the 
maximum heights \zm\ of the CR confinement halo in the 
models.}\label{fig:emgalprop}              
\end{center}\end{figure*}
{\changes The model predictions are calculated for distances along the line of sight approximately corresponding to the peak gas column densities for each ROI}\footnote{{\changes This accounts for the different ranges of Galactocentric radius in each line of sight.} The 
effect is clearly visible for ROI~A in Figure~\ref{fig:emgalprop} where the 
profiles are concave, and the curves for $z_\mathrm{max}=10$~kpc bifurcate into 
two families for $R_\mathrm{max}=20$~kpc and $R_\mathrm{max}=30$~kpc due to 
particle escape at the radial boundary.}. We normalized the model emissivities 
to the 
value in the disk at the solar circle. The model emissivities at the solar 
circle integrated from 300~MeV to 10~GeV in \g-ray energy were found to be 
within $\sim 7\%$ from the measurement in \citet{casandjian2014}. {\changes Note that the calculation of the emissivities in \citet{LATdiffpapII} only accounts for interactions between CR and interstellar gas nuclei with atomic number $\leq 2$. Rescaling to the locally measured emissivity takes into account the contribution from metals in the targets and heavier CR species. Variations of the target gas metallicities are not taken into account in these models, and could produce a further decrease of emissivity up to at most $5\%$ with increasing $z$ \citep{mori2009}.}

The model parameter with the largest impact on the vertical gradient of CR 
densities, or, equivalently, \g-ray emissivities, is the maximum height of the 
confinement halo \zm. Models in \citet{LATdiffpapII} assume values of \zm\ 
between 4~kpc and 10~kpc. {\changes While there is a broad agreement between models and measurements, the rapid decrease of the emissivities within 2~kpc from the disk inferred from the upper limit for the upper IV Arch seems to favor smaller values of \zm. The lower and upper IV Arch have similar altitude brackets, $0.4-1.7$~kpc and $0.7-1.7$~kpc, respectively, but their emissivities are consistent with local for the first and $<50\%$ of local for the latter. This is not necessarily contradictory {\chbis because} the distance brackets are pairs of upper and lower limits on the distances; hence the two clouds may be separated by a physical distance up to $\sim~1$~kpc.}

To obtain a crude estimate of \zm\ from our results we 
fit to the emissivity scaling factors of IVCs and HVCs from 
Table~\ref{hiscaling} a function
\begin{equation}\label{emgradientfun}
 \varepsilon(z)=1-\left(\frac{z}{z_\mathrm{max}}\right)^\beta
\end{equation}
where $\varepsilon$ is the emissivity expressed as a fraction of the emissivity 
at $z=0$, \zm\ is the maximum height of the confinement halo and $\beta$ is an 
index which modifies the steepness of the vertical gradient. 
Eq.~\ref{emgradientfun} with $\beta=1$ is {\changes an accurate $z$-dependent part of the solution of the diffusion equation in the plane-parallel geometry (infinitely thin Galactic plane) with a uniform source distribution when only ionization losses are assumed \citep[e.g.,][]{jones2001}.} The GALPROP models in 
\citet{LATdiffpapII} have vertical emissivity gradients at given 
Galactocentric radius characterized by $1 < 
\beta < 1.5$. Values of $\beta$ greater than 1 result from energy dependent escape at the 
radial boundary of the propagation model and, to a lesser extent, {from} energy 
losses of nuclei.  In any case, we stress that the functional form in 
Eq.~\ref{emgradientfun} holds only in diffusion-dominated scenarios where it is 
required that the CR densities go to zero at $z=z_\mathrm{max}$.

We perform the fit by using a maximum likelihood method where we neglect the 
different Galactocentric distances 
of the targets. We assume for the emissivity scaling 
factors a Gaussian probability distribution with $\sigma$ given by the 
quadratic sum of statistical uncertainties, {jackknife uncertainties, and the largest brackets from model input variations} 
from Table~\ref{hiscaling}. Upper limits including systematic uncertainties 
from Table~\ref{hiscaling} are assumed to be the 95\% containment value of a 
Gaussian probability distribution. Finally, 
we assume that the probability 
distribution of target distances is uniform within the brackets in 
Table~\ref{targettable}. Owing to the paucity of the data points, rather 
than 
fitting the index $\beta$ to the data we assume the extreme values $\beta=1$ 
and $\beta=1.5$ {\changes as found from the fits to GALPROP predictions}. The best fit to the data is obtained for 
$z_\mathrm{max}=(2.2\pm1.9)$~kpc, and $z_\mathrm{max}=(1.8\pm2.1)$~kpc for the 
two values of $\beta$. Note that the profile would be steeper, hence 
\zm\ smaller, if the emissivities were biased by the presence of unaccounted 
ionized gas {in HVCs/IVCs}. 

{\chbis This method provides an} estimate of $z_\mathrm{max}<6$~kpc ($2\sigma$ upper containment value from the fits above). Most of the constraining power on the \zm\ parameter, however, at present comes from the upper limit on the 
emissivity of the upper IV Arch. The latter hints at a possible tension with indications from the modeling of other observables related to CRs. {\chbis Differences with respect to studies of synchrotron emission \citep{orlando2013} can be understood if the densities of CR electrons beyond the boundaries of the confinement halo are sufficient to produce the emission observed in the domain from radio to microwaves. However,} the interpretation of the flat radial 
profile of \g-ray emission produced by interactions of CR nuclei in the outer disk of the Milky Way in terms of a large 
\zm\ of the order of $\sim 10$~kpc in the 
context of GALPROP models similar to those considered here \citep{abdo2010cascep,ackermann20113quad} seems to be disfavored.

{\changes  
The observed emissivities of the HVC and IVCs were compared to predictions from the limited set of  diffusive reacceleration models for a simplified axisymmetric model of the Galaxy in \citet{LATdiffpapII}. Other models proposed in the literature include CR-driven Galactic winds and anisotropic diffusion \citep[e.g.,][]{breitschwerdt2002}, and a non-uniform diffusion coefficient that increases with the Galactocentric radius and the distance from the Galactic plane \citep[e.g.,][]{shibata2007}, as well as {\chbis more sophisticated ways of modeling the Milky Way structure} \citep[e.g.,][]{werner2015}, but further comparison to theoretical models is beyond the scope of this article.


Finally, we note that the CR propagation equations are often solved assuming that the CR particle number densities go to zero at the boundaries of the propagation volume. This assumption has little effect on the predicted CR fluxes for the Galaxy as a whole and at a considerable distance from the boundaries. However, close to the halo boundaries one may expect significant deviations from the model predictions and tracing the CR distribution in the halo can provide unique model-independent information about its structure and extent. Studies of the CR outflow from the Milky Way into the intergalactic space that are still in their infancy \citep[e.g.,][]{everett2012} could also benefit from the method illustrated in this work.}

\section{CONCLUSIONS}\label{conclusions}

We have searched for high-energy \g-ray emission from a sample of HVCs and IVCs 
in the halo of the Milky Way at different distances from the 
Galactic disk using 73~months of data from the LAT in the energy range 
between 300 MeV and 10 GeV.

We have achieved the first detections of IVCs in \g~rays, the low-latitude IV Arch, the lower IV Arch, and the IV Spur,  and set constraints on 
the emission from the remaining targets, the upper IV Arch and the HVC Complex~A (Table~\ref{hiscaling}). The spectra of 
the detected complexes were all consistent within statistical uncertainties 
with that of gas in the local interstellar space.  

We find evidence at 97.5\% confidence level that the \g-ray emissivity per H 
atom of the clouds 
decreases as a function of distance from the 
disk. This corroborates the {\chbis notion} that CRs are accelerated in 
the Galactic disk and then propagate in a surrounding halo, for the 
first time in a direct way.

The \g-ray emissivity per H atom {\changes of the upper IV Arch hints at a 50\% decline of 
the CR densities within 2~kpc from the disk}. The upper limit on 
the emissivity of Complex~A at 22\% of 
the local value gives the most stringent constraint to date on the fluxes of 
CRs at few kpc from the Milky Way disk.


Our results 
can {\changes be compared to CR} propagation models and inform
further development to 
more realistically take into account
the escape of CRs from the halo of the 
Milky Way and the outflow of particles merging into the local intergalactic 
medium.
At the same time, 
the observational constraints can be improved if distance brackets for more HVCs and IVCs in the 
mass-distance range detectable by the LAT are measured {\changes or existing brackets are made more constraining}, e.g., 
by the ongoing  \textit{Gaia} survey
\citep[e.g.,][]{debruijne2012}.

\acknowledgments

The \textit{Fermi} LAT Collaboration acknowledges generous ongoing support
from a number of agencies and institutes that have supported both the
development and the operation of the LAT as well as scientific data analysis.
These include the National Aeronautics and Space Administration and the
Department of Energy in the United States, the Commissariat \`a l'Energie 
Atomique
and the Centre National de la Recherche Scientifique / Institut National de 
Physique
Nucl\'eaire et de Physique des Particules in France, the Agenzia Spaziale 
Italiana
and the Istituto Nazionale di Fisica Nucleare in Italy, the Ministry of 
Education,
Culture, Sports, Science and Technology (MEXT), High Energy Accelerator Research
Organization (KEK) and Japan Aerospace Exploration Agency (JAXA) in Japan, and
the K.~A.~Wallenberg Foundation, the Swedish Research Council and the
Swedish National Space Board in Sweden.
 
Additional support for science analysis during the operations phase is gratefully acknowledged from the Istituto Nazionale di Astrofisica in Italy and the Centre National d'\'Etudes Spatiales in France.

This work was partially funded by NASA grant NNX13O87G. {\changes I.~V.~M., {\chbis E.~O.,} and T.~A.~P. acknowledge NASA's support of GALPROP development through grant NNX13AC47G}.

The authors acknowledge the use of HEALPix 
\citep[\url{http://healpix.jpl.nasa.gov/},][]{gorski2005} in some parts of the 
analysis. { We made use of data products based on observations obtained with \P\ (\url{http://www.esa.int/Planck}), an ESA science mission with instruments and contributions directly funded by ESA Member States, NASA, and Canada.}

{\chbis L.~T. thanks B.~P.~Wakker for an insightful discussion on the determination 
of distance brackets for IVCs and HVCs, as well as E.~Charles and P.~Mertsch for stimulating conversations.}



{\it Facilities:} \facility{Fermi}, \facility{Planck}.



\appendix

\section{ADDITIONAL RESULTS ON THE INTERSTELLAR MEDIUM}\label{dustresults}
The analysis described in this paper is aimed at determining the CR content of 
IVCs and HVCs. Nevertheless, we obtained results relevant to the  
properties of interstellar gas and dust in such clouds, as well as in the local 
complexes seen in their foregrounds. This appendix summarize{s} these results, 
including some methodological aspects.

\subsection{Example of Results from the Iterative Dust Fitting Procedure}
In this section we take as an example ROI~A to illustrate some details of the 
iterative fitting procedure applied to the dust maps in Section~\ref{dustfit}. 
We 
follow the fit of the radiance map, $R$, for the assumption that the errors are 
proportional to the model $M$ from Eq.~\ref{eq:dustmodel} in calculating 
the $\chi^2$. The results from all 
iterations 
are shown in {\chbis Figures}~\ref{fig:dustfit_pars}, \ref{fig:dustfit_parvar}, 
and~\ref{fig:dustfit_chi2}.
\begin{figure}[!htbp]\begin{center}
\plotone{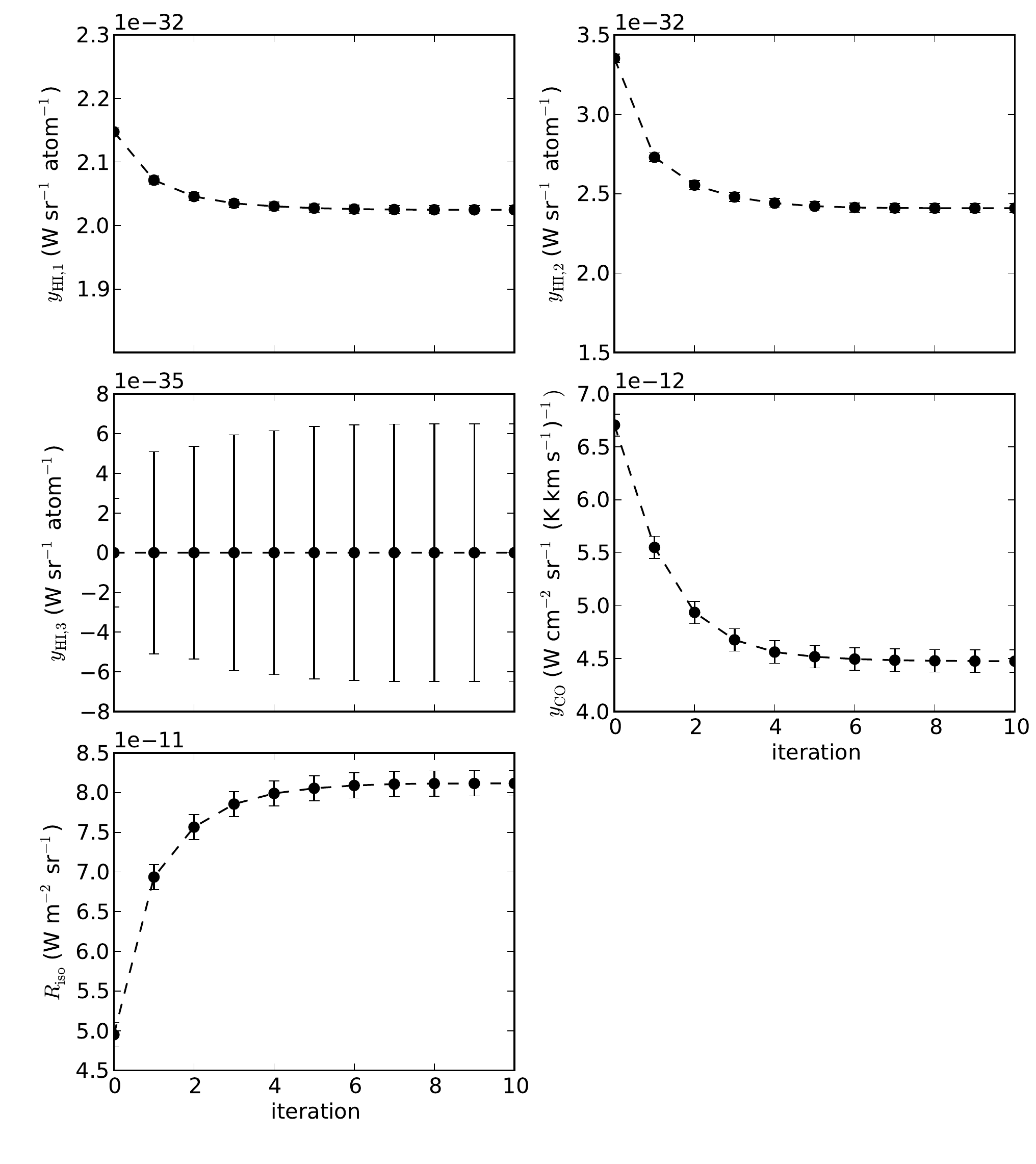}
\caption{Best-fit parameters of the model described in 
Eq.~\ref{eq:dustmodel} for ROI~A in each iteration of the dust analysis. 
We show as an example the case for which the radiance map 
$R$ is considered and the errors in the $\chi^2$ definition are assumed to be 
proportional to $M$.}\label{fig:dustfit_pars}
\end{center}\end{figure}
\begin{figure}[!htbp]\begin{center}
\plotone{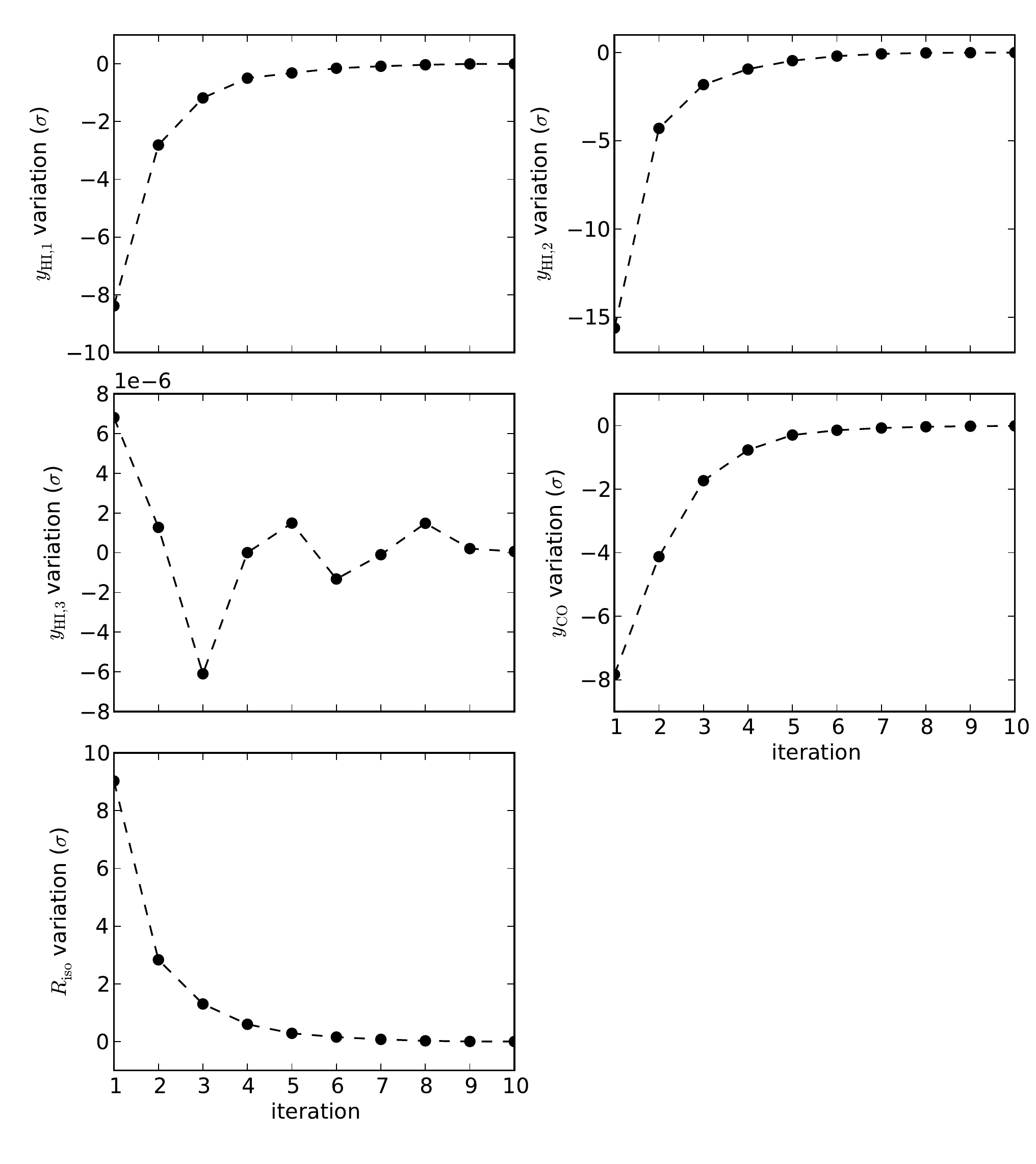}
\caption{Variation of the best-fit parameters of the model described in 
Eq.~\ref{eq:dustmodel} for ROI~A in each iteration of the dust analysis with 
respect to the previous one (expressed as the number of $1\sigma$ statistical 
uncertainties from the fits). 
We show as an example the case for which the radiance map 
$R$ is considered and the errors in the $\chi^2$ definition are assumed to be 
proportional to $M$.}\label{fig:dustfit_parvar}
\end{center}\end{figure}
\begin{figure}[!htbp]\begin{center}
\plottwo{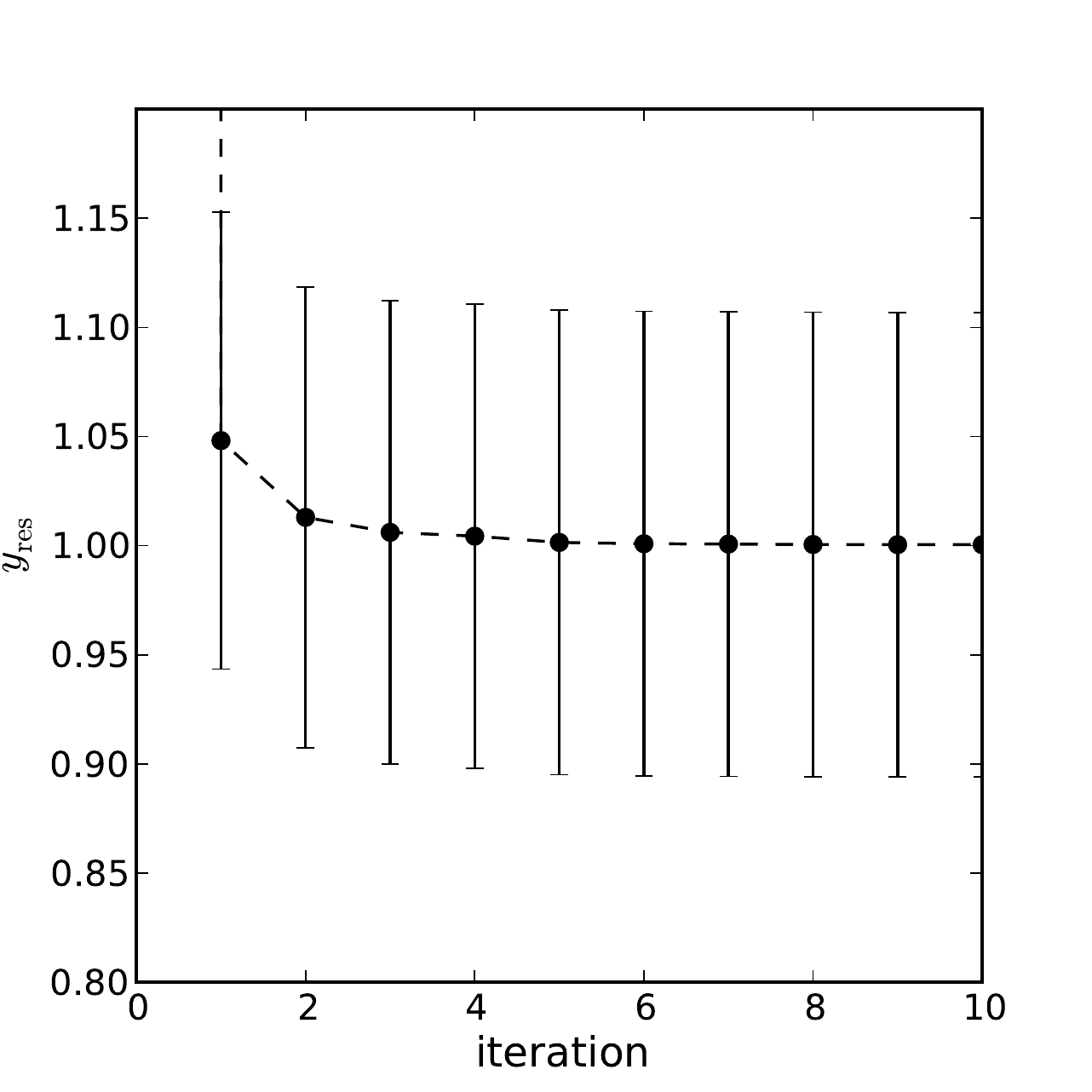}{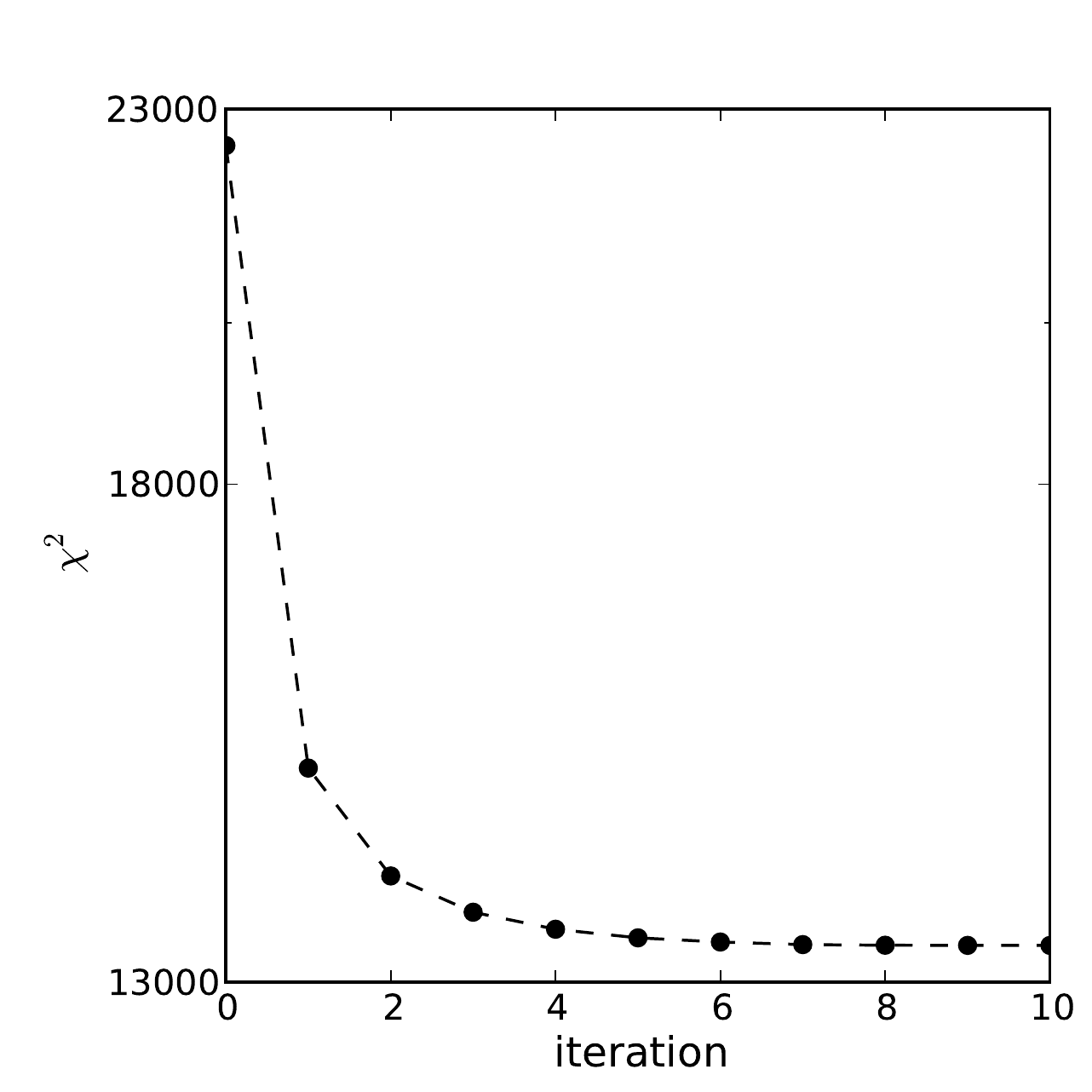}
\caption{Normalization of the residual dust component, $y_\mathrm{res}$, 
(left), and $\chi^2$ value (right) for ROI~A in each iteration of the 
dust analysis. 
We show as an example the case for which the radiance map 
$R$ is considered and the errors in the $\chi^2$ definition are assumed to be 
proportional to $M$.}\label{fig:dustfit_chi2}
\end{center}\end{figure}

Figures~\ref{fig:dustfit_pars} and~\ref{fig:dustfit_parvar} show how the scaling 
coefficients of the $\hi$ and CO maps {\chbis change significantly} during the iterative 
procedure and how they stabilize at a final value at the end. This can 
be 
interpreted as 
removing
the bias due to the missing DNM component in the 
initial iteration, which is alleviated as the missing component is recovered 
based on the data themselves.
%


Figure~\ref{fig:dnmiter} shows the difference between the filtered dust residual 
maps from the last iteration and the initial one. It is evident that borders exist 
between the $\hi$-dominated and DNM-dominated, and between the 
DNM-dominated and CO-dominated regions, where the missing component is not 
recovered from the residuals in the initial iteration. This can be understood 
because the $\hi$ and CO components in the initial fit 
compensating for 
the missing DNM component.
\begin{figure}[!htbp]
\begin{center}
\includegraphics[width=0.5\textwidth]{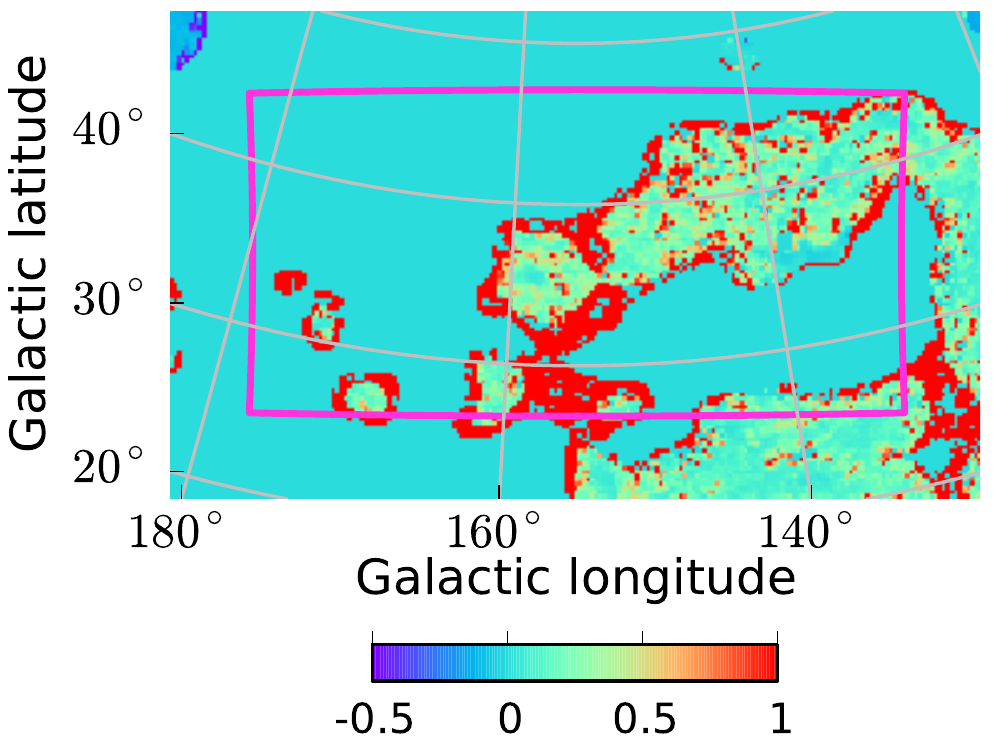}
\caption{Difference of the filtered dust residual map 
from the last iteration minus the map from iteration 
zero, divided by the former. We show as an example in ROI~A the case for which the 
radiance map 
$R$ is considered and the errors in the definition of the $\chi^2$ are assumed 
to be 
proportional to $M$ as defined in Eq.~\ref{eq:dustmodel}. The map is 
shown in the \textit{plate carr\'ee} projection used in the 
dust and \g-ray fits for ROI~A, and the magenta line shows 
the 
border of the region considered in the fit.}\label{fig:dnmiter}
\end{center}
\end{figure}

Finally, we note that while in the initial iterations of the fits there are 
often both positive and negative significant residuals, only positive ones are 
found at the end of the iterative procedure. This is consistent with our 
working hypothesis that the missing DNM component traced by the dust maps is 
recovered by the template determined from the iterative procedure.

\subsection{Fitting Coefficients for the Dust Maps}
The final fit coefficients, reduced $\chi^2$ values, and error scaling factors for 
all the fits performed are given in Tables~\ref{dustfit:R_A}, 
\ref{dustfit:tau_A}, \ref{dustfit:R_B}, {\ref{dustfit:tau_B}}, \ref{dustfit:R_C}, and
\ref{dustfit:tau_C}. In ROI~B there are no significant positive residuals 
when fitting $\tmap$, so the iterative procedure was not performed. The number 
of iterations varies depending on the region 
and the number of free parameters considered, ranging from 
at least five in ROI~B up to at most 30 in ROI~A.

\begin{deluxetable}{lcc}
\tabletypesize{\scriptsize}
\tablecaption{Results from $R$ fit in ROI~A\label{dustfit:R_A}}
\tablewidth{0pt}
\tablehead{
\colhead{}&\colhead{$\sigma \propto R$}&\colhead{$\sigma \propto M$}
}
\startdata
$\sigma$ scaling factor & 15\% & 12\%\\
Number of iterations & 12& 11\\
$\chi^2$/n.d.f. & 1.006 & 1.000\\
$y_{\mathrm{HI},1}$\tablenotemark{a}  & $2.160\pm0.008$ & 
$2.025\pm0.007$\\
$y_{\mathrm{HI},2}$\tablenotemark{a}  & $3.19\pm0.04$ & 
$2.41\pm0.03$\\
$y_{\mathrm{HI},3}$\tablenotemark{a}  & $0.00\pm0.01$ & 
$0.000\pm0.006$\\
$y_\mathrm{CO}$\tablenotemark{b}  & $3.84\pm0.18$ & 
$4.48\pm0.10$\\
$y_\mathrm{res}$ & $1.001\pm0.002$ & $1.000\pm0.007$\\
$R_\mathrm{iso}$\tablenotemark{c}  & $0.731\pm0.019$ & 
$0.811\pm0.016$\\
\enddata
\tablenotetext{a}{10$^{-32}$ W sr$^{-1}$ H$^{-1}$}
\tablenotetext{b}{10$^{-12}$ W cm$^{-2}$ sr$^{-1}$ (K \kms)$^{-1}$}
\tablenotetext{c}{10$^{-10}$ W m$^{-2}$ sr$^{-1}$}
\end{deluxetable}

\begin{deluxetable}{lccc}
\tabletypesize{\scriptsize}
\tablecaption{Results from $\tmap$ fit in ROI~A\label{dustfit:tau_A}}
\tablewidth{0pt}
\tablehead{
\colhead{}& \colhead{$\sigma \propto \sigma_\tau$} & \colhead{$\sigma \propto 
\tmap$}&\colhead{$\sigma \propto M$}
}
\startdata
$\sigma$ scaling factor & 4.9 & 23\% & 24\%\\
Number of iterations & 11 & 9 & 5\\
$\chi^2$/n.d.f. & 0.993 & 1.003 & 1.000\\
$y_{\mathrm{HI},1}$\tablenotemark{a}  & $0.738\pm0.002$ & $0.728\pm0.002$ & 
$0.806\pm0.003$\\
$y_{\mathrm{HI},2}$\tablenotemark{a}  & $0.879\pm 0.013$ & $0.859\pm0.016$ & 
$0.899\pm0.021$\\
$y_{\mathrm{HI},3}$\tablenotemark{a}  & $0.01\pm0.03$ & $0.00\pm0.01$ & 
$0.000\pm0.006$\\
$y_\mathrm{CO}$\tablenotemark{b}  & $2.30\pm0.08$ & $2.31\pm0.09$ & 
$3.2\pm0.2$\\
$y_\mathrm{res}$ & $0.995\pm0.014$ & $0.997\pm0.013$ & $1.001\pm0.001$\\
$\tau_\mathrm{iso}$\tablenotemark{c}  & $0.00\pm0.01$ & $0.000\pm0.012$ & 
$0.000\pm0.005$\\
\enddata
\tablenotetext{a}{10$^{-26}$ cm$^2$ H$^{-1}$}
\tablenotetext{b}{10$^{-6}$  (K \kms)$^{-1}$}
\tablenotetext{c}{10$^{-9}$ mag}
\end{deluxetable}

\begin{deluxetable}{lcc}
\tabletypesize{\scriptsize}
\tablecaption{Results from $R$ fit in ROI~B\label{dustfit:R_B}}
\tablewidth{0pt}
\tablehead{
\colhead{}&\colhead{$\sigma \propto R$}&\colhead{$\sigma \propto M$}
}
\startdata
$\sigma$ scaling factor & 14\% & 17\%\\
Number of iterations & 6& 5\\
$\chi^2$/n.d.f. & 0.996 & 1.008\\
$y_{\mathrm{HI},1}$\tablenotemark{a}  & $2.187\pm0.010$ & 
$2.246\pm0.013$\\
$y_{\mathrm{HI},2}$\tablenotemark{a}  & $1.982\pm0.011$ & 
$2.168\pm0.014$\\
$y_{\mathrm{HI},3}$\tablenotemark{a}  & $0.46\pm0.03$ & 
$1.07\pm0.04$\\
$y_\mathrm{res}$ & $0.986\pm0.015$ & $0.98\pm0.06$\\
$R_\mathrm{iso}$\tablenotemark{b}  & $0.935\pm0.012$ & 
$1.039\pm0.016$\\
\enddata
\tablenotetext{a}{10$^{-32}$ W sr$^{-1}$ H$^{-1}$}
\tablenotetext{b}{10$^{-10}$ W m$^{-2}$ sr$^{-1}$}
\end{deluxetable}

 \begin{deluxetable}{lccc}
 \tabletypesize{\scriptsize}
 \tablecaption{Results from $\tmap$ fit in ROI~B\label{dustfit:tau_B}}
 \tablewidth{0pt}
 \tablehead{
 \colhead{}& \colhead{$\sigma \propto \sigma_\tau$} & \colhead{$\sigma \propto 
 \tmap$}&\colhead{$\sigma \propto M$}
 }
 \startdata
 $\sigma$ scaling factor & 7.0 & 36\% & 32\%\\
 Number of iterations & 5 & 6 & 8\\
 $\chi^2$/n.d.f. & 0.994 & 1.003 & 0.994\\
 $y_{\mathrm{HI},1}$\tablenotemark{a}  & $0.729\pm0.005$ & $0.680\pm0.003$ & 
 $0.708\pm0.006$\\
 $y_{\mathrm{HI},2}$\tablenotemark{a}  & $0.390\pm 0.005$ & $0.351\pm0.003$ & 
 $0.401\pm0.007$\\
 $y_{\mathrm{HI},3}$\tablenotemark{a}  & $0.146\pm0.012$ & $0.152\pm0.012$ & 
 $0.221\pm0.018$\\
 $y_\mathrm{res}$ & $1.03\pm0.02$ & $0.98\pm0.02$ & $1.07\pm0.06$\\
 $\tau_\mathrm{iso}$\tablenotemark{b}  & $0.50\pm0.06$ & $0.00\pm0.01$ & 
 $2.60\pm0.08$\\
 \enddata
 \tablenotetext{a}{10$^{-26}$ cm$^2$ H$^{-1}$}
 \tablenotetext{b}{10$^{-9}$ mag}
 \end{deluxetable}

\begin{deluxetable}{lcc}
\tabletypesize{\scriptsize}
\tablecaption{Results from $R$ fit in ROI~C\label{dustfit:R_C}}
\tablewidth{0pt}
\tablehead{
\colhead{}&\colhead{$\sigma \propto R$}&\colhead{$\sigma \propto M$}
}
\startdata
$\sigma$ scaling factor & 11\% & 13\%\\
Number of iterations & 6& 7\\
$\chi^2$/n.d.f. & 1.058 & 1.027\\
$y_{\mathrm{HI},1}$\tablenotemark{a}  & $2.045\pm0.011$ & 
$2.259\pm0.013$\\
$y_{\mathrm{HI},2}$\tablenotemark{a}  & $2.124\pm0.019$ & 
$2.32\pm0.02$\\
$y_\mathrm{res}$ & $1.000\pm0.014$ & $1.00\pm0.04$\\
$R_\mathrm{iso}$\tablenotemark{b}  & $1.38\pm0.02$ & 
$1.25\pm0.02$\\
\enddata
\tablenotetext{a}{10$^{-32}$ W sr$^{-1}$ H$^{-1}$}
\tablenotetext{b}{10$^{-10}$ W m$^{-2}$ sr$^{-1}$}
\end{deluxetable}

\begin{deluxetable}{lccc}
\tabletypesize{\scriptsize}
\tablecaption{Results from $\tmap$ fit in ROI~C\label{dustfit:tau_C}}
\tablewidth{0pt}
\tablehead{
\colhead{}& \colhead{$\sigma \propto \sigma_\tau$} & \colhead{$\sigma \propto 
\tmap$}&\colhead{$\sigma \propto M$}
}
\startdata
$\sigma$ scaling factor & 5.4 & 23\% & 26\%\\
Number of iterations & 8 & 7 & 6\\
$\chi^2$/n.d.f. & 1.018 & 1.018 & 0.994\\
$y_{\mathrm{HI},1}$\tablenotemark{a}  & $0.745\pm0.003$ & $0.682\pm0.003$ & 
$0.865\pm0.004$\\
$y_{\mathrm{HI},2}$\tablenotemark{a}  & $0.320\pm 0.006$ & $0.300\pm0.005$ & 
$0.377\pm0.007$\\
$y_\mathrm{res}$ & $1.001\pm0.004$ & $0.992\pm0.011$ & $1.033\pm0.039$\\
$\tau_\mathrm{iso}$\tablenotemark{b}  & $0.000\pm0.005$ & $0.000\pm0.003$ & 
$0.000\pm0.006$\\
\enddata
\tablenotetext{a}{10$^{-26}$ cm$^2$ H$^{-1}$}
\tablenotetext{b}{10$^{-9}$ mag}
\end{deluxetable}

The scaling factors for the errors on $\tmap$ given by 
the \citet{planck2013dustmaps} are $\sim 5$. Possible reasons for this are 
discussed in Section \ref{dustfit}. The scaling factors 
obtained when assuming $\sigma \propto R$ are 10\% to 15\%; when assuming 
$\sigma 
\propto \tmap$ they are $\sim 25\%$. It can be expected that the dispersion on $R$ is 
smaller because this is an integrated quantity rather than 
referring to a specific frequency.

We note that the $y_{\mathrm{HI},\imath}$ coefficients can have a physical 
interpretation in terms of average specific power, i.e., power radiated by 
dust per H atom, per unit solid angle, $\overline{R/N(\mathrm{H})}$, and 
average specific opacity, i.e., optical depth of the dust per H atom, 
$\overline{\tmap/N(\mathrm{H})}$. {Non-zero coefficients for IVCs are consistent with previous measurements of dust thermal emission from this class of objects \citep[e.g.,][Chapter 9, and references therein]{hvcassl2005}, and, for ROIs~A and B, with the measurements of metallicities close to solar (Table~\ref{targettable}). On the other hand, the $y_{\mathrm{HI},3}$ coefficient for Complex A being compatible with~0 in Tables~\ref{dustfit:R_A} and~\ref{dustfit:tau_A} is consistent with the low metallicity measured in this cloud \citep{wakker2001}.}

\subsection{Differences between Alternative Determinations of the DNM Column 
Density Templates}
In Figure~\ref{fig:dnmdiff} we show the relative difference between different 
determinations of the DNM column density templates for ROI~A, namely the 
difference between the baseline determination of the DNM column densities based 
on $R$ assuming $\sigma \propto M$  (see Eq.~\ref{eq:dustmodel} for a definition of the quantities), and from two alternative determinations 
based 
on $R$ assuming $\sigma \propto R$ (left), and based 
on $\tmap$ assuming $\sigma \propto M$ (right). For the second case we 
calculated the average $R/\tmap$ ratio in non-empty pixels of the two maps and 
used it to convert $\tmap$ in equivalent $R$.

The figure illustrates how the different assumptions influence the final DNM 
template, for example how some regions have 
DNM only for one of the 
cases and not the other, and how the contrast within the map can change. For 
these reasons we considered the different determinations of the DNM template in 
the assessment of the systematic uncertainties for the \g-ray analysis 
in Section~\ref{sys_eval}.

\begin{figure}[!htbp]
\begin{center}
\plotone{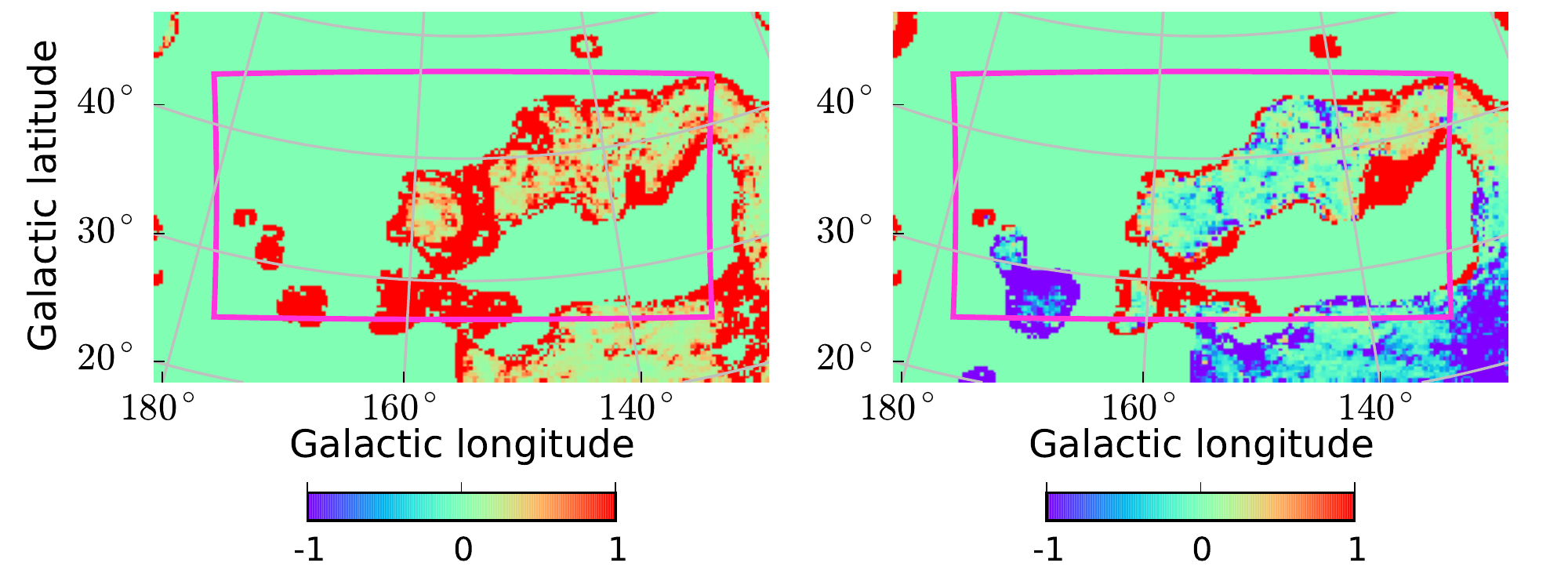}
\caption{Differences between alternative determinations of the DNM column 
density templates for ROI~A. Left: difference between the 
DNM template derived from $R$ assuming $\sigma \propto M$ 
minus the 
DNM template derived from $R$ assuming $\sigma \propto R$, 
divided by the former. Right: difference between the 
DNM template derived from $R$ assuming $\sigma \propto M$ minus the DNM
template derived from $\tmap$ assuming $\sigma \propto M$ (converted into 
$R$ by multiplying by $1.88 \times 10^{-2}$ W m$^{-2}$ sr$^{-1}$ mag$^{-1}$, the 
average of the ratios between non-zero pixels in the two maps), divided by the 
former. The magenta line shows the border of ROI~A where the 
dust fit was performed. The maps are 
shown in the \textit{plate carr\'ee} projection used for the dust and \g-ray fits in ROI~A.}\label{fig:dnmdiff}
\end{center}
\end{figure}

\subsection{$\xco$, Dust Specific Opacity and Power in the Local DNM from the 
\g-Ray Analysis}

The results in Section~\ref{analgamma} can constrain some properties of the local
ISM seen in the foreground of the HVCs and IVCs, in particular 
the $\xco=\nhd/\wco$ ratio in the Ursa Major molecular clouds, and the average specific power radiated by dust per H atom and unit solid 
angle, $\overline{R/N(\mathrm{H})}$, and the average specific opacity per 
H atom, $\overline{\tmap/N(\mathrm{H})}$ in the DNM for the other regions.   

These quantities can be extracted from the parameters in Eq.~\ref{eq:model} 
under the assumption that the same CR fluxes illuminate the atomic gas traced 
by $\hi$ and the molecular gas traced by CO or the gas in the DNM. Then, 
$\xco=x_\mathrm{CO}/2x_\mathrm{HI}$, and 
$\overline{R/N(\mathrm{H})}=x_\mathrm{HI}/x_\mathrm{DNM}$ (when the DNM is 
traced using $R$), or  
$\overline{\tmap/N(\mathrm{H})}=x_\mathrm{HI}/x_\mathrm{DNM}$ (when the DNM is 
traced using $\tmap$). In ROI~B we do not detect any significant \g-ray emission associated with the sparsely populated DNM templates, therefore we exclude it from the following discussion.

First, we check the assumption that the spectrum of CRs illuminating the 
different phases in the local gas complexes is the same in each ROI. 
Figure~\ref{fig:codnmspec} shows the $x_\mathrm{CO}$ and $x_\mathrm{DNM}$ 
coefficients as a function of energy from the analysis in energy bins 
in Section~\ref{cloudspectra}.
\begin{figure}\begin{center}
 \includegraphics[width=0.5\textwidth]{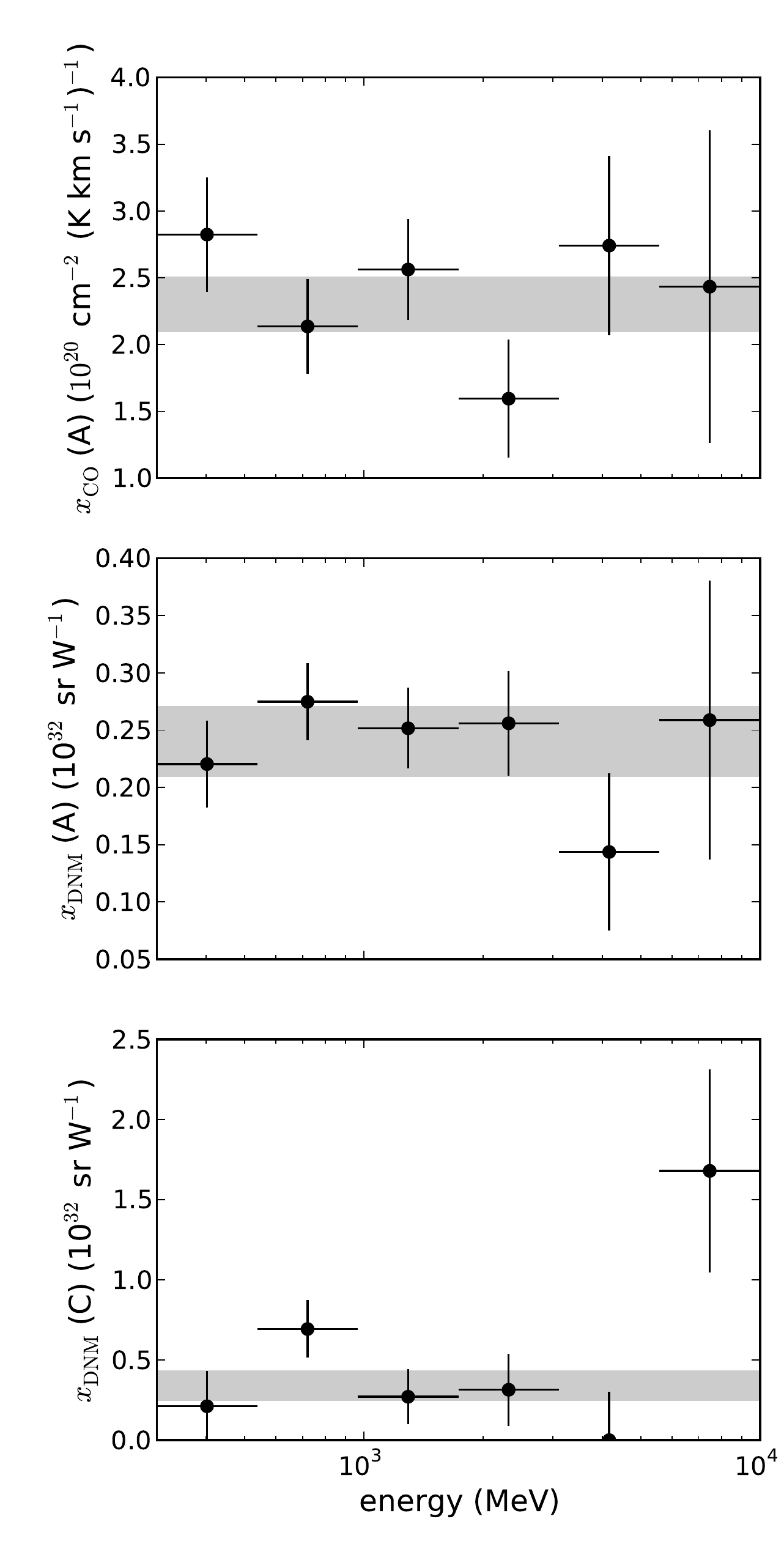}
 \caption{Best-fit values for $x_\mathrm{CO}$ and 
$x_\mathrm{DNM}$ (Eq.~\ref{eq:model}) from the analysis in 
multiple energy bins (see Sec.~\ref{cloudspectra}) in ROI~A and C. In each panel the points show the values as a 
function of energy with their statistical uncertainties. 
The shaded bands show the $\pm 1\sigma$ interval from the 
analysis over the entire energy range (see Sec.~\ref{basefit}).}\label{fig:codnmspec}
\end{center}\end{figure}
Within the large statistical uncertainties the spectra of the emissivities in 
each ROI are consistent with those of $\hi$ in Figure~\ref{fig:hiloc_spec}, all 
of them being consistent with the spectrum of gas in the local interstellar 
medium from \citet{casandjian2014}. There is no clear analog of the 
softening in Figure~\ref{fig:hiloc_spec} for ROI~A for either CO-traced gas or 
the DNM. However, the softening may be present and hidden by the large 
statistical uncertainties.

Given the fact that the spectra are consistent within the statistical 
uncertainties, we extract from the fit coefficients the properties {\chbis of the local ISM discussed above}.
We take into account 
the systematic uncertainties related to inputs to the interstellar emission 
model using the results of the analysis in Section~\ref{modelunc} {\changes and from the jackknife tests presented in Section~\ref{jackknife}. The ISM properties are reported in Table~\ref{gammafit:ism}. They are} similar to those inferred from LAT data for other nearby 
interstellar complexes \citep[e.g.,][]{planckfermi2014cham}, demonstrating that 
the modeling of the foregrounds in our analysis is robust. The $\xco$ ratio in 
the Ursa Major molecular clouds that we obtain is consistent with other 
estimates from alternative methods in the literature 
\citep[e.g.,][]{devries1987}. 
\begin{deluxetable}{lccc}
\tabletypesize{\scriptsize}
\tablecaption{Local ISM properties from the 
\g-ray analysis\label{gammafit:ism}}
\tablewidth{0pt}
\tablehead{
\colhead{}& \colhead{ROI A} & \colhead{ROI C}
}
\startdata
$\xco$\tablenotemark{a} & $1.13\pm0.08^{+0.20}_{-0.73} \pm 0.10 $ &  \nodata\\
$\overline{R/N(\mathrm{H})}$\tablenotemark{b} & $4.1\pm 0.3^{+0}_{-1.0}  \pm 0.3$ & $3.2\pm 0.3^{+0.7}_{-1.0} \pm 0.3$ \\ 
$\overline{\tmap/N(\mathrm{H})}$\tablenotemark{c} &  $2.06\pm 
0.06^{+0.11}_{-0.65}$ & \nodata \\
\enddata
\tablecomments{No CO emission is detected in ROI~C. The \g-ray scaling 
coefficients for the DNM maps derived from $\tmap$ in ROI~C were always consistent 
with 0.}
\tablecomments{The statistical uncertainties take into account the covariances 
of the fit parameters.}
\tablenotetext{a}{10$^{20}$ cm$^{-2}$ (K \kms)$^{-1}$}
\tablenotetext{b}{10$^{-32}$ W sr$^{-1}$ H$^{-1}$}
\tablenotetext{c}{10$^{-26}$ cm$^2$ H$^{-1}$}
\end{deluxetable}

\section{EVALUATION OF ERRORS AND UPPER LIMITS WITH THE JACKKNIFE METHOD}\label{jackknifeerrs}
{\changes
In this appendix we summarize the method used to evaluate uncertainties and upper limits from the jackknife tests in {\chbis Section}~\ref{jackknife}. Following \citet{dudewicz1988}, given the jackknife estimates $\theta_\imath$ with $\imath=1,\ldots,N$ let
\begin{equation}
\tilde{\theta}=\frac{1}{N} \sum_{\imath=1} ^N \theta_\imath
\end{equation}
be their arithmetic mean. A quantity known as $J_\imath$ is defined as
\begin{equation}
J_\imath = N\tilde{\theta}-(N-1)\theta_\imath. 
\end{equation}
If we define
\begin{equation}
J=\frac{1}{N} \sum_{\imath=1} ^N J_\imath
\end{equation}
the $1\sigma$ uncertainty of $\theta$ is
\begin{equation}
\sigma_\theta = \sqrt{\frac{1}{N(N-1)}\sum_{\imath=1}^N \left( J_\imath - J\right)^2}.
\end{equation}
The $(1-\alpha)$ c.l. upper limit on $\theta$ is
\begin{equation}
\mathrm{UL}=J+t_{\alpha,N-1}\sigma_\theta
\end{equation}
where $t_{\alpha,N-1}$ is the upper $\alpha^\mathrm{th}$ quantile of the Student's $t$-distribution with $N-1$ degrees of freedom.
}



\end{document}